%% file: jhep.tex
\documentclass[11pt,a4paper]{article}
\pdfoutput=1
\usepackage{jheppub}

\input{definitions}

\begin{document}
\begin{flushright}
Belle II Preprint 2024-013 \\
KEK Preprint 2024-7
\end{flushright}

\title{\input{title}}

\input{pub074-orcid}
\abstract{\input{abstract}}

\maketitle
\flushbottom
\input{body_v9}

\input{acknowledgements}

\bibliographystyle{JHEP}
\bibliography{reference_JHEP}
\newpage
\appendix
\input{07-appendix}

\end{document}

%% file: definitions.tex
\usepackage{graphicx} % Include figure files
\usepackage{epstopdf} % Allow to include eps files
\usepackage{dcolumn} % Align table columns on decimal point
\usepackage{xcolor} % Color support
\usepackage{amsmath,amssymb} % Extended symbol collection
\usepackage{hyperref} % Generate pdfs with hyperlinks
\usepackage[utf8]{inputenc} % Allow UTF8 characters
\usepackage[english]{babel} % All publications are in English
\usepackage[displaymath, mathlines]{lineno} % Line numbers for drafts
\usepackage{blindtext} % For testing
\usepackage{ifthen} % for conditional statements
\usepackage{orcidlink} % for ORCIDs in author list
\usepackage{multirow}
\usepackage{appendix}
\usepackage{longtable}
\graphicspath{{figures/}} % Use figures directory for figures

\newboolean{articletitles}
\setboolean{articletitles}{true} % False removes titles in references

\input{belle2-symbols}

\def\eetautau     {\ensuremath{e^+e^-\to\tau^+\tau^- (\gamma)}\xspace}

\def\taulks    {\ensuremath{\tau^{-} \to \ell^{-}K_{S}^{0}}\xspace}

\def\Mlk {\ensuremath{M(\ell K_S^0)}\xspace}
\def\dE {\ensuremath{\Delta E(\ell K_S^0)}\xspace}

%% file: title.tex
% WRITE THE TITLE IN THIS FILE.
Search for lepton-flavor-violating $\tau^- \to \ell^- K_s^0$ decays at Belle and Belle II 

%% file: pub074-orcid.tex
%%% Paper:    tau to ell KS
%%% Journal:  JHEP
%%% Contacts: K. Lautenbach, L. Zani, J. Serrano
%%% ====================================================================
%%% Use \input{pub074-orcid} to insert this material into your latex file.
\collaboration{The Belle and Belle II Collaborations}
  \author{I.~Adachi\,\orcidlink{0000-0003-2287-0173},} % 2590
% \author{K.~Adamczyk\,\orcidlink{0000-0001-6208-0876},} % 2239
  \author{L.~Aggarwal\,\orcidlink{0000-0002-0909-7537},} % 10083
% \author{P.~Ahlburg\,\orcidlink{0000-0002-9832-7604},} % 2367
  \author{H.~Ahmed\,\orcidlink{0000-0003-3976-7498},} % 11323
% \author{J.~K.~Ahn\,\orcidlink{0000-0002-5795-2243},} % 7423
  \author{Y.~Ahn\,\orcidlink{0000-0001-6820-0576},} % 14363
  \author{H.~Aihara\,\orcidlink{0000-0002-1907-5964},} % 2223
  \author{N.~Akopov\,\orcidlink{0000-0002-4425-2096},} % 9443
  \author{S.~Alghamdi\,\orcidlink{0000-0001-7609-112X},} % 27804
  \author{M.~Alhakami\,\orcidlink{0000-0002-2234-8628},} % 28103
  \author{A.~Aloisio\,\orcidlink{0000-0002-3883-6693},} % 2194
% \author{S.~Al~Said\,\orcidlink{0000-0002-4895-3869},} % 6823
  \author{N.~Althubiti\,\orcidlink{0000-0003-1513-0409},} % 21524
% ...
% 13403
% \author{S.~Zou\,\orcidlink{0000-0003-3377-7222}} % 19363

%% file: abstract.tex
% WRITE THE ABSTRACT IN THIS FILE.
We present the results of a search for charged-lepton-flavor violating decays 
$\tau^{-} \rightarrow \ell^{-}K_{S}^{0}$ \footnote{Charge conjugation is implied throughout this document.}, where $\ell^{-}$ is either an electron or a muon.
%produced in association  with a neutral $K_S^0$ meson. 
%The analysis is performed with an exclusive $3\times 1$ prong strategy, reconstructing the 
%signal-side $\tau$ with 3 tracks and tag-side $\tau$ with one track, using a boosted decision tree.
We combine $e^+e^-$ data samples recorded by the Belle II experiment at the SuperKEKB collider (428\invfb) with samples recorded by the Belle experiment at the KEKB collider (980 \invfb) to obtain a sample of 1.3 billion $e^+e^-\to\tau^+\tau^-$ events. We observe 0 and 1 events and set $90\%$ confidence level upper 
limits of 
$0.8 \times  10^{-8}$ and $1.2 \times  10^{-8}$ on the branching fractions of the decay modes 
$\tau^{-} \rightarrow e^{-}K_{S}^{0}$ and $\tau^{-} \rightarrow \mu^{-}K_{S}^{0}$, respectively. These are the most stringent upper limits to date. 

%% file: body_v9.tex
\input{01-intro}

\input{02-belleII}
\input{03-selection}

\input{04-sys_v9}

\input{05-UL}
\input{06-summ}

%\input{body_october10}

%% file: 01-intro.tex
\section{Introduction}
\label{sec:intro}

The conservation of charged-lepton flavor is built into the Standard Model (SM) at tree level.
The observation of neutrino oscillations implies that neutrinos have mass, so charged-lepton-flavor violation (LFV) can occur via internal loops, manifesting in processes such as $\mu\to e$, $\tau\to e$, and $\tau\to \mu$ conversions. All LFV amplitudes are suppressed by the squared ratio of the neutrino mass to the $W$-boson mass 
$(m_{\nu}/m_W)^2$. 
Consequently, predicted branching fractions are of the order of $10^{-50}$~\cite{ref:SM_cLFV, ref:tau_lfv, Blackstone:2019njl}, well below the 
sensitivities of current experiments. The detection of charged LFV decays would unequivocally indicate physics beyond the SM.
Due to the large mass of the $\tau$ lepton, a wide range of possible LFV $\tau$ decays can be probed. In particular, in an effective field theory approach, the decays $\tau\to\ell M$, where $M$ is a meson, can be used to constrain different types of operators \cite{Petrov:2013vka, Husek:2020fru, Husek:2021isa}, for example the two-lepton- and two-quark-operators that can arise in leptoquark models
\cite{Carpentier:2010ue}. 

Over the past four decades, experiments such as CLEO at CESR and the first-generation $B$-factory experiments, BaBar %\cite{PhysRevD.79.012004} 
at SLAC and Belle at KEK, have tested
LFV in $\tau$-lepton decays \cite{ref:hflav}. A total of 52 LFV $\tau$ decay modes involving neutrinoless two-body and three-body final states have been investigated. 
The most stringent limits on the decays \taulks, where $\ell=e,\mu$, were obtained by the Belle collaboration, which found branching fraction upper limits at 90~\% confidence level (C.L.) of $2.6\times10^{-8}$ for the electron mode and $2.3\times10^{-8}$ for the muon mode 
using a 671\invfb data sample, or $617$ million $e^+e^-\to\tau^+\tau^-$ events~\cite{Belle:2010rxj}.

We present the results of a search for the LFV decays \taulks using a combined data sample of 1.3 billion $e^+e^-\to\tau^+\tau^-$ events recorded with 
the Belle ($980~$fb$^{-1}$) and $\belletwo{}$ ($428~$fb$^{-1}$) detectors at the asymmetric-energy \epem{} KEKB and SuperKEKB colliders~\cite{ABASHIAN2002117, Akai:2018mbz}. 
Candidate \taulks decays are selected in events where the second $\tau$ (\emph{tag}) is reconstructed in a one-prong (single charged track) topology.
The background rejection is optimized separately for Belle and $\belletwo{}$ datasets, and for each tag-side category, electronic, muonic or pionic.  The optimization, which does not use data in the kinematic region where signal events are expected to peak, places loose requirements on kinematic and global-event variables and refines the selection using a boosted decision tree (BDT).
The expected background yield is obtained from a fit to the reconstructed $\tau$-mass sidebands, and an upper limit on the branching fraction is obtained using a frequentist method. The rest of this paper is structured as follows.
Section~\ref{sec:BelleII} provides an 
overview of the Belle (II) detector and the data samples utilized. Section~\ref{sec:selection} outlines the candidate reconstruction and selection process, and
Section~\ref{sec:systematics} addresses the systematic uncertainties.  Section~\ref{sec:results} presents the branching fraction measurement and limit computation, and Section~\ref{sec:summ} summarizes the results.

%% file: 02-belleII.tex
\section{The Belle and Belle II detectors, simulation and data samples}
\label{sec:BelleII}

The Belle~II experiment is located at SuperKEKB~\cite{Akai:2018mbz},
which collides electrons and positrons near the $\Upsilon(4S)$
resonance. The $\belletwo{}$
detector~\cite{Abe:2010gxa} has a cylindrical geometry and includes a
two-layer silicon-pixel detector~(PXD) surrounded by a four-layer
double-sided silicon-strip detector~(SVD)~\cite{Belle-IISVD:2022upf}
and a 56-layer central drift chamber~(CDC). These detectors
reconstruct the trajectories (tracks) of charged particles and provide energy loss measurements.
Only one sixth of the second
layer of the PXD was installed for the data analyzed here.   Surrounding the CDC are a
time-of-propagation detector~(TOP)~\cite{Kotchetkov:2018qzw} in the
central region and an aerogel-based ring-imaging Cherenkov
detector~(ARICH) in the forward region, which corresponds to the electron beam direction.  These detectors provide information for
identifying charged particles.  Surrounding the TOP and ARICH is an
electromagnetic calorimeter~(ECL) based on CsI(Tl) crystals that
%primarily
provides energy and timing measurements, primarily for photons and
electrons. Outside the ECL is a superconducting solenoid
magnet that provides a 1.5~T axial field. Its flux return is instrumented with resistive-plate chambers
and plastic scintillator modules to detect muons, $K^0_L$ mesons, and
neutrons. 
The symmetry axis of the magnet, which is almost coincident with the direction of the electron beam, is used to
define the $z$ axis.
The Belle detector was located at the interaction point of the KEKB collider~\cite{Kurokawa:2001nw}.  It shares a similar 
structure to $\belletwo{}$ but lacks a silicon pixel detector and uses aerogel threshold
Cherenkov counters (ACC) and a barrel-like arrangement of time-of-flight scintillation counters (TOF) for particle identification.
Charged particle trajectories are reconstructed using the Belle SVD and CDC. A detailed description of the Belle detector can be found in Ref.~\cite{ABASHIAN2002117}.
The Belle dataset used in this search was recorded between 2000 and 2010, 
and comprises $711~$fb$^{-1}$ collected at the $\Upsilon(4S)$ resonance, $121~$fb$^{-1}$ at the $\Upsilon(5S)$ resonance, $89~$fb$^{-1}$ recorded $60~$MeV below 
the $\Upsilon(4S)$ (off-resonance), $28~$fb$^{-1}$ in energy scans above the $\Upsilon(4S)$ resonance, and the remainder at and near the $\Upsilon(1S, 2S, 3S)$ resonances. The $\belletwo{}$
dataset, recorded between 2019 and 2022, includes $366~$fb$^{-1}$ at the $\Upsilon(4S)$, $42~$fb$^{-1}$ $60~$\mev below it, and  $19~$fb$^{-1}$ from a scan around a center-of-mass energy of 
$10.75~$GeV. 
The resulting combined data sample has an integrated luminosity of  $1408~$fb$^{-1}$, corresponding to 1.3 billion
$e^+e^-\to\tau^+\tau^-$ events ~\cite{Banerjee:2007is}.

Monte-Carlo (MC) simulated events are used to optimize the selection and background rejection and to measure the signal efficiency.
To study the signal process in Belle ($\belletwo{}$), we use 400 thousand (1 million) \eetautau events, where both initial- and final-state photon radiation is included. The signal tau decays via a phase space model to an electron or muon and a
$K_S^0$ and the other tau decays to a SM-allowed decay.
The potential background processes studied using simulation include
$\epem \to \qqbar$ events, where $q$ indicates a $u$, $d$, $c$, or $s$ quark;
 $\epem \to b\Bar{b}$ events; 
$\epem\to \ell^+ \ell^- (\gamma)$, where $\ell=e,\mu$; the two-photon processes
$\epem \to e^+e^-+$hadrons,   where the hadrons can be $h^+h^-$ ($h = \pi, K$ or proton) or the fragmentation products from a $q\bar q$ pair;
and four-lepton processes: $\epem \to e^+e^-e^+e^-$,\ $\mu^+\mu^-\mu^+\mu^-$,\ $\mu^+\mu^-e^+e^-$,\ $e^+e^-\tau^+\tau^-$, $\mu^+\mu^-\tau^+\tau^- $. 
A full list of simulated processes used for Belle and $\belletwo{}$ can be found in Table~\ref{tab:simMC} in Appendix~\ref{sec:appA}.
We simulate the \eetautau{} process using the KKMC generator~\cite{Jadach:1999vf, Jadach_2000}, with subsequent tau decays  simulated by the TAUOLA~\cite{Jadach:1990mz} package, with final state radiation (FSR) added
by the PHOTOS~\cite{Barberio:1990ms} package. KKMC is also used to simulate $\mu^+\mu^-(\gamma)$ and \qqbar production.
Fragmentation of \qqbar pairs is simulated using the PYTHIA~\cite{Sjostrand:2014zea} package.
 For the production and decay of $\epem \to b\Bar{b}$ events, we use
PYTHIA interfaced with the EvtGen~\cite{Lange:2001uf} generator.
Different versions of the same generator packages are deployed for Belle and Belle II, which results in different samples for the simulated background processes for the two experiments and therefore different starting points for selection optimization.
In Belle (\belletwo{}) the BHLUMI~\cite{Jadach:1991by} (BabaYaga@NLO~\cite{Balossini:2006wc, Balossini:2008xr, CarloniCalame:2003yt, CarloniCalame:2001ny,CarloniCalame:2000pz}) generator is used to
simulate $\epem \to \epem (\gamma)$ events. 
Two-photon processes are simulated using the
AAFH~\cite{BERENDS1985421,BERENDS1985441,BERENDS1986285} and
TREPS~\cite{Uehara:1996bgt} packages.

The \belletwo{} analysis software framework (basf2)~\cite{Kuhr:2018lps, basf2-zenodo} uses the GEANT4~\cite{Agostinelli:2002hh} package to simulate the detector response to  particles traversing the active volume.
For the simulation of the detector response in Belle, GEANT3~\cite{Brun:1994aa} is used. Belle collision and simulation data are converted into the Belle II format for basf2 compatibility using the B2BII framework 
~\cite{Gelb:2018agf}.
The online event selection (hardware trigger) for Belle and \belletwo{} data is based on the energy deposits (clusters) and their topologies in the ECL,  or on an independent trigger selection based on the number of charged particles reconstructed in the CDC.
Most of the events are selected by requiring a total ECL energy larger than 1~GeV and a topology incompatible with Bhabha events.

%% file: 03-selection.tex
\section{Event selection and background rejection}
\label{sec:selection}

\subsection{Candidate reconstruction and event selection}
\label{sec:cand_selection}
We search for  $e^+e^-\to\tau^+\tau^-$ events where one tau decays into the LFV channel  \taulks{} and the other into a one-prong final state.
In the center-of-mass (c.m.) frame of the electron-positron collision, the $\tau$ leptons are emitted in opposite directions, with the decay 
products of each $\tau$ confined to opposite hemispheres. The partition into hemispheres is delineated by the plane perpendicular to the estimated flight axis of the $\tau$ pair, which is determined experimentally as the direction $\mathbf{\hat{t}}$ that maximizes the thrust value:
\begin{equation}
\label{eq:thrust}
    T = \max_{\mathbf{\hat{t}}} \left(\dfrac{\sum_{i} \left|\mathbf{p^*}_i \cdot \mathbf{\hat{t}}\right|}{\sum_{i} \left|\mathbf{p^*}_i\right|} \right),
\end{equation}
where $\mathbf{p}_i^*$ represents the momentum of the final state particle $i$ in the \epem{} c.m. frame~\cite{Brandt:1964sa,Farhi:1977sg}, including 
both charged and neutral particles. Quantities in the \epem{} c.m. frame are marked with an asterisk throughout this paper.

We designate the signal hemisphere as the one containing the \taulks{} decay candidate, formed by combining a lepton with a $K_S^0$.
We reconstruct $K_S^0\to\pi^+\pi^-$ candidates using two oppositely charged tracks assumed to be pions. The invariant mass of the
$K_S^0$ candidates must satisfy $0.45 < M_{\pi^+\pi^-} < 0.55~$GeV/c$^2$.
The pions are then fit to a common vertex,
%after which the same requirement on the mass is applied
%which is also required by the applied vertex fit, 
and the significance of the distance between the pion vertex and the interaction point,
defined as the flight length divided by its uncertainty, must exceed 3.
%Additionally, they are selected by imposing requirements on the impact parameter of the
%pions and the azimutal angle between the flight direction and  $K_S^0$ momentum.

Lepton candidates, as well as all remaining charged particles
not used for $K_S^0$ reconstruction, % taken as pions,
are required to have a transverse momentum $p_T>0.1~$\gevcc and must 
originate within 3~\cm along the $z$ axis and 1~\cm in the transverse plane from the $e^{+}e^{-}$ interaction point.
Charged particles not identified as electrons or muons are assumed to be pions.
Muons in $\belletwo{}$ are identified using the discriminator 
$\mathcal{P}_{\mu} = {\cal L}_{\mu} / ({\cal L}_{e} + {\cal L}_{\mu} + {\cal L}_{\pi} + {\cal L}_{K} + {\cal L}_{p} + {\cal L}_{d})$, where
the likelihoods ${\cal L}_i$ for each charged-particle hypothesis ($i = e, \mu, \pi, K$, proton or deuteron) combine particle-identification information 
from CDC, TOP, ARICH, ECL, and KLM subdetectors. For $\belletwo{}$ electrons, the output of a classifier based on 
a BDT, $\mathcal{P}_e$,  is used. 
%This method has demonstrated enhanced performance when compared to the purely likelihood-based approach, especially in distinguishing between pions and electrons.
The BDT incorporates likelihoods from individual subdetectors and supplementary ECL observables, including variables that are sensitive to shower development \cite{Milesi:2020esq}.
For $\belletwo{}$, we retain electrons and muons with $\mathcal{P}_{e,\mu}>0.95$, which correspond to identification efficiencies of 
$96.3\,\%$ for electrons and $91.4\,\%$ for muons. The corresponding pion misidentification probabilities in $\belletwo{}$ are 
$0.3\,\%$  and $2.9\%$, respectively.
The muon identification in Belle uses information from the KLM and extrapolated tracks to form a likelihood-based discriminator
$\mathcal{P}^\prime_{\mu} = {\cal L}_{\mu} / ({\cal L}_{\mu} + {\cal L}_{\pi} + {\cal L}_{K})$, \cite{ABASHIAN200269}.
Electrons in Belle are identified using a likelihood ratio $\mathcal{P}^\prime_{e} = {\cal L}_{e} / ({\cal L}_{\mu} + {\cal L}_{non-e})$\cite{Hanagaki_2002} based on information from the CDC, ACC and ECL.
We retain candidates with $\mathcal{P}^\prime_{e,\mu}>0.9$, resulting in identification efficiencies in Belle of $93.5\,\%$ for electrons 
and $88.6\,\%$ for muons and corresponding pion misidentification rates of $0.5\,\%$ for electrons and $2.7\,\%$ for muons.

The vertices of $\tau$ and $K_S^0$ candidate are fitted with the TreeFitter tool~\cite{Krohn:2019dlq}, which updates the momenta of the reconstructed parent particles and the vertex positions of the individual tracks in the fit.  Candidates with a vertex fit $\chi^2 > 0$ and a $K_S^0$ candidate mass $0.45 < M_{\pi^+\pi^-} < 0.55~$GeV/c$^2$  are selected.

The tag tau is reconstructed as an electron, muon or pion in the hemisphere opposite to the signal tau candidate. The total number of tracks in the event must be four, and their assigned charges must sum to zero.

In addition to the signal and tag tau reconstruction, all charged and neutral particles in the events are used to compute event-based observables that can be used to reduce the backgrounds.
Photons are reconstructed from ECL clusters within the CDC acceptance with no tracks in the vicinity of the cluster. For \piz reconstruction,  photons that leave an energy deposit of at least 0.1~\gev are combined in pairs and required to have an invariant mass in the range $0.115 < M_{\gamma\gamma}< 0.152$~\gevcc, which corresponds to approximately $\pm 2.5$ units of resolution about the known $\pi^0$ mass~\cite{ParticleDataGroup:2022pth}. Photons that contribute to the reconstructed $\pi^0$ candidates, as well as all photons with energies exceeding 0.1 GeV and all tracks, are used to calculate variables related to event kinematics, such as the missing momentum, missing mass, or the thrust axis. 

%For electrons, energy loss via bremsstrahlung is accounted for and
%corrected by adding back to their energy all the photons within 8.6 degrees and with an
%energy higher than 20 MeV. 

A dedicated correction for electron bremsstrahlung energy loss is applied during reconstruction.  All photons with energies $E_{\gamma}>0.02~$GeV within a cone of 0.05 (0.15) radians around the direction of the electron momentum for Belle ($\belletwo{}$) data are added to the electron energy.

\begin{figure}[htb]
    \centering
    \includegraphics[width=0.49\columnwidth]{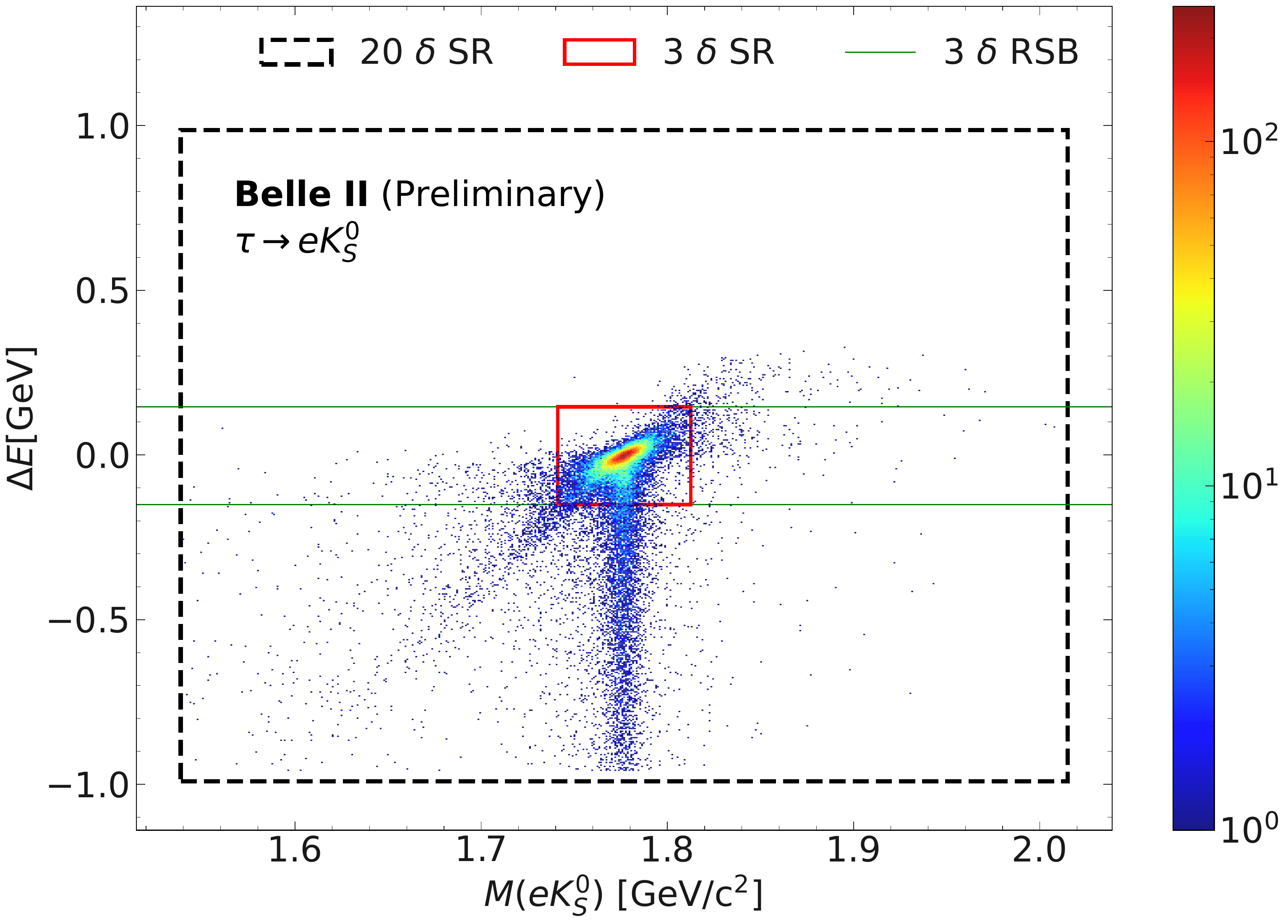}
    \includegraphics[width=0.49\columnwidth]{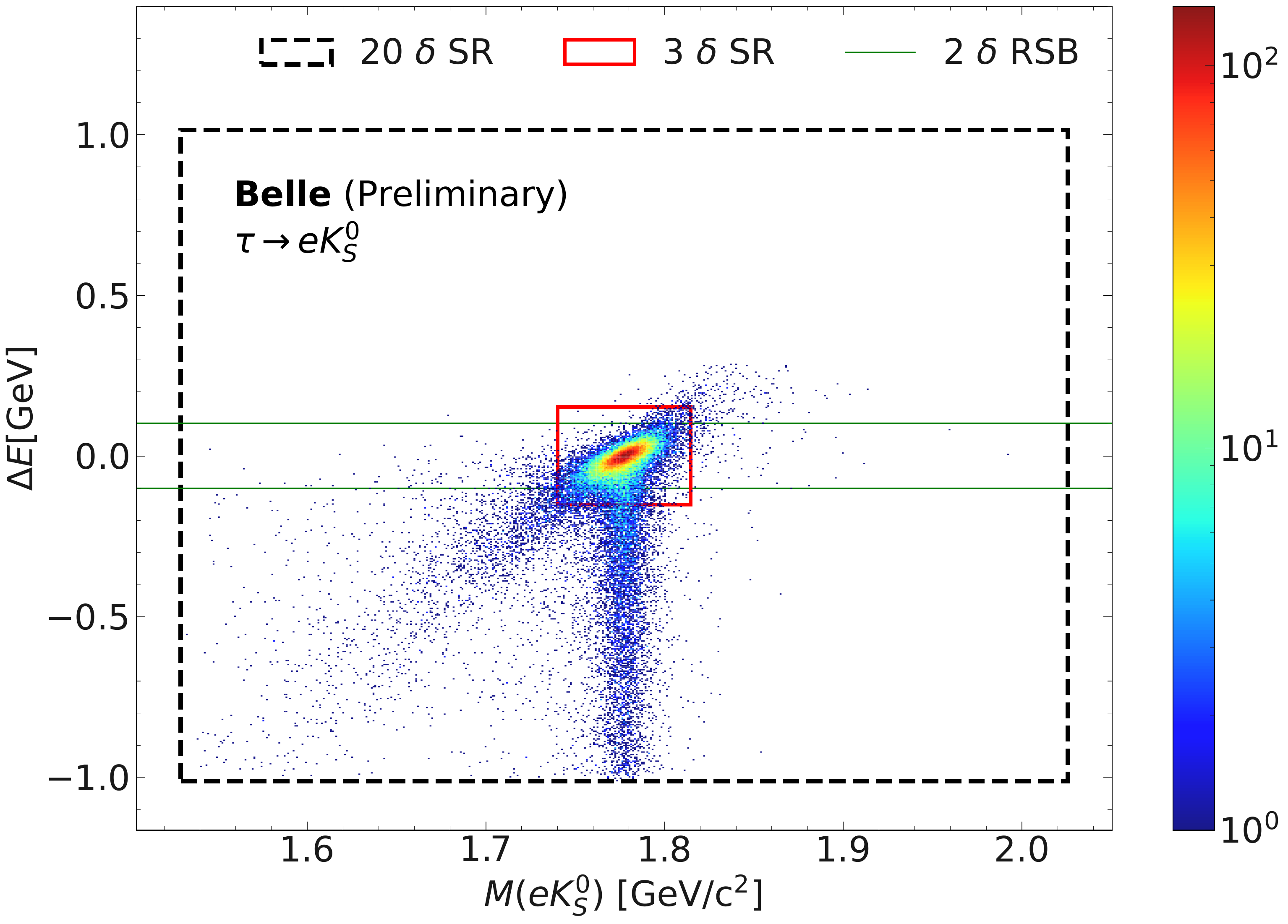}\\
    
    \includegraphics[width=0.49\columnwidth]{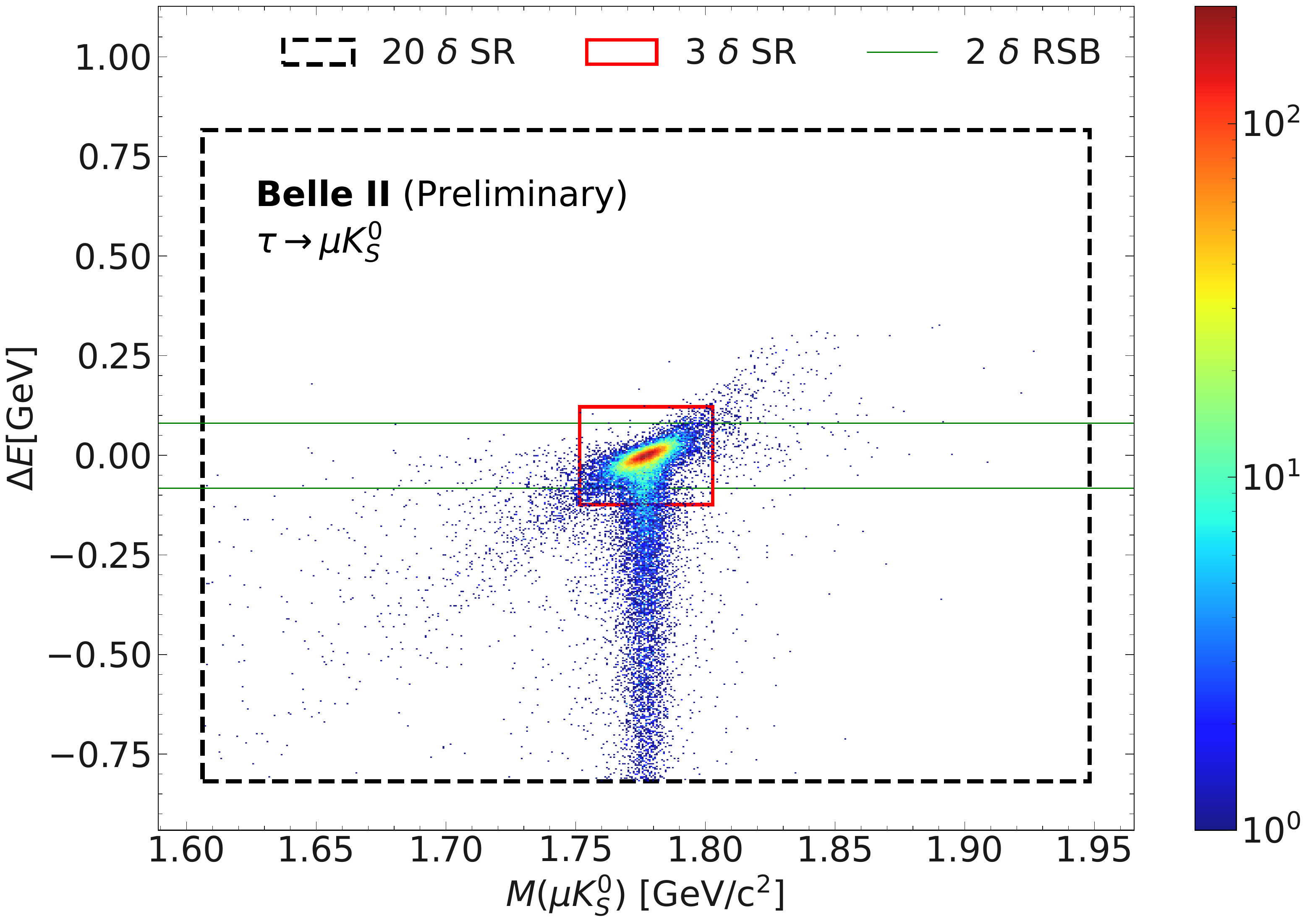}
    \includegraphics[width=0.49\columnwidth]{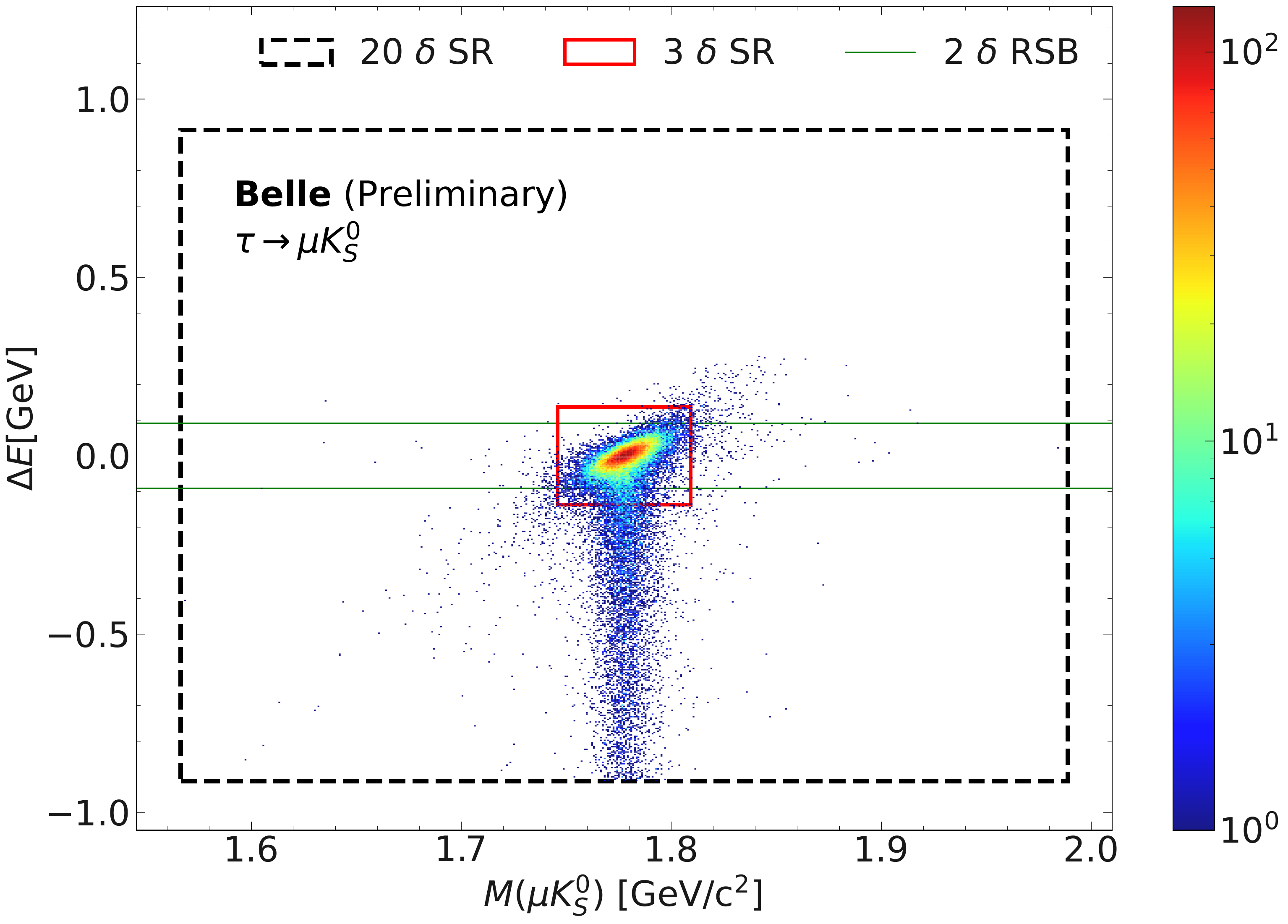}
    \caption{Distributions of simulated $\tau^-\rightarrow e^-K_S^0$ events for $\belletwo{}$ (left) and $\tau^-\rightarrow e^-K_S^0$ for Belle (right) in the (\Mlk, \dE) plane. The lower row shows the $\tau^-\rightarrow \mu^- K_S^0$ channel. The large, dashed black box represents the $20\,\delta$ region, the small red rectangle is the SR and the green bands define the RSB.}
    \label{fig:signalRegion}
\end{figure}

Since  the \taulks{} decay is a neutrinoless process, the invariant mass \Mlk of the reconstructed $\tau$ decay products should
coincide with the $\tau$ lepton mass~\cite{ParticleDataGroup:2022pth},
except for decays affected by final state radiation (FSR).
In the c.m.\ system, the $\tau$ energy $E^*_{\tau}$ should  be half 
of the \epem energy, $\sqrt{s}/2$, apart from corrections due to initial state radiation (ISR) from the $e^\pm$ beams and FSR.
%, considering initial state radiation (ISR) from the $e^\pm$ beams and FSR effects.
Thus, the energy difference 
$\dE = E^*_{\tau} - \sqrt{s}/2  $ should be near zero. 
%Leveraging these characteristic features, we define the signal region and refine selection criteria to enhance signal efficiency while mitigating background contributions.
The distribution of signal candidates in the (\Mlk, \dE) plane 
(see Fig.~\ref{fig:signalRegion}) is broadened by detector resolution and radiative effects. Initial state photon emission yields a tail towards lower \dE values,  
while  photons from FSR create a diagonal band, primarily at lower \Mlk and \dE values.

The analysis is performed in the (\Mlk, \dE) plane.  We define several rectangular regions for use in optimizing the selection. 
These rectangular boxes are centered around  the expected signal peak in the (\Mlk, \dE) plane, with side lengths proportional to their corresponding resolutions, $\delta$.
For each variable, $\delta$ is approximated as the standard deviation of the sum of two Gaussians and a Crystal Ball function~\cite{Skwarnicki:1986xj} fitted 
to the simulated signal distribution. The fitted means and resolutions are shown in Table~\ref{tab:sig_resol}.
\begin{table}[htbp]
    \centering
     \caption{Fitted means and resolutions for \Mlk and \dE for both signal channels on $\belletwo{}$ and Belle simulation.}
\begin{tabular}{cc|cc|cc}

                                                   &            & \multicolumn{2}{c}{Belle II}                                             & \multicolumn{2}{c}{Belle}                                          \\  
                                                   &            & \multicolumn{1}{c}{mean}             & \multicolumn{1}{c}{$\delta$} & \multicolumn{1}{c}{mean}             & \multicolumn{1}{c}{$\delta$} \\ \hline
\multicolumn{1}{c|}{\multirow{2}{*}{$eK_S^0$}}    & $M$(\mevcc)        & \multicolumn{1}{c|}{$1776.79\pm 0.36$} & $11.91\pm 0.82$               & \multicolumn{1}{c|}{$1777.39\pm 1.62$} & $12.41\pm 0.16$               \\ %\cline{2-6} 
\multicolumn{1}{c|}{}                             & $\Delta E$(\mev) & \multicolumn{1}{c|}{$-2.25\pm 0.20$}    & $49.43\pm 1.12$               & \multicolumn{1}{c|}{$-1.25\pm 0.22$}   & $50.68\pm 2.57$               \\ \hline
\multicolumn{1}{c|}{\multirow{2}{*}{$\mu K_S^0$}} & $M$(\mevcc)        & \multicolumn{1}{c|}{$1777.15\pm 0.22$} & $8.55\pm 0.95$                & \multicolumn{1}{c|}{$1777.75\pm 0.80$}  & $10.56\pm 3.07$               \\ %\cline{2-6} 
\multicolumn{1}{c|}{}                             & $\Delta E$ (\mev)& \multicolumn{1}{c|}{$-1.05\pm 0.17$}   & $40.87\pm 0.91$               & \multicolumn{1}{c|}{$0.55\pm 2.39$}    & $45.64\pm 2.51$                
\end{tabular}   
    \label{tab:sig_resol}
\end{table}

Only events that lie within a rectangular region of width $\pm 20\delta$ are retained for the optimization.
%For the optimization of background rejection, a rectangular region of width $\pm 20\delta$ is used. All reconstructed events falling outside this box in 
%the (\Mlk, \dE) plane are rejected.
The final yield extraction is carried out within an elliptical signal region (SR) whose orientation and size is optimized for signal efficiency and 
background rejection. The optimized widths used for the final signal yields are $2\delta$ for both semi-axes, for all channels, except for the Belle II electron channel, whose major semi-axis is $3\delta$ wide (described in Section~\ref{sec:bdt}).
% is optimized along with the  BDT classifier selection.
Events falling inside this elliptical SR in the data are masked during the selection optimization to avoid experimental bias.

A sideband region (SB), corresponding to the events falling within the $\pm 20\,\delta$ rectangle but outside the $\pm 3\,\delta$ rectangle
% $\pm[3-20]\,\delta$ range
in the (\Mlk, \dE) plane, is defined to validate the background rejection.
The final number of expected background events 
is extracted from a fit to the  data in the reduced sideband (RSB), defined as the region within $\pm 3 \delta$ ($\pm 2 \delta$)
of the \dE peak for the Belle II electron mode (the other signal modes).
A larger RSB is chosen for the $\belletwo{}$ electron mode since it has lower expected background levels than the other modes.
The data yields in the SB and RSB are consistent with the expected yields in simulation.
These regions are shown for the Belle and $\belletwo{}$ electron channels in Fig.~\ref{fig:signalRegion}.

After the candidate reconstruction, the most important backgrounds for the electron channel arise from low-multiplicity processes such as $e^+e^-\rightarrow e^+e^-(\gamma)$, $e^+e^-e^+e^-$, $e^+e^-\mu^+\mu^-$, $e^+e^-h^+h^-$ and $\mu^+\mu^-(\gamma)$, while the background for the muon channel is dominated by $e^+e^-\rightarrow q\bar{q}$ processes.

\begin{figure}[h!]
    \centering    
        \includegraphics[width=0.47\textwidth,page=9]{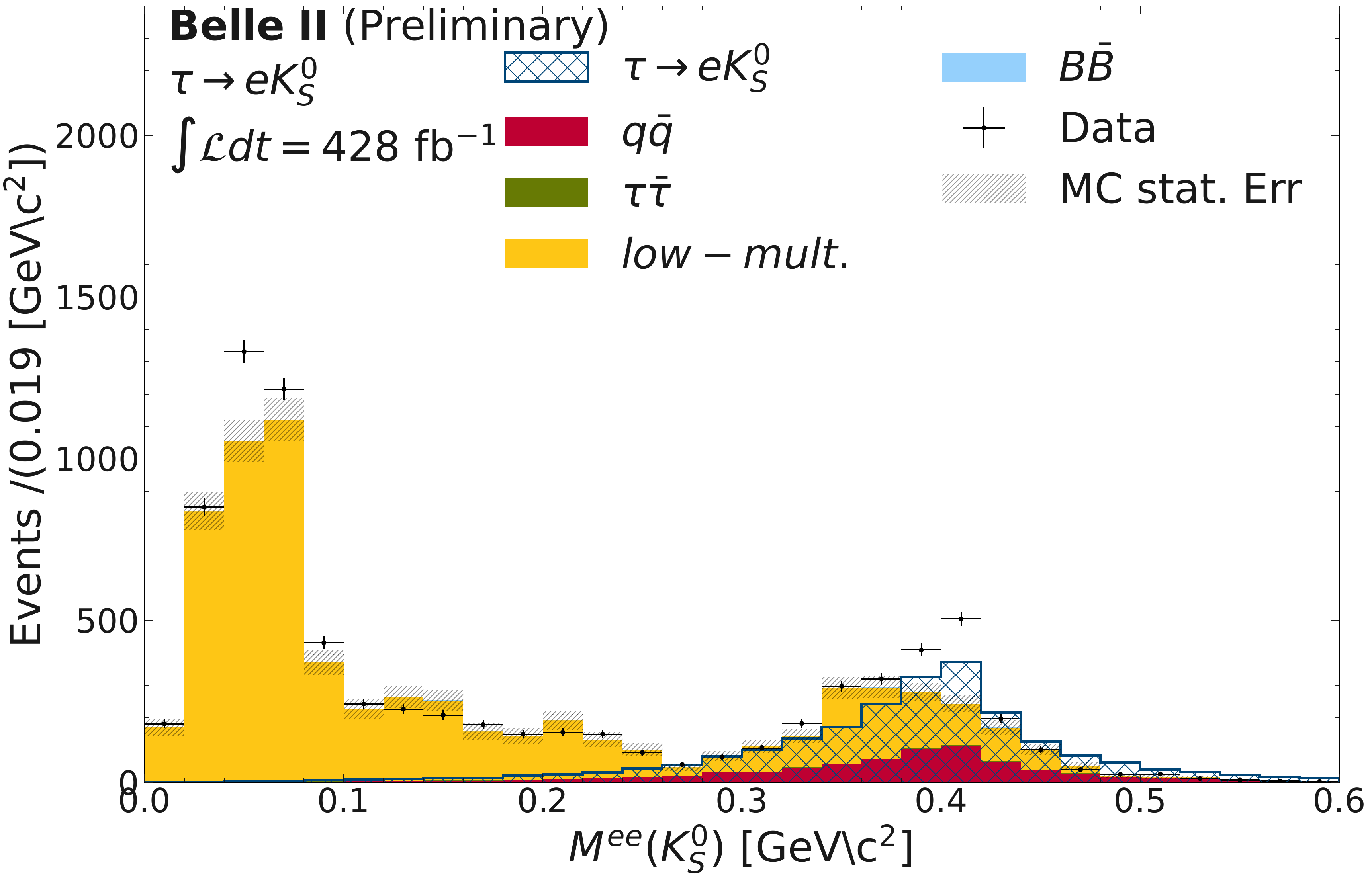}
        \includegraphics[width=0.47\textwidth,page=9]{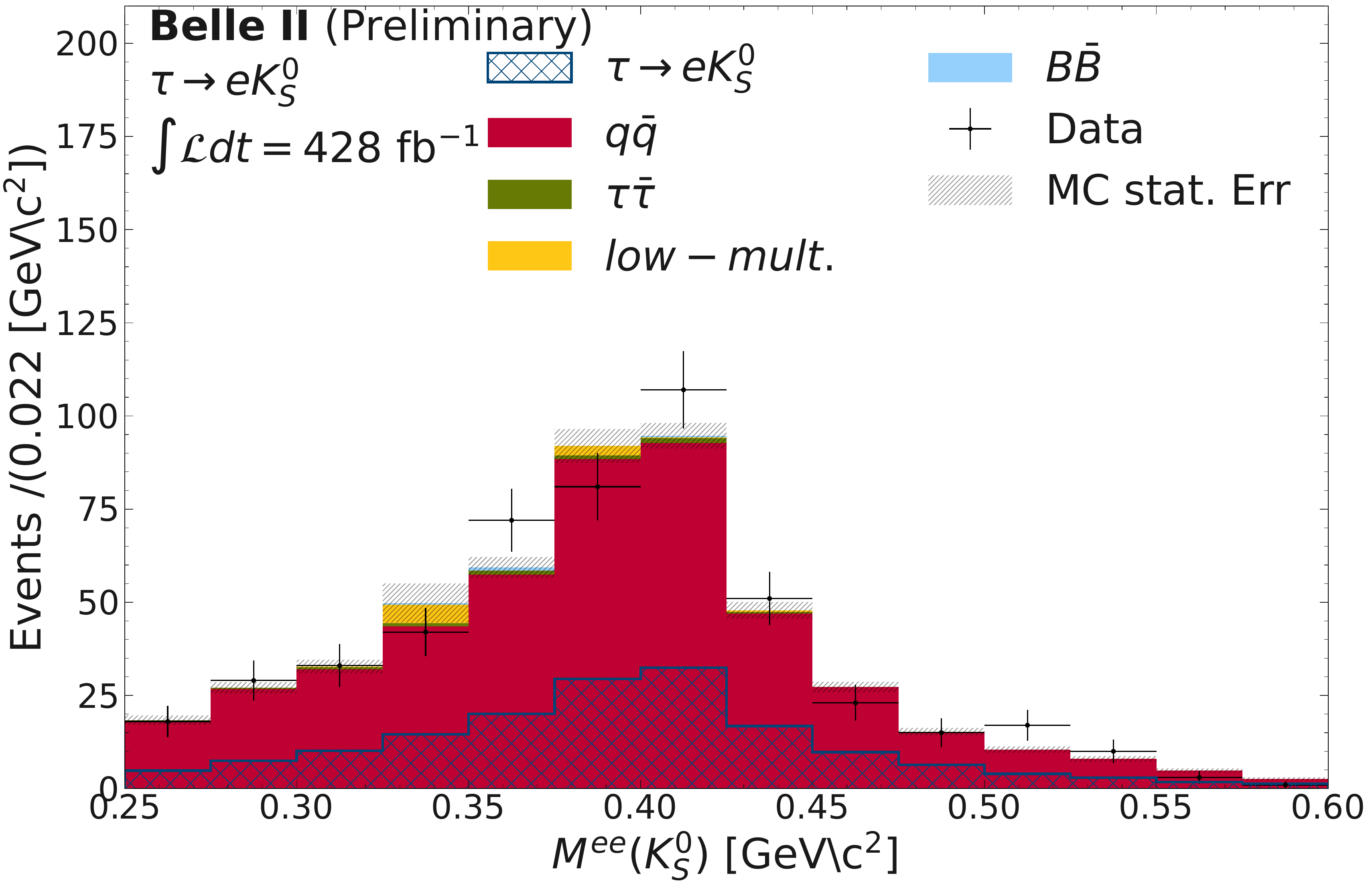}\\

        \includegraphics[width=0.46\textwidth,page=9]{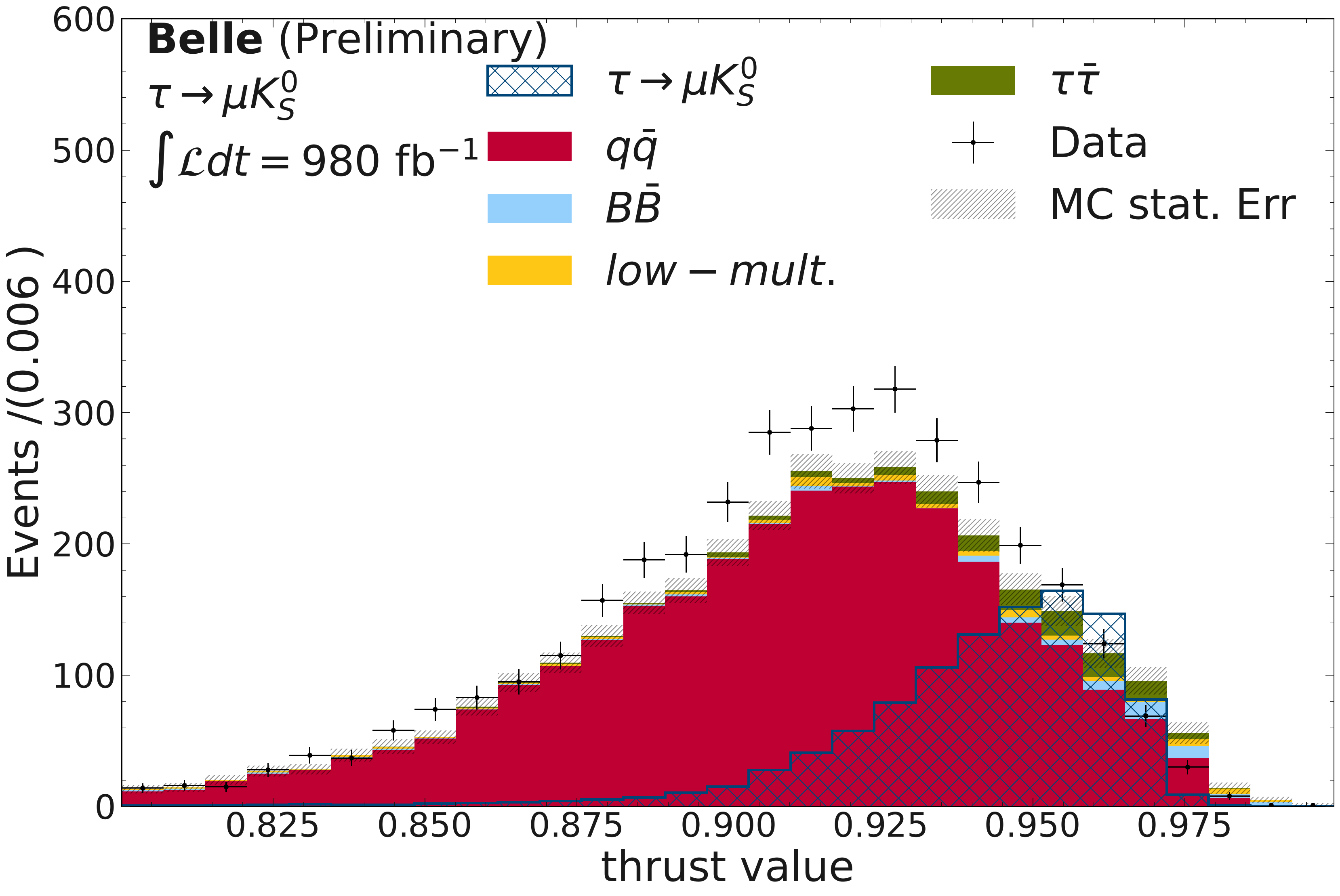}
        \includegraphics[width=0.46\textwidth,page=9]{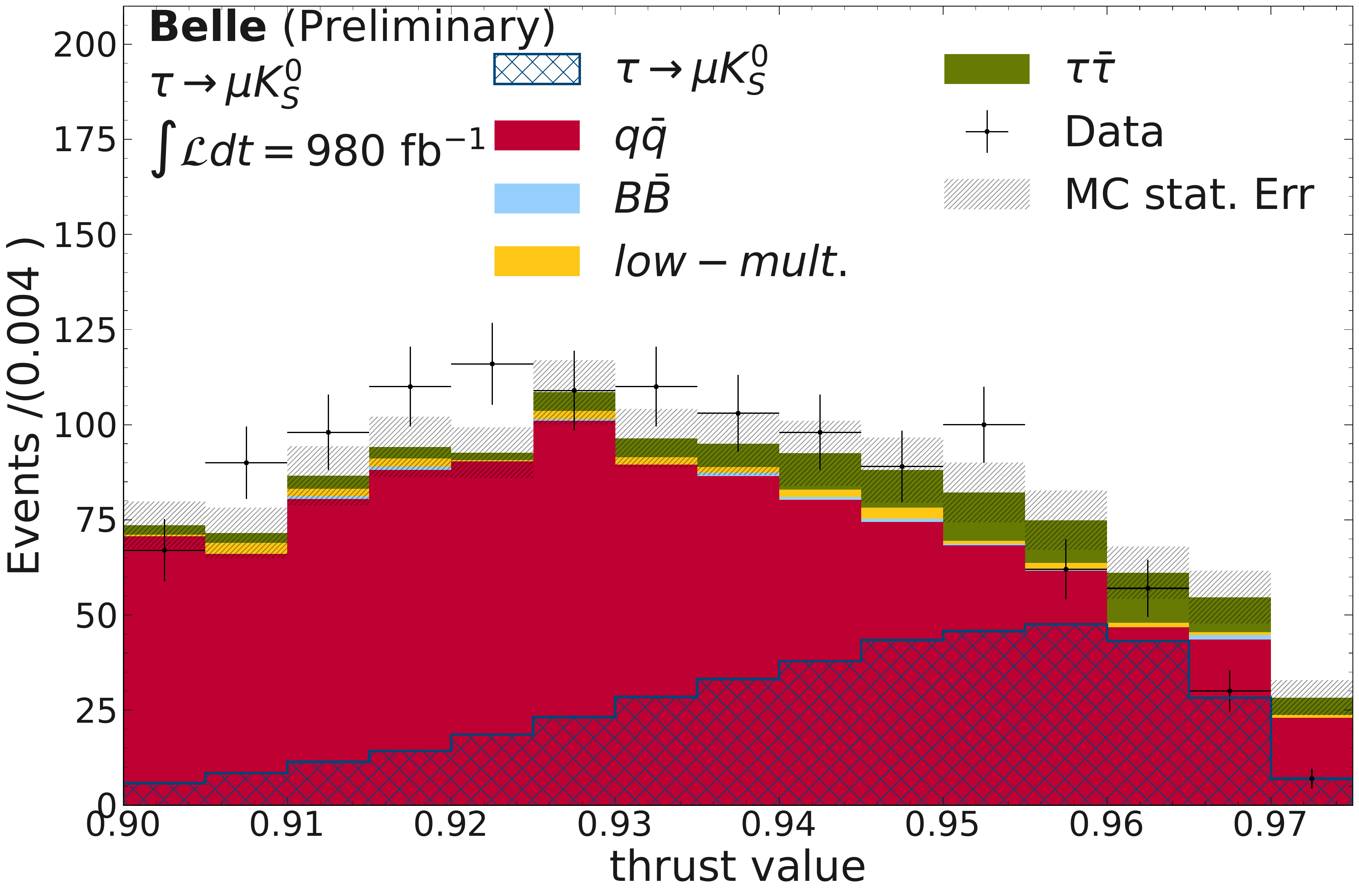}
    \caption{Comparison between data (black points with error bars) and background simulation (solid-filled histograms) in the sidebands before (left) and after (right) pre-selections are applied to reject 
      low-multiplicity events.  The signal in the $20\delta$ region is shown as a blue hatched histogram with an arbitrary scale. The statistical uncertainties are displayed in light gray shading.
    %The left (right) plots show candidates before (after) applying preselection requirements.
    The upper row displays the $K_S^0$ mass reconstructed using the electron mass hypothesis for the pions, $M^{ee}(K_S^0)$, for the $\belletwo{}$ electron channel. The lower row displays the thrust value for the Belle muon channel. %The lower panel in each plot shows the per-bin-ratio of data over simulation.
%    RVK: GENERAL QUESTION - WHY WOULD YOU CHOOSE SUCH ODD BIN SIZES? CAN'T YOU ROUND THEM TO THE NEAREST 0.02 GEV (TOP) OR 0.005 (BOT)?
    }
    \label{fig:presel_e}
\end{figure}

We first apply a set of selection requirements (preselection) to
remove obvious background events and then a dedicated 
BDT-based selection requirement to further suppress remaining 
background. The preselection requirements are chosen by
comparing the simulated signal distributions with the background distributions obtained from the sideband regions in data and in simulation, separately for Belle and $\belletwo{}$ and for each tag-side decay type. %\textcolor{red}{
As mentioned previously in Section~\ref{sec:BelleII}, the Belle and Belle II experiments use different MC simulations. As a consequence, selection optimization results in different preselection requirements and differences in input features for the BDT selections, as detailed in Section~\ref{sec:bdt}.
In order to remove low-multiplicity events with converted photons, we impose a minimum requirement on the $K_S^0$ mass reconstructed using the electron mass hypothesis for the pions, $M^{ee}(K_S^0)$. 
%as illustrated on Fig.~\ref{fig:presel_e}(upper row). 
Since
%in general all particles from
low-multiplicity events
consist only of detectable particles,
%are used to reconstruct the signal,
these events can be discarded using variables related to the missing energy and momentum, the tag-side mass and $\Delta E$. The projections of the lepton and tag-side track momentum on the $z$ axis are also used, as the background from low-multiplicity events tends to be aligned more closely with the  beam axis than are signal events.
Continuum background events are rejected using event shape properties like the thrust value,  the numbers of photons and neutral pions, or characteristics of the decay dynamics, like the energy of the lepton. The full set of preselections is shown in Appendices~\ref{sec:electron_presel} and \ref{sec:muon_presel} for the electron and muon channels, respectively, comparing the different requirements applied for Belle and $\belletwo{}$. The distributions of the pair-converted invariant mass $M^{ee}(K_S^0)$ (upper row) and event thrust (lower row) are shown before (left) and after (right) applying the preselection requirements in Fig.~\ref{fig:presel_e}.
%RVK: I FIND IT ODD TO PUT ESSENTIAL INFORMATION LIKE THE PRESELECTION REQUIREMENTS IN THE APPENDIX - PLEASE CONSIDER INCLUDING A TABLE WITH THIS INFORMATION IN THE MAIN BODY OF THE PAPER.

\begin{figure}[h!]
    \centering    
        \includegraphics[width=0.45\textwidth,page=2]{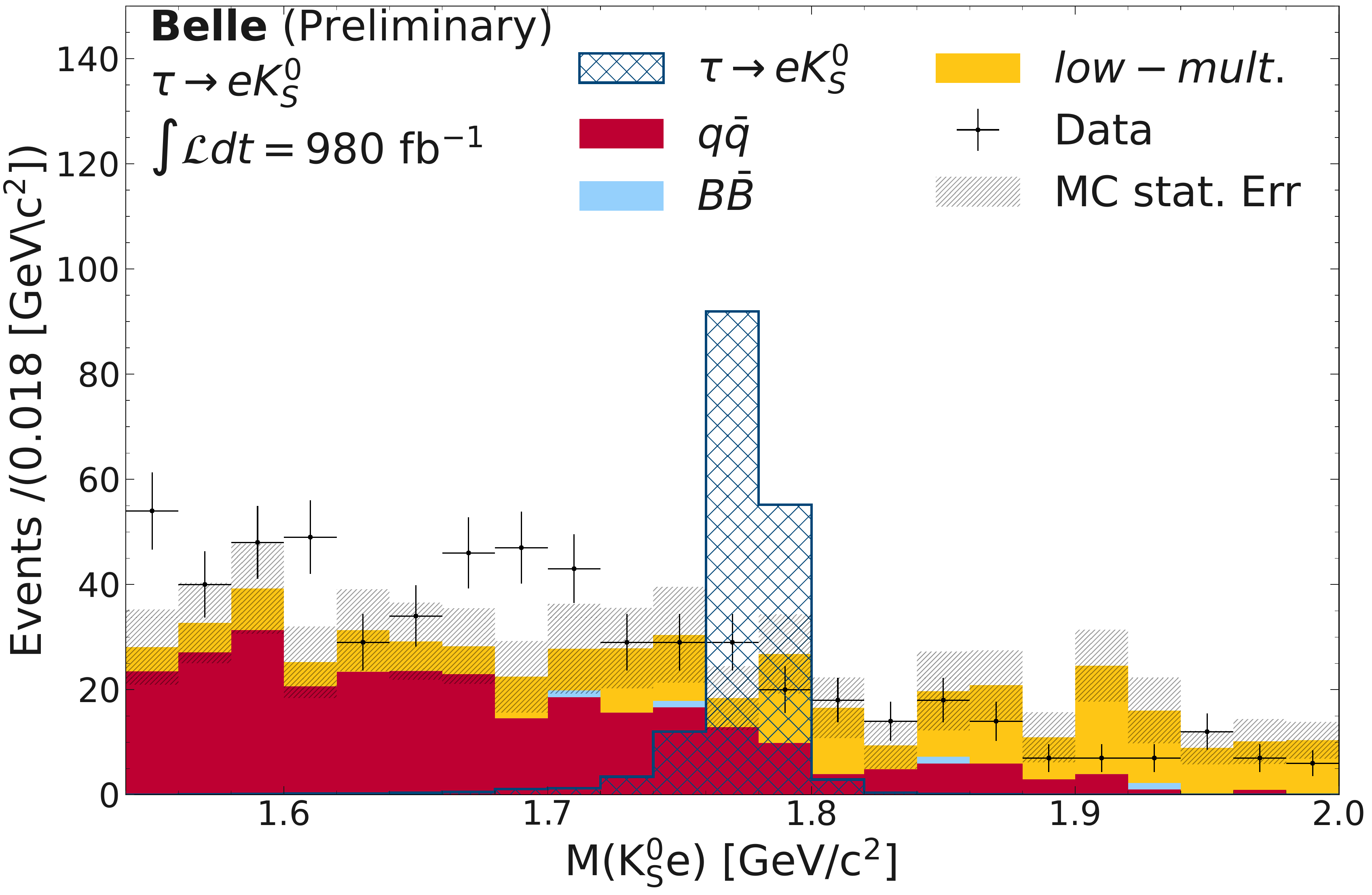}
        \includegraphics[width=0.45\textwidth,page=2]{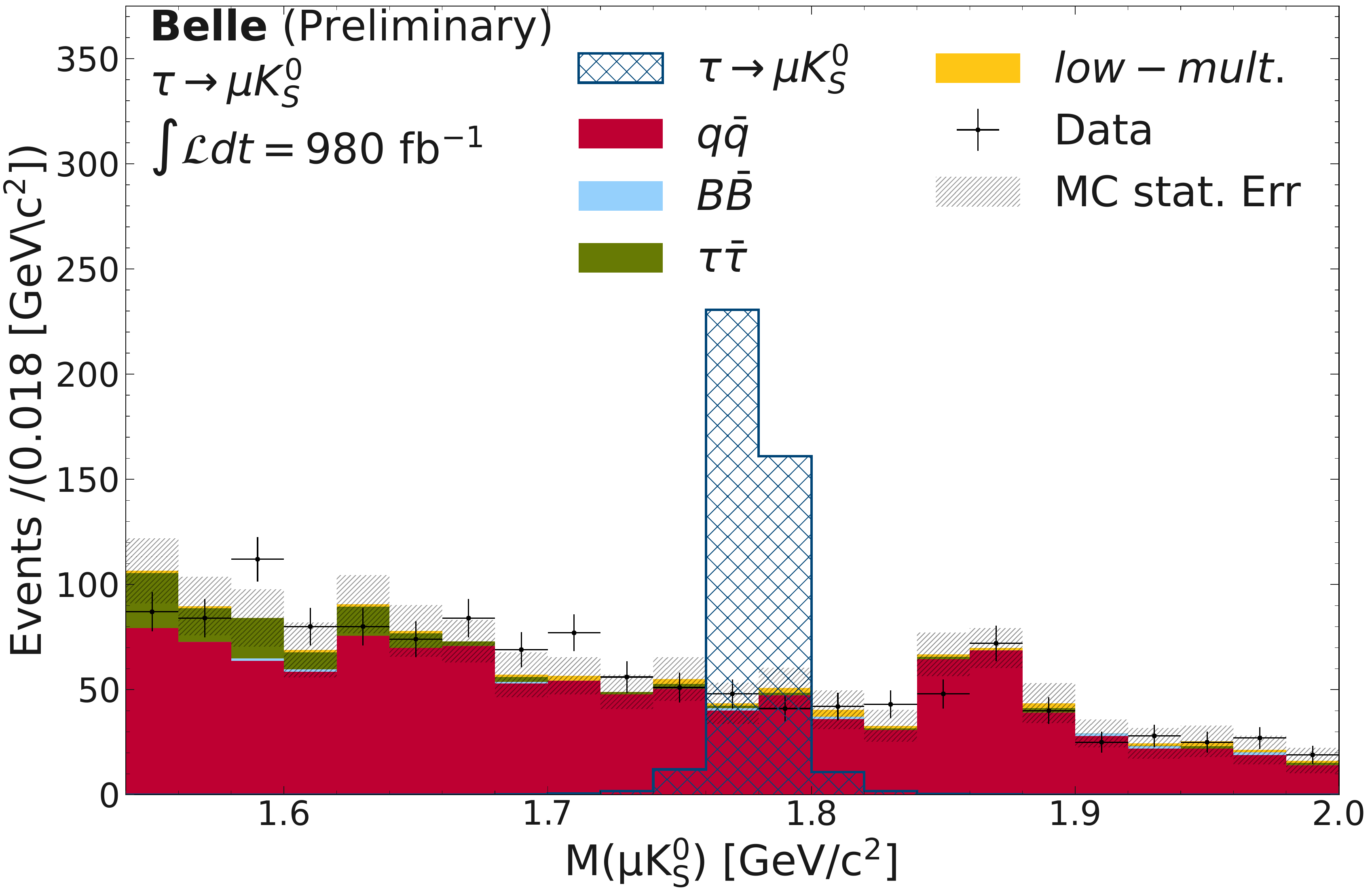}\\

        \includegraphics[width=0.45\textwidth,page=3]{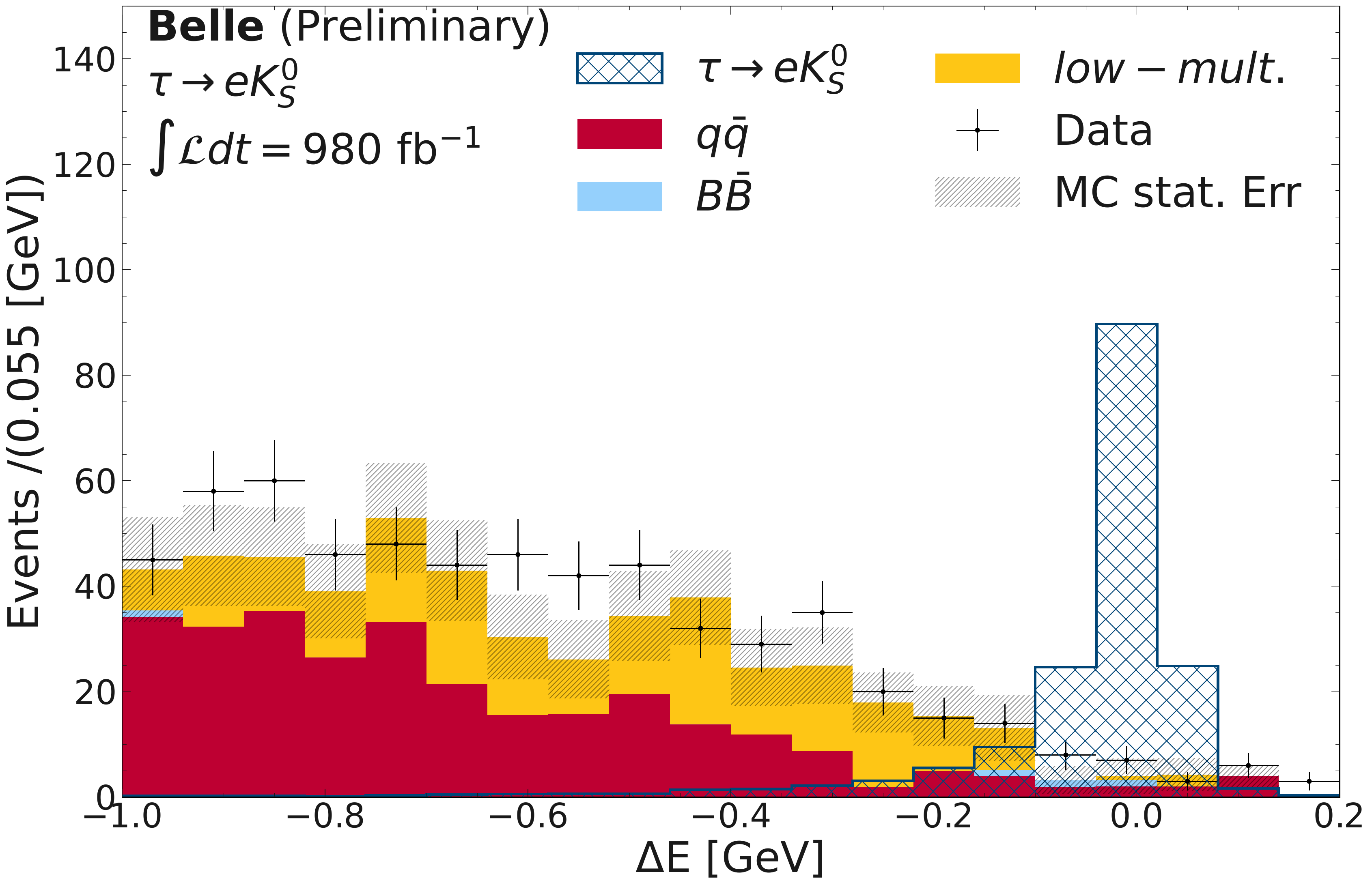}
        \includegraphics[width=0.45\textwidth,page=3]{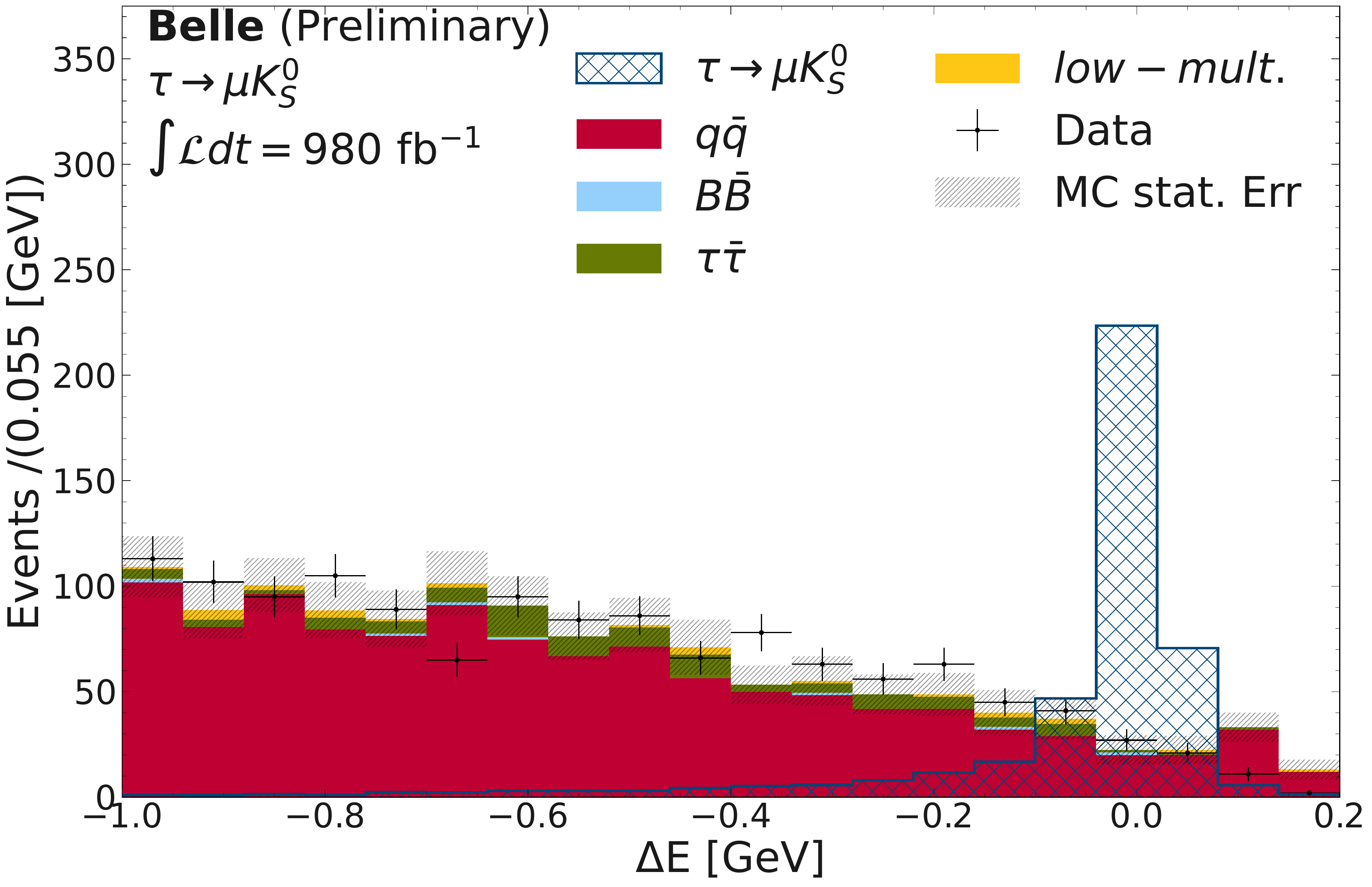}\\
        
        \includegraphics[width=0.45\textwidth,page=2]{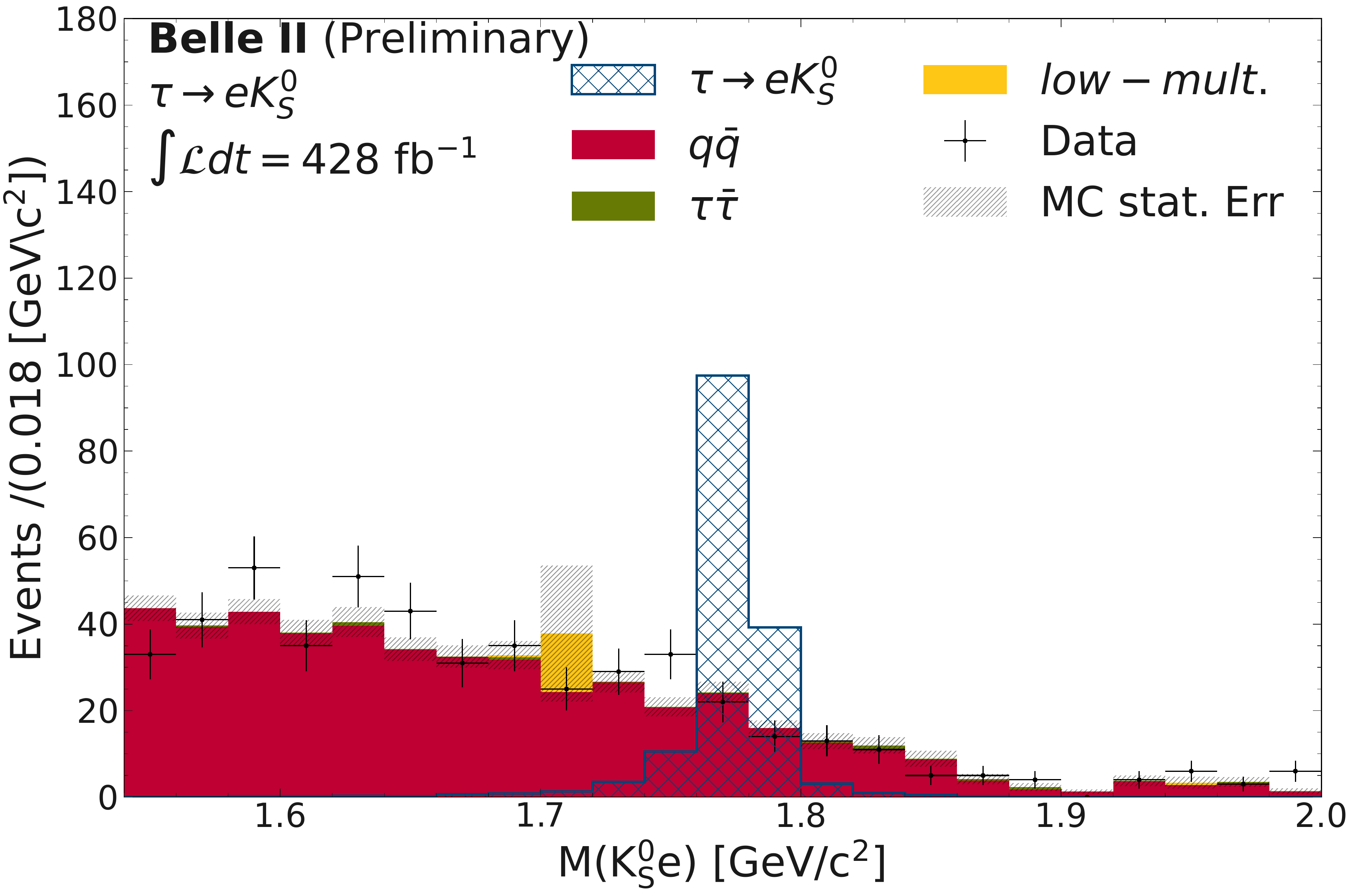}
        \includegraphics[width=0.45\textwidth,page=2]{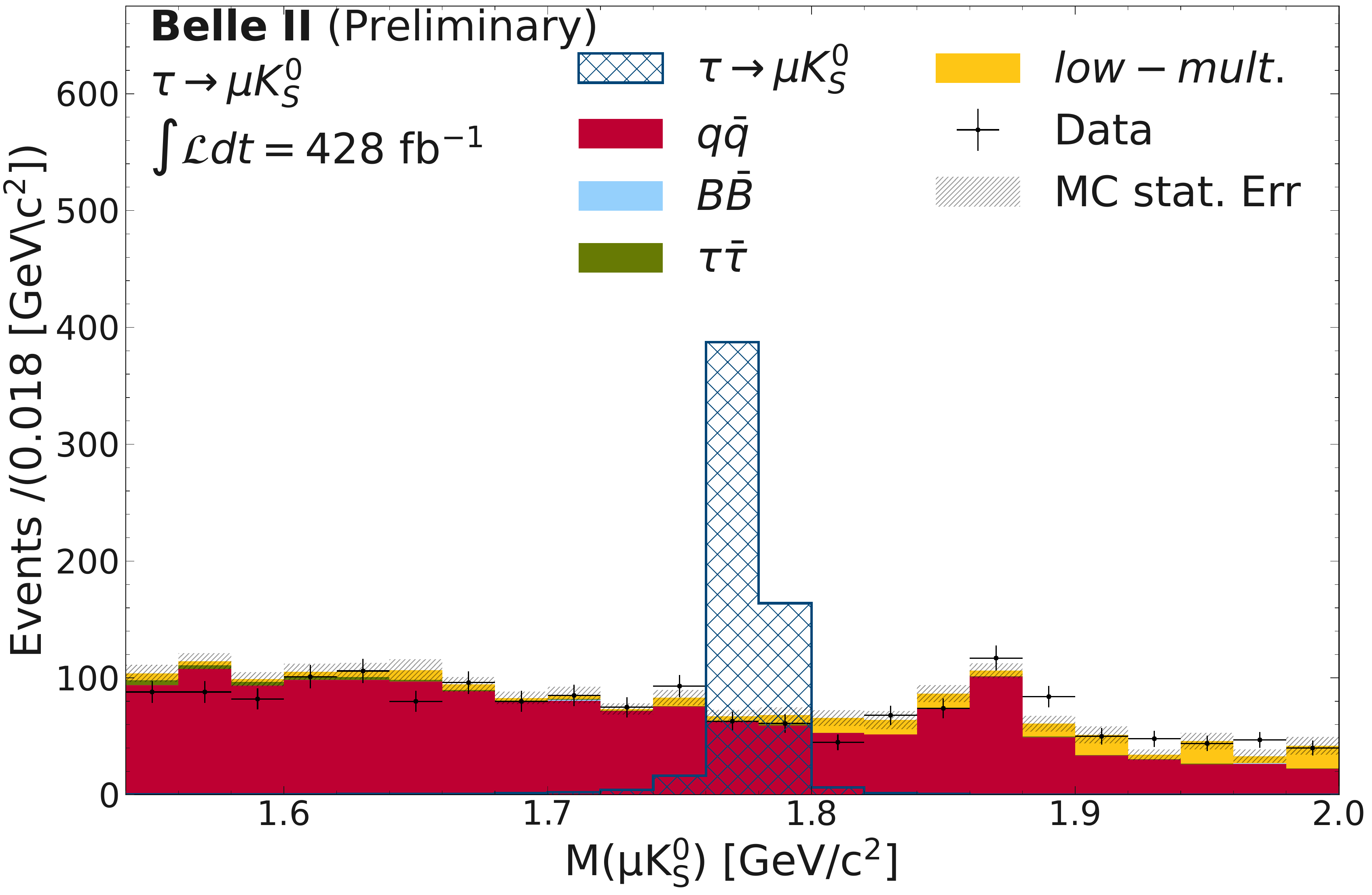}\\

        \includegraphics[width=0.45\textwidth,page=3]{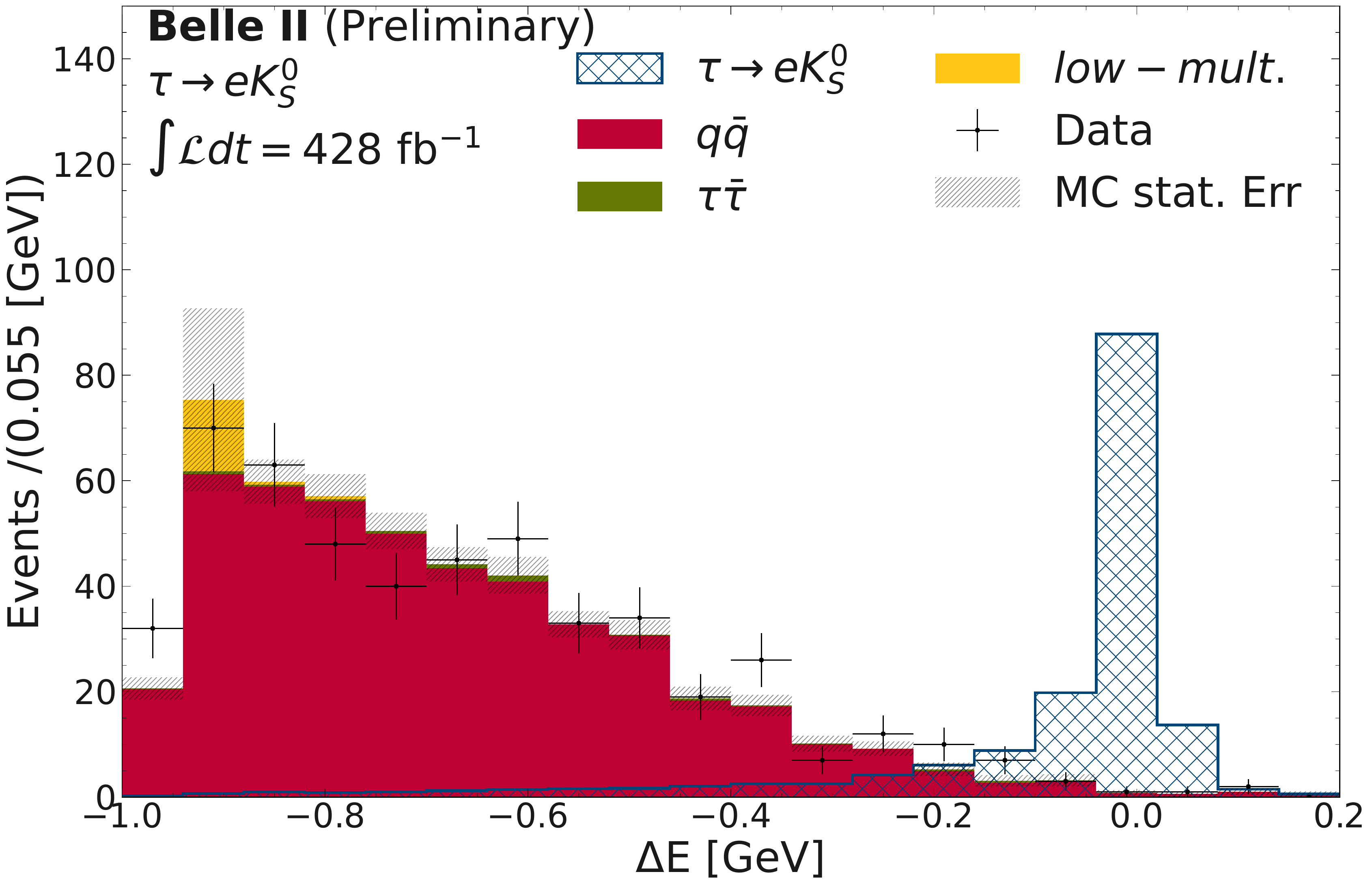}
        \includegraphics[width=0.45\textwidth,page=3]{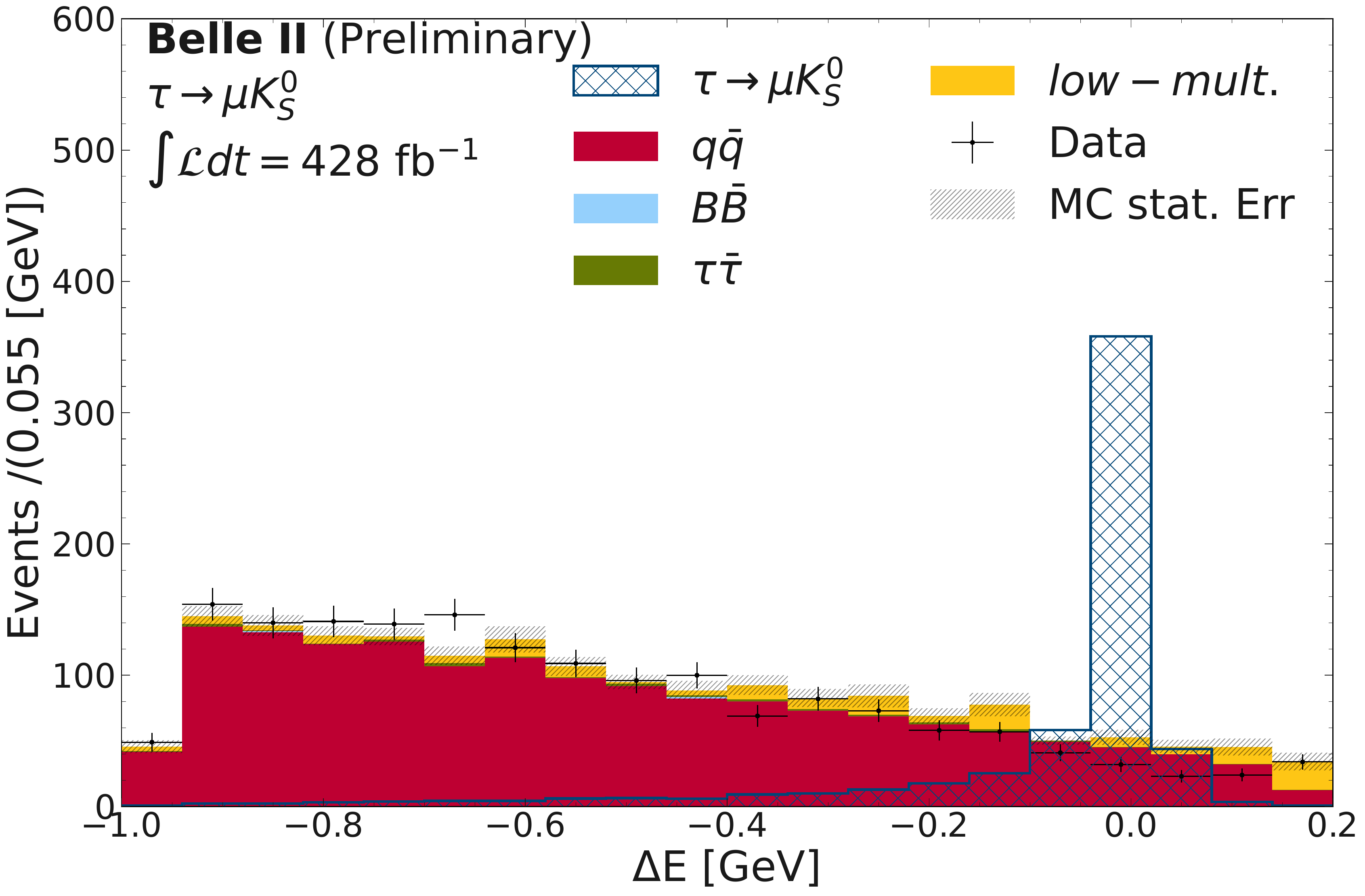}%22
        \caption{Comparison between data (black points with error bars) and background simulation (solid-filled histograms) in the sideband regions after preselections for the \dE and \Mlk distributions, for both Belle (first and second rows) and Belle II (third and fourth rows) data sets. Only backgrounds with a non-null residual contribution after the selection is applied are shown in the plots.  The signal in the $20\delta$ region is shown as a blue hatched histogram with an arbitrary scale. The statistical uncertainties are displayed in light gray shading. The left column shows the  distributions for electron channels and the right column for the muon channels. 
    }
    \label{fig:presel}
\end{figure}

We show the \Mlk and \dE distributions for the four different categories 
after preselection in Fig. \ref{fig:presel}.
The overall signal efficiencies in the $20\,\delta$ region after reconstruction and preselection obtained from Belle ($\belletwo{}$) simulation are 13.6 (13.9)\% and 17.2 (16.6)\% for the electron and muon channels, respectively. These values are obtained after correcting the simulation to account for the mismodelling of the detector response, which affects the lepton identification.

\input{032-exclusive_tagging}

%% file: 032-exclusive_tagging.tex
%Identification of muons relies mostly on charged-particle penetration depth in the KLM for momenta larger than 0.7\gevc and on information from the CDC and ECL otherwise. Particle-identification likelihoods are obtained by combining information from the relevant subdetectors.
%

\subsection{BDT-based background rejection}\label{sec:bdt}
To reduce the residual background contamination,
%mitigate the presence of residual background events,
primarily due to $e^+e^-\to q\bar{q}$ processes, a BDT is trained using the XGBoost library~\cite{XGBoostPaper} on simulated signal and background samples. The BDT incorporates 34 variables (listed in Appendix \ref{sec:bdtvar}) %\footnote{the most powerful variables are highlighted by an asterisk}
 pertaining to both signal- and tag-side $\tau$ kinematics, $K_S^0$ and 
track kinematics, event shape properties, and variables associated with photons and neutral pions in the event.

%The first set of variables encompasses the reconstructed mass of the tag-side $\tau$, along with flight time and distance parameters, including their respective errors.
The first set of variables includes the $\tau$ properties: the beam-constrained mass of the tag-side $\tau$, the invariant mass hypothesis for the track in the tag side, as well as the  invariant mass of the tag-side particles (the track and any tag-side photons), while for the signal $\tau$ it consists of the transverse momentum ($p_T$), flight time and distance and their respective measurement uncertainties.
Additionally, it includes the angle between the lepton and the $K_S^0$. Variables related to the $K_S^0$ candidate are the flight time and distance with their uncertainties, the momentum and energy in the c.m. system, and two hypothetical
invariant masses computed under the assumption that either the first or second daughter is a proton. Track-related variables consist of the lepton momentum and the ranked $p^T$ values of the
three signal-side tracks, all in the c.m.\ system. The event characteristics included are the missing mass, missing energy, missing momentum and missing transverse momentum of the event in the c.m.\ frame.  We also use the cosine of the angle between the missing momentum and the tag-side track, and between the missing momentum and the signal-side lepton, as well as the  visible energy
%(RVK: YOU USE BOTH THE MISSING ENERGY AND THE VISIBLE ENERGY - AREN'T THEY FULLY CORRELATED?)
in the c.m.\ and the thrust value. The last category of variables contributing to the BDT classifier  includes the number of photons and their total energy
on the signal- and tag-sides, along with the total photon energy in the event and the number of $\pi^0$ candidates.

Four BDTs are separately trained, one for each of the two final states in Belle and Belle II,
%Four BDT trainings are conducted separately for Belle and Belle II and electron or muon final states, 
on  samples corresponding to 1/3 of the Belle (Belle II) simulation statistics within the 20$\delta$ region for the background and about 12000 simulated signal events. The BDT parameters are optimized using the \texttt{Optuna} library~\cite{Akiba:2019lwq}, which minimizes the logarithmic loss function.
%the criterion for the BDT output is chosen to maximize the Punzi figure-of-merit~\cite{Punzi:2003bu} within the 3$\delta$ region. This figure-of-merit is defined as $\frac{\varepsilon_{\ell K_S^0}}{\alpha /2 + \sqrt{B}}$, where $\varepsilon_{\ell K_S^0}$ represents the signal efficiency, $B$ denotes the number of background events, and $\alpha$ is set to 3, 
%corresponding to $3~\sigma$ signal significance.
%to a search sensitivity at the 90\% confidence level.
%Following the training, 
The BDT is validated using an independent data sample, ensuring that the signal retention and background rejection rates remain comparable to those of the training sample, thus mitigating the risk of overfitting. Both the training and validation samples are merged into a combined sample.
%used to optimize the selection on the BDT output for each tag-side category. 
This combined sample is then used to optimize simultaneously the size of the elliptical SR in units of $\delta$ and
the BDT classifier requirement for each tag-side category using the Punzi figure of merit.
%, using elliptical regions optimized by the Punzi figure-of-merit on simulation beforehand. 
\begin{figure}[h!]
    \centering    
    \includegraphics[width=0.48\textwidth,page=1]{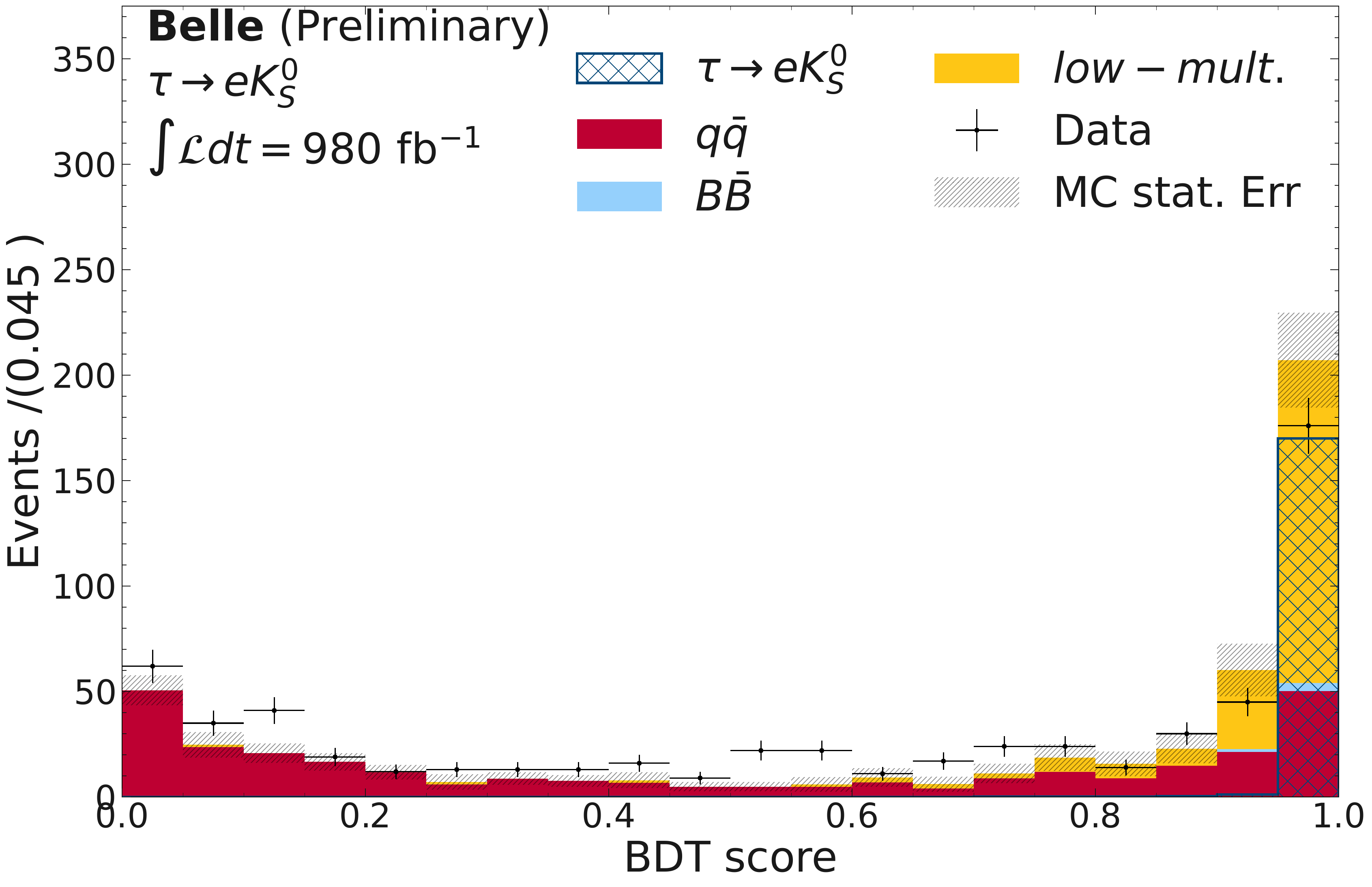}
    \includegraphics[width=0.48\textwidth,page=1]{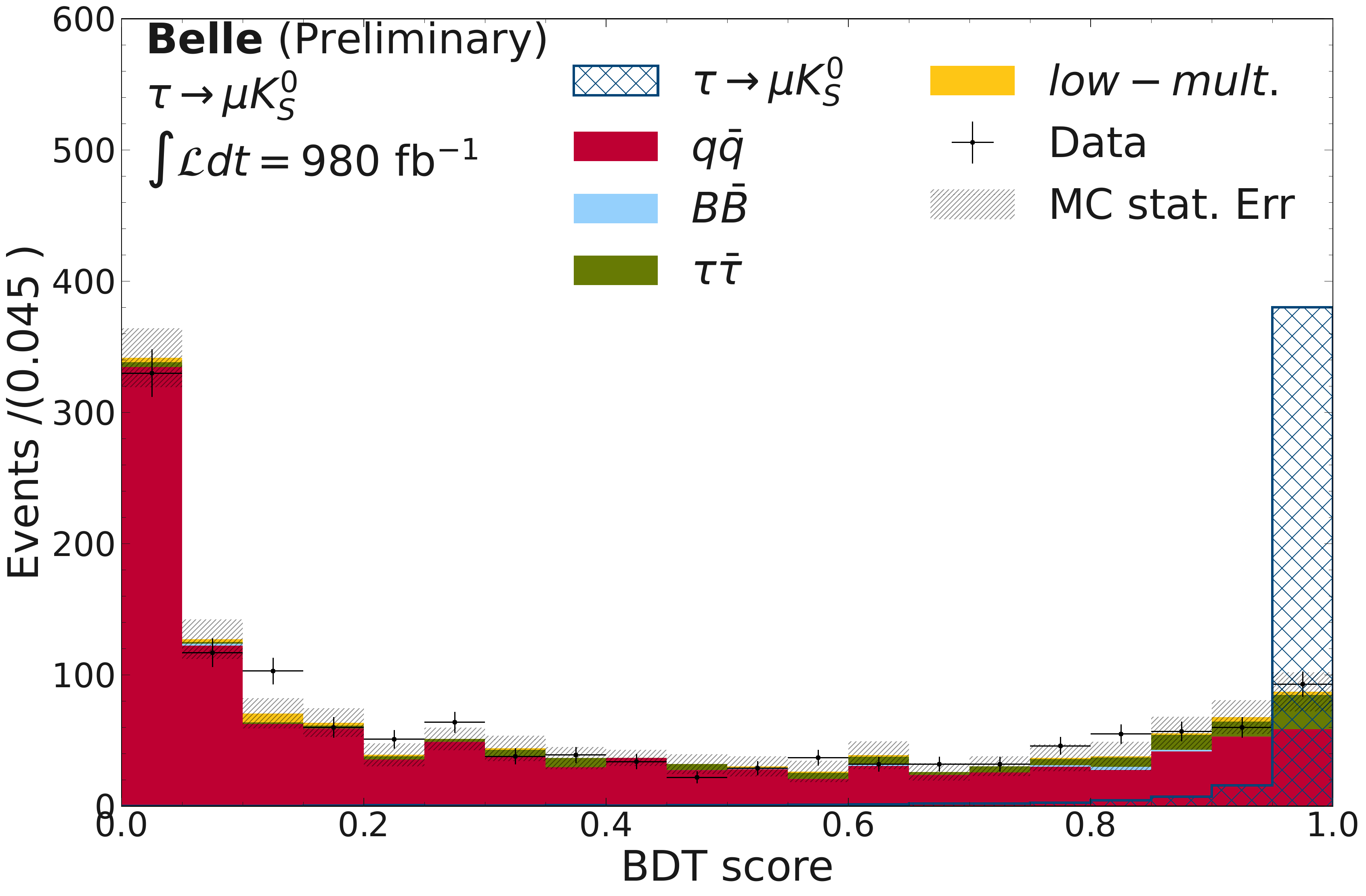}\\
    
    \includegraphics[width=0.48\textwidth,page=1]{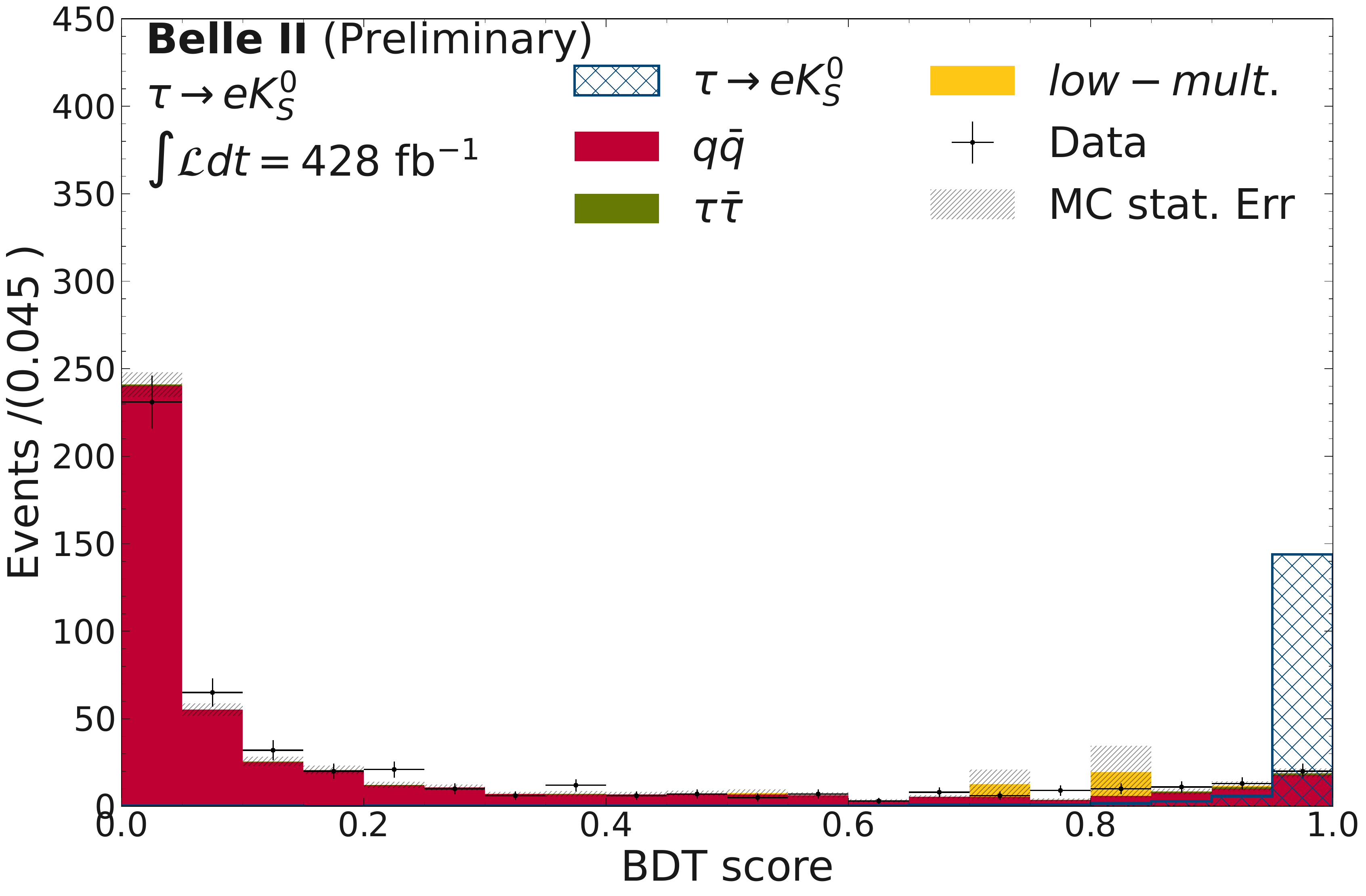}
    \includegraphics[width=0.48\textwidth,page=1]{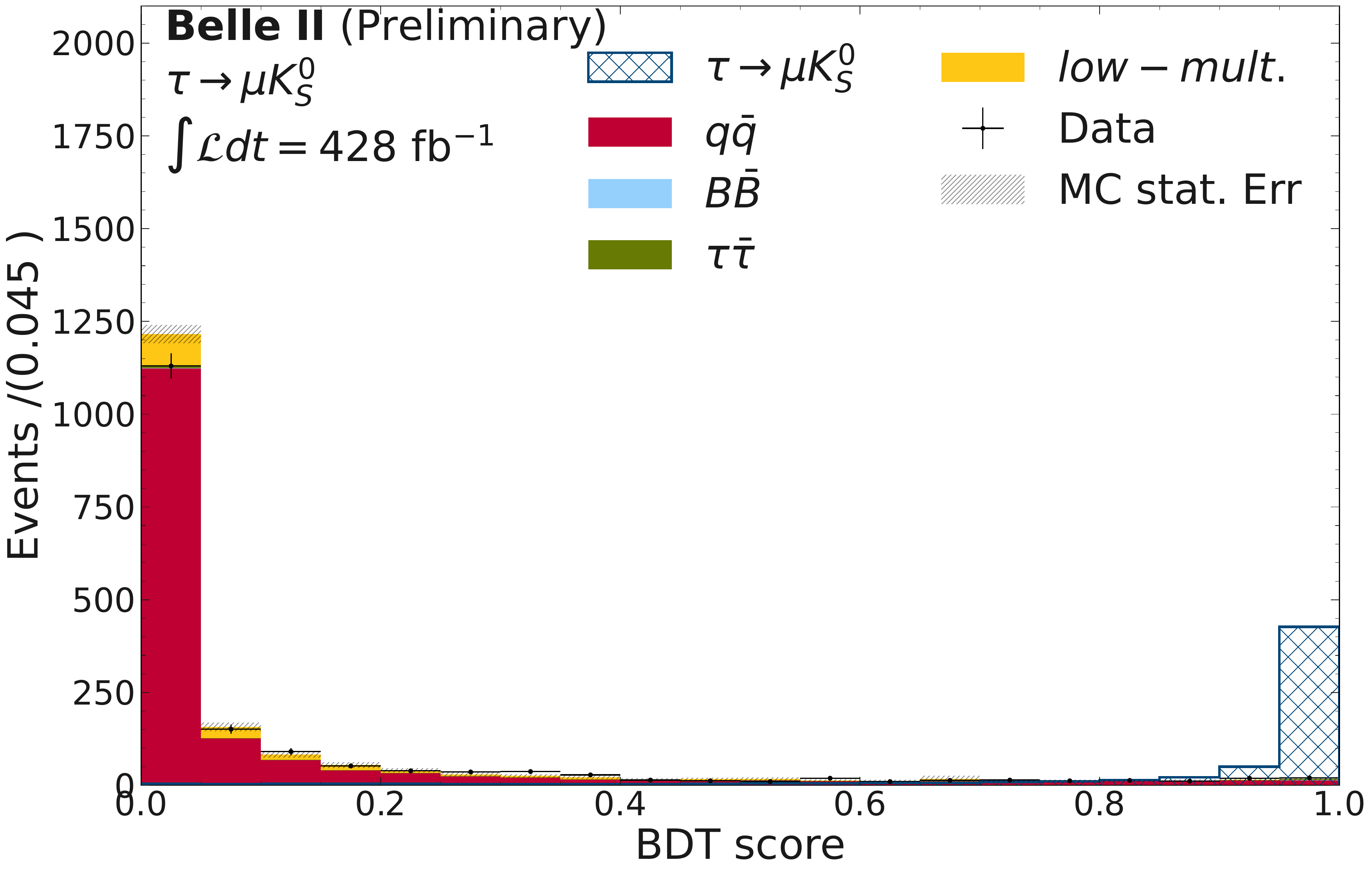}
    \caption{Comparison between data (black points with error bars) and simulation (solid-filled histograms) for the BDT output distribution for events in the sideband region after preselection. The signal is shown as a blue hatched histogram with arbitrary scale. The statistical uncertainties are displayed as gray shaded areas. Top row shows the electron (left) and muon (right) channel for 
      Belle, the bottom row shows electron (left) and muon (right) channel for Belle II.% The lower panel in each plot shows the per-bin-ratio of data over simulation.
%      RVK: AGAIN, BIN WIDTH CHOICE IS WEIRD - WHY NOT JUST HAVE 20 BINS FROM 0 TO 1? 
} 
    \label{fig:BDT_SB}
\end{figure}
This figure of merit~\cite{Punzi:2003bu} is defined as $\frac{\varepsilon_{\ell K_S^0}}{\alpha /2 + \sqrt{B}}$, where $\varepsilon_{\ell K_S^0}$ is the signal efficiency and $B$  is the number of background events in the elliptical SR region, respectively. We set $\alpha$  to 3, 
corresponding to $3\sigma$ signal significance.
 The applied selections are listed in Table \ref{tab:bdt_sel} for all channels and experiments. %\textcolor{red}{
%As noted in Section 3.1, the BDT working points differ for each experiment and channel due to the different reference simulations used in Belle and Belle II.
 As noted in Section~\ref{sec:cand_selection}, the input features and requirements for the BDT classifiers differ for each channel and experiment since the working points for each BDT are different, due to the different MC simulations used in Belle and Belle II for the optimization of the pre-selections that are applied before the BDT training. 
 The BDT rejects 87\% (72\%) of the background events in the electron channel according to Belle (Belle II) simulation, and  96\% (89\%) of the background events in the muon  channel.
The overall signal efficiencies after the BDT requirements obtained from Belle (Belle II) simulation in the signal region are 10.4\% (10.1\%) and 10.2\% (10.2\%) for the electron and muon channels, respectively. 

\begin{table}[htbp]
\caption{Optimized BDT classifier requirement values for the different tag-side categories for the electron and muon channels.}
    \centering
\begin{tabular}{l|lll|lll}

& \multicolumn{3}{c|}{$eK_S^0$}                                             & \multicolumn{3}{c}{$\mu K_S^0$}                                          \\  
& \multicolumn{1}{c|}{$e$-tag} & \multicolumn{1}{c|}{$\mu$-tag} & $\pi$-tag & \multicolumn{1}{c|}{$e$-tag} & \multicolumn{1}{c|}{$\mu$-tag} & $\pi$-tag \\ 
\hline
\multicolumn{1}{c|}{Belle}    & \multicolumn{1}{c|}{0}       & \multicolumn{1}{c|}{0.99}         & 0.99      & \multicolumn{1}{c|}{0}       & \multicolumn{1}{c|}{0.91}         & 0.99      \\ 
\multicolumn{1}{c|}{Belle II} & \multicolumn{1}{c|}{0.95}       & \multicolumn{1}{c|}{0}         & 0.89      & \multicolumn{1}{c|}{0.5}       & \multicolumn{1}{c|}{0.6}       & 0.95      \\ 
\end{tabular}
    
    \label{tab:bdt_sel}
\end{table} 

\subsection{Simulation validation and expected background}
The agreement between data and simulation is checked using the SB region, defined as the black dashed rectangle in Fig.~\ref{fig:signalRegion} without the SR box, and the RSB region (green bands in Fig.~\ref{fig:signalRegion}).
%table 3 below:

\begin{table}[htbp]
 \caption{Signal efficiencies in the elliptical SR, observed yields in data and 
    expected yields in simulation in the RSB for the electron and muon channels in Belle and Belle II data.}
    \label{tab:dmc_RSB}
    \centering
\begin{tabular}{c|ccc|ccc}

& \multicolumn{3}{c|}{$eK_S^0$}                                                                              & \multicolumn{3}{c}{$\mu K_S^0$}                                                                           \\  
& \multicolumn{1}{c|}{$\epsilon_{e K_S^0}$ {[}\%{]}} & \multicolumn{1}{l|}{$N_{RSB}^{data}$} & $N_{RSB}^{MC}$          & \multicolumn{1}{c|}{$\epsilon_{\mu K_S^0}$ {[}\%{]}} & \multicolumn{1}{c|}{$N_{RSB}^{data}$} & $N_{RSB}^{MC}$          \\ \hline
\multicolumn{1}{c|}{Belle}    & \multicolumn{1}{c|}{10.4}               & \multicolumn{1}{c|}{6}               & $8.3_{-2.8}^{+4.0}$ & \multicolumn{1}{c|}{10.2}               & \multicolumn{1}{c|}{8}               & $9.4_{-3.0}^{+4.2}$  \\
\multicolumn{1}{c|}{Belle~II}    & \multicolumn{1}{c|}{10.1}               & \multicolumn{1}{c|}{6}               & $4.1_{-1.9}^{+3.2}$ & \multicolumn{1}{c|}{10.2}               & \multicolumn{1}{c|}{5}               & $10.5_{-3.2}^{+4.3}$ \\ 
\end{tabular}
\end{table}
\begin{figure}[h!]

    \centering    
        \includegraphics[width=0.46\textwidth,page=1]{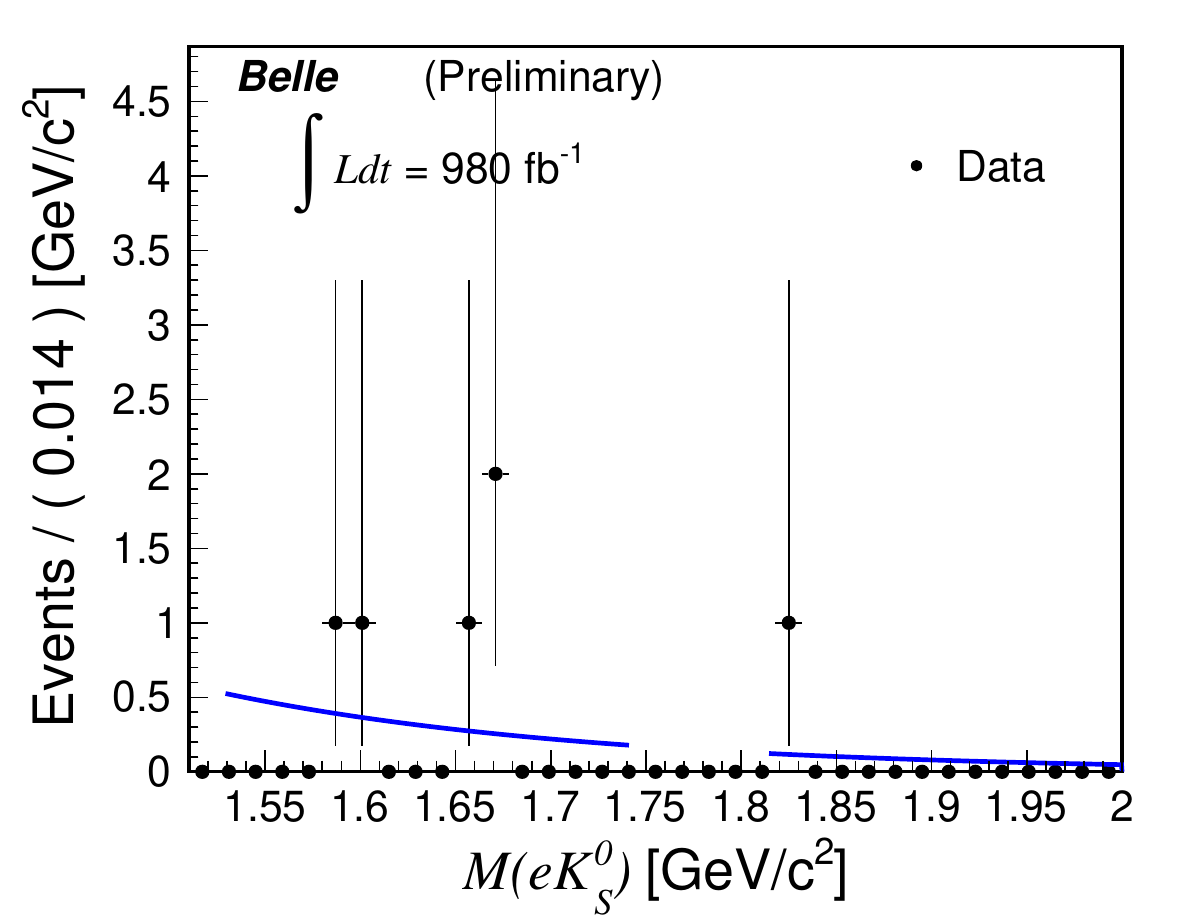}
        \includegraphics[width=0.46\textwidth,page=1]{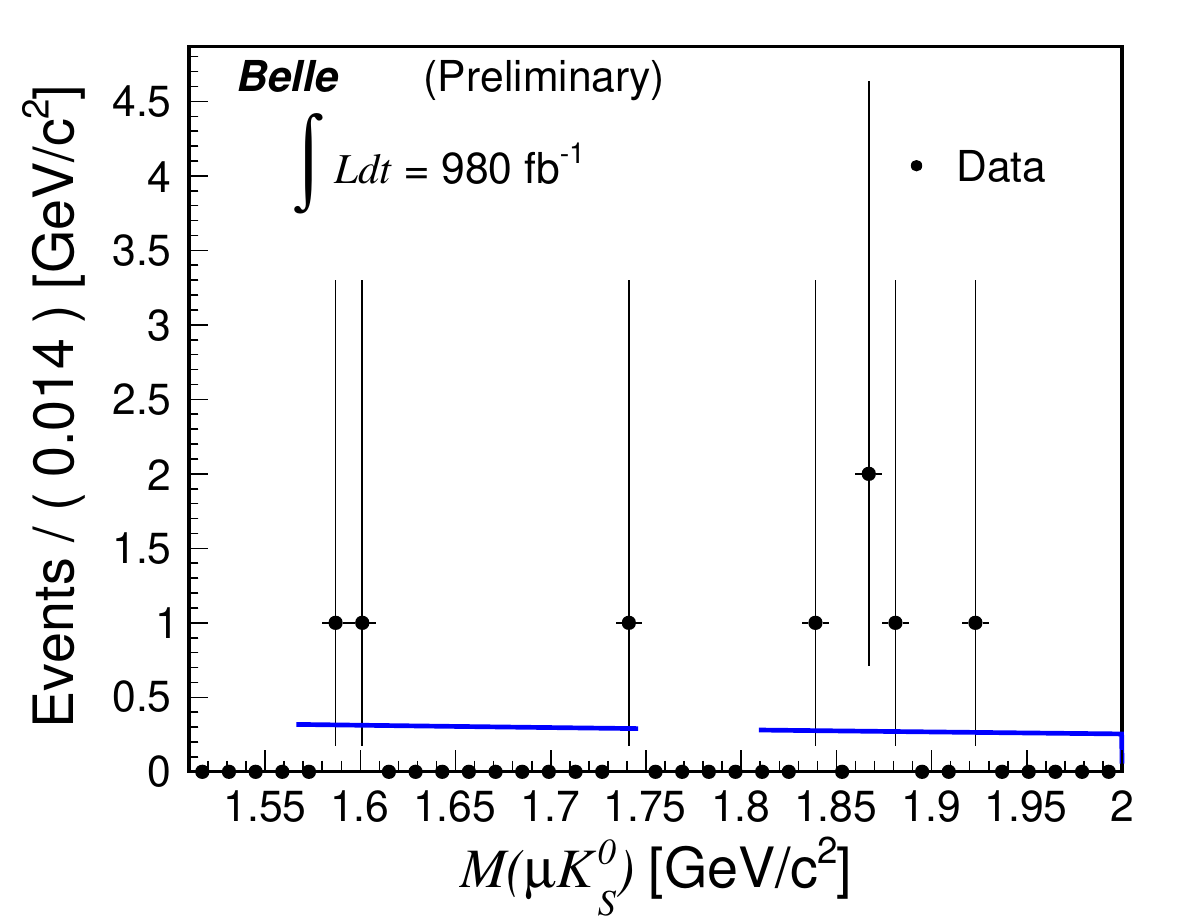}\\

        \includegraphics[width=0.46\textwidth,page=1]{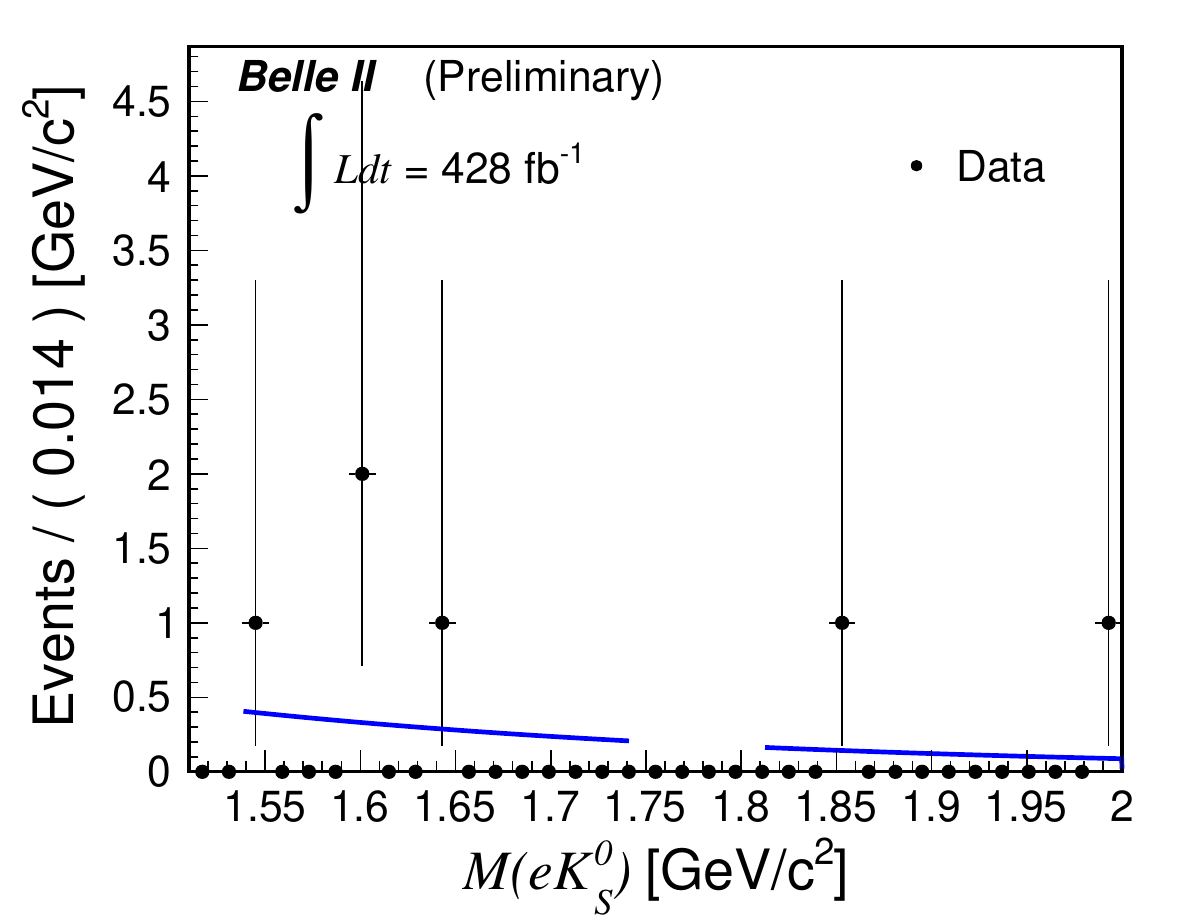}
        \includegraphics[width=0.46\textwidth,page=1]{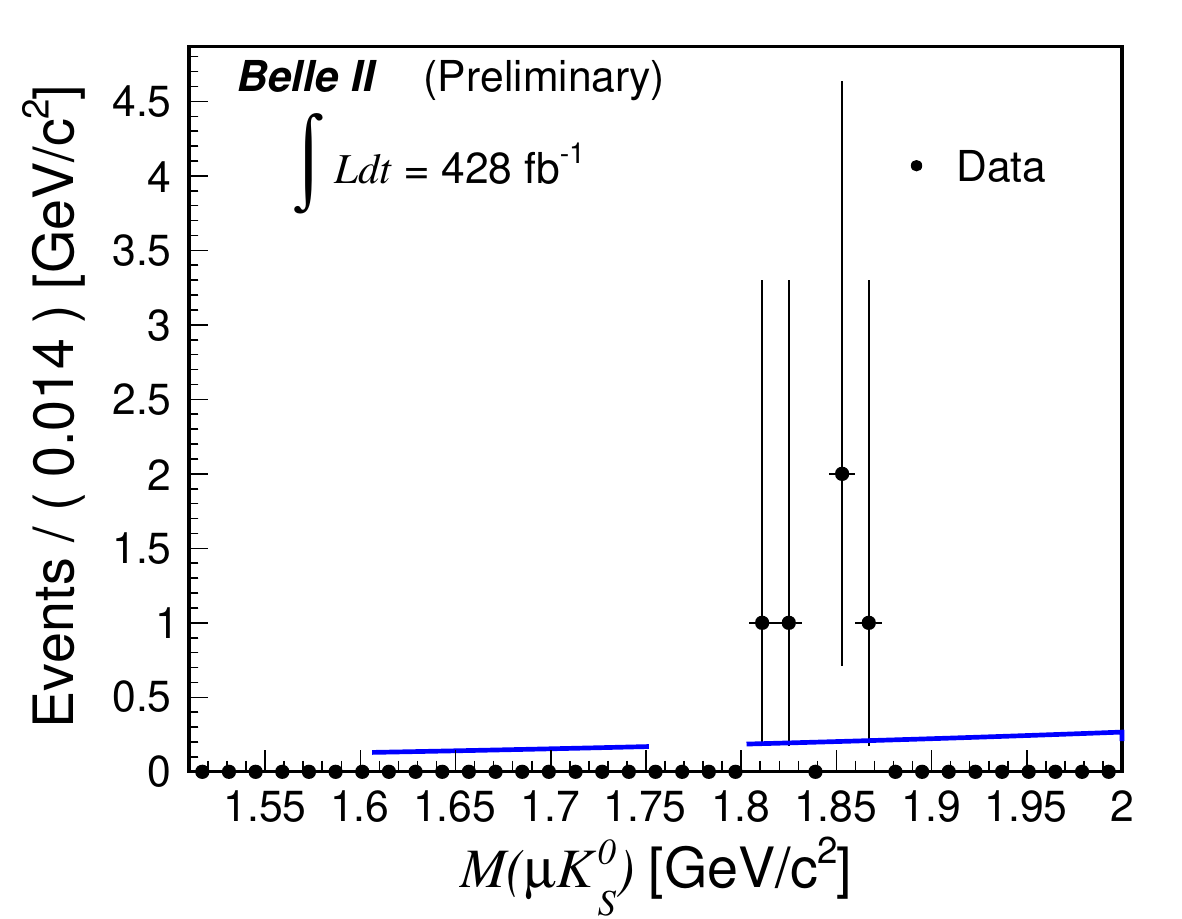}
    \caption{Fits (blue solid line) to selected events (points with error bars) in the RSB region as function of \Mlk. Top row shows the electron (left) and muon (right) channels for Belle, the bottom row shows the electron (left) and muon (right) channels for Belle II. }
    \label{fig:exp_fit}
\end{figure}

The outputs of the BDTs in the SB region for all channels are shown in Fig.~\ref{fig:BDT_SB}, 
with the simulated signal overlaid as a hatched blue histogram. The signal efficiency and the yields in data and simulation in the RSB
are given in Table~\ref{tab:dmc_RSB}. Differences in the background components between 
the Belle and Belle II electron channels arise from different simulated 
backgrounds (see Table~\ref{tab:simMC} in Appendix~\ref{sec:appA}) and preselections. 
%One can note from Fig.~\ref{fig:presel} that only the tail of the low-multiplicity background enter the SR.
%For Belle, we used simulations of $ee\to ee q\bar{q}$ backgrounds, with  $q\in {u,d,s,c}$. Those components are a subsample of the yellow low multiplicity  component in Fig. \ref{fig:BDT_SB}, which are not simulated for Belle II.

To estimate the number of expected events in the signal region after applying the BDT selection, we perform an extended unbinned maximum likelihood fit to the signal tau invariant mass distribution of the events retained in the data RSB region. We use an exponential function to model the background shape,  $f(m) = N_{\mathrm{bkg}}e^{Cm} $, where the free parameters in the fit are the shape parameter $C$ and the background normalization  $N_{\mathrm{bkg}}$ in the range 1.5 $<M_{lK_S^0}<$2.0~\gevcc.  The results for each channel and experiment are shown in Fig.~\ref{fig:exp_fit}.
The expected yields $N_{\mathrm{exp}}$ in the elliptical SR are obtained by integrating the fitted functions over the $\pm 3\delta$ intervals in $M_{lK_S^0}$ based on signal resolutions as defined in Table~\ref{tab:sig_resol} and approximated by multiplying the integral $N^{\mathrm{SR}}_{\mathrm{bkg}}$ by the ratio of the 
elliptical and rectangular areas, $f_{\mathrm{ell}}$, so that $N_{\mathrm{exp}}= f_{\mathrm{ell}} N^{\mathrm{SR}}_{\mathrm{bkg}} $. 
The inputs and the resulting yields with their statistical uncertainties propagated from the fits are listed in Table~\ref{tab:expected_data}.

%To account for the statistical uncertainty in the number of expected events in the SR, we consider the 68 $\%$ confidence interval of the fit. Table \ref{tab:expected_data} displays the expected number of events in the SR obtained through this method. 
%
\begin{table}[htbp]
  \caption{Number of background events in the signal regions $N^{\mathrm{SR}}_{\mathrm{bkg}}$ as integrated from fits to Belle and Belle II data RSB regions, scaling factors $f_{\mathrm{ell}}$ and the number of expected background yields in the elliptical SR for the electron (top) and muon (bottom) channels. The uncertainties are statistical only.}
    
     \centering
\begin{tabular}{cc|ccc}
& & $N^{\mathrm{SR}}_{\mathrm{bkg}}$& $f_{\mathrm{ell}}$ & $N_{\mathrm{exp}}$  \\
\hline
\multirow{2}{*}{$e K_S^0$} & Belle &$0.78^{+0.36}_{-0.28}$ & $0.557$ & $0.43^{+0.20}_{-0.16}$\\
 & Belle II & $0.94^{+0.44}_{-0.33}$ & 0.453&$0.42^{+0.20}_{-0.15}$ \\ 
\hline
\multirow{2}{*}{$\mu K_S^0$}  & Belle &$1.29^{+0.51}_{-0.40}$  & 0.554&$0.71^{+0.28}_{-0.22}$ \\
 & Belle II &$0.65^{+0.34}_{-0.25}$ &  0.555 &$0.36^{+0.19}_{-0.14}$\\
\end{tabular}
    \label{tab:expected_data}
\end{table}

%% file: 04-sys_v9.tex
\section{Systematic uncertainties}
\label{sec:systematics}
%\textcolor{red}{
Systematic uncertainties can affect the branching fraction measurements through the modeling of the background used to extract $N_{\mathrm{exp}}$, possible differences between experimental data and simulation that could impact the signal efficiency, and external inputs, such as the data luminosity and the tau pair cross-section. %}

%detector effects not taken into account in the simulation and
% due to possible mismodeling in the generation and reconstruction of the simulated samples.

%moved bkg model impact as first since it is additive, as per Jim's request
The background function used in the RSB fits was modified from the exponential model to a linear function 
%Studies on the background estimation using a linear function instead of the exponential model in the RSB fits have been conducted
to test the impact of the assumed background shape on the background yield estimate.
The differences between the yields obtained with the two functions is 
%We found the systematic impact on the background estimation due to the choice of the model is
small compared to the statistical uncertainty of the fitted yields and has a negligible effect on the expected upper limits. Therefore, no systematic uncertainty in $N_{\mathrm{exp}}$ is assigned for the background modeling.

We take into account the systematic uncertainty associated with the  corrections to the simulated lepton-identification efficiencies, derived from auxiliary measurements in data using $\jpsi\to \mup\mun$, $\epem\to \ell^+\ell^-\gamma$, and $\epem\to\epem\mup\mun$ events. 
These corrections are obtained as functions of momentum, polar angle and charge, and  applied to events reconstructed from simulation.
The systematic uncertainty is obtained by varying the corrections by their statistical and systematic uncertainties and estimating the impact of these variations on the selection efficiency.  %The applied corrections are then varied within their statistical and systematic uncertainties, and the final selection efficiency for each configuration is calculated. 
Adding the statistical and systematic variations in quadrature, the result is a relative uncertainty in the signal efficiency of 2.35(2.41)\% for the electron (muon) channel for Belle and 0.72(1.34)\% for the electron (muon) channel for Belle II.

%A track reconstruction uncertainty is furnished by a tag-and-probe study with $e^+ e^-\to \tau^+ \tau^-$ events. This results in an uncertainty of $0.24~\%$ per track for Belle II, 
%which linearly accumulates to a relative uncertainty of $0.96~\%$. Conversely, for Belle, the per-track uncertainty amounts to $0.35~\%$, 
%leading to a linear accumulation of $1.05~\%$.

In Belle II the difference between data and simulation in the track-reconstruction efficiency is measured in  $e^{+} e^{-} \to \tau^{+} \tau^{-}$ events with $\tau^{-} \to e^{-}\nu_{e}\nu_{\tau}$ and $\tau^{-} \to \pi^{-}\pi^{+}\pi^{-}\nu_{\tau}$ to yield a 0.24\% uncertainty per track for a total 0.96\% relative uncertainty.
%The difference between data and simulation in track-reconstruction efficiency is measured in $e^{+} e^{-} \to \tau^{+} \tau^{-}$ events, selecting decays $\tau^{+} \to e^{+}\nu_{e}\nu_{\tau}$ and $\tau^{-} \to \pi^{-}\pi^{+}\pi^{-}\nu_{\tau}$.  
%No discrepancy is observed in tracking efficiencies between data and simulation within the 0.24~\% uncertainty per track for Belle II, resulting in a relative systematic uncertainty of 0.96\% on the signal efficiency.
%A discrepancy of 0.24\% per track is observed, resulting in a relative systematic uncertainty of 0.96\%. 
For Belle, a $0.35\%$ per-track uncertainty is assigned using 
$D^{*+}\to\pi^+ D^0, D^0 \to\pi^+\pi^- K^0_{S}$ decays, resulting in a total relative signal efficiency uncertainty 
of 1.4\%.

%For Belle, the per track discrepancy was measured in 
%$D^*\to\pi D^0, D^0 \to\pi\pi K^0_{S}$ and is assessed to be 0.35\%, resulting in a total %relative uncertainty in the signal efficiency of 1.4\%.

%To correct for the non-perfect trigger simulation, the trigger efficiency difference 
%between data and MC is calculated on the $\tau^-\rightarrow K_S^0 \pi^-\nu$ reference channel. 
%For this purpose, the momentum distribution of the 3-prong signal side tracks are used.
%The trigger efficiency ratio between data and MC is fitted with a constant over all three 
%distributions and then averaged. As relative uncertainties, we derive $0.68~\%$ for Belle II and $0.9~\%$ for Belle, for both electron and muon channel.

For the Belle II experiment, we use triggers provided by the ECL and CDC sub-detectors. In data, the trigger efficiency is evaluated using independent trigger selections: the efficiency of the ECL-based trigger selection is obtained using events triggered by the CDC, while the efficiency of the CDC-based trigger selection is evaluated using events passing the ECL trigger requirements. The level of agreement between data and simulation efficiencies is 0.5\% for the ECL trigger selection and 4.3\% for the CDC trigger selection. Given that the efficiency of trigger selections based on the ECL only is 88\%, the weighted average of the data-simulation efficiency differences is computed to be 0.68\%, which is taken as the systematic uncertainty. For Belle, we use 
a trigger efficiency uncertainty of $0.9\%$ from reference~\cite{Belle:2023ziz}.

To correct for differences in $K_S^0$ reconstruction between data and simulation for Belle II, we compare $K_S^0$ yields in
data and MC using the decay 
$\tau^{-}\rightarrow K_S^0 \pi^-\nu$. 
The $K_S^0$  yields are obtained from fits to the $\pi^+\pi^-$ invariant mass in ten bins of the flight distance. The yield ratio is then fitted with a linear function, which is used to reweight the 
%then applied to the mean distance of the
reconstructed $K_S^0$ in signal MC to evaluate the systematic uncertainty.
%The ratio against the distance, normalized to the first bin. The uncertainty is determined by the slope of the linear fit through these points multiplied by the mean distance of  the reconstructed $K_S^0$ in signal MC. 
Values of $5.96\%$ and $5.31\%$ are obtained for the electron and muon channels respectively. 
For Belle,
$D^{*+}\to\pi^+ D^0, D^0 \to\pi^+\pi^- K^0_{S}$ decays are used, leading to a correction of the simulated signal efficiency by 
a factor of $0.9789$ per $K_S^0$ candidate  and  a $0.73~$\% contribution to the systematic uncertainty.% on the $K_S^0$ efficiency.
%to be added per $K_S^0$ candidate to the total systematic uncertainty. 

We obtain a systematic uncertainty on the signal efficiency in the $20\,\delta$ plane due to the BDT selection by applying the BDTs trained for the electron and muon channels to the reconstructed standard model decay $\tau^{-}\rightarrow K_S^0 \pi^{-}\nu$.
%on which we apply the BDTs trained for the electron and muon channels.
Employing the same BDT requirement as optimized for the
respective signal channels, we calculate BDT selection efficiencies on both simulation and data. Their relative difference is used as a systematic uncertainty, giving 
$1.49\%$ ($1.59\%$) and $5.06\%$ ($5.37\%$) for the electron and muon channels, respectively, for
Belle (Belle II).

To assess a systematic uncertainty due to potential mismodeling of the signal resolutions in \dE and \Mlk variables,
the size of the elliptical SR, whose values are given in Tab. \ref{tab:sig_resol}, is varied by $\pm 1\delta$ and the corresponding change in signal efficiency is used as systematic uncertainty. The check is performed on the final selected signal samples, where the signal peaks are mostly contained in the $1\delta$-wide SR and the signal leakage of non-Gaussian tails parametrized by the $\delta$ resolutions is below 6\% for all channels. We find relative systematic uncertainties of $(+3.15, -5.75)\%$ and $(+2.98, -5.22)\%$
for the Belle electron and muon channels, respectively.  The corresponding uncertainties for Belle II are $(+3.22, -4.51)\%$ and $(+2.63, -4.23)\%$.

The luminosities are determined independently using Bhabha and diphoton events.  The differences between these determinations are taken as systematic uncertainties, yielding an average uncertainty of $0.5\%$ for Belle II and $1.4$\% for Belle~\cite{ref:lumi_new, ref:belle_lumi}. %~\cite{Belle-II:2019usr}
The uncertainty on the production cross-section of tau pairs is evaluated in Ref.~\cite{Banerjee_2008} to be $0.003~$nb.

In Table \ref{tab:sys}, systematic uncertainties for Belle and Belle II analyses are summarized. 
%All sources entering the signal efficiency $\epsilon_{lK_S^0}$ are summed in quadrature.

\begin{table}[]
    \centering
    \caption{Relative systematic uncertainties in \% for the Belle and Belle II datasets.}
\begin{tabular}{cl|ll|ll|}
\multicolumn{1}{l}{}                                      &                       & \multicolumn{2}{c|}{$eK_S^0$ [\%]}                      & \multicolumn{2}{c|}{$\mu K_S^0$ [\%]}                     \\ \cline{3-6} 
\multicolumn{1}{l|}{Quantity}                             & Source       & \multicolumn{1}{l|}{Belle}          & Belle II       & \multicolumn{1}{l|}{Belle}          & Belle II       \\ \hline
\multicolumn{1}{c|}{\multirow{6}{*}{$\epsilon_{lK_S^0}$}} & Lepton identification & \multicolumn{1}{l|}{2.35}           & 0.72           & \multicolumn{1}{l|}{2.41}           & 1.34           \\ \cline{2-6} 
\multicolumn{1}{c|}{}                                     & Tracking efficiency   & \multicolumn{1}{l|}{1.40}            & 0.96           & \multicolumn{1}{l|}{1.4}            & 0.96           \\ \cline{2-6} 
\multicolumn{1}{c|}{}                                     & Trigger efficiency    & \multicolumn{1}{l|}{0.90}            & 0.68           & \multicolumn{1}{l|}{0.9}            & 0.68           \\ \cline{2-6} 
\multicolumn{1}{c|}{}                                     & $K_S^0$ efficiency    & \multicolumn{1}{l|}{0.73}           & 5.96           & \multicolumn{1}{l|}{0.73}           & 5.31           \\ \cline{2-6} 
\multicolumn{1}{c|}{}                                     & BDT efficiency        & \multicolumn{1}{l|}{1.49}           & 1.59           & \multicolumn{1}{l|}{5.06}           & 5.37           \\ \cline{2-6} 
\multicolumn{1}{c|}{}                                     & Signal region         & \multicolumn{1}{l|}{$^{+3.15}_{-5.75}$} & $^{+3.22}_{-4.51}$ & \multicolumn{1}{l|}{$^{+2.98}_{-5.22}$} & $^{+2.63}_{-4.23}$ \\ \hline
%\multicolumn{1}{c|}{$N_{exp}$}           & Momentum scale        & \multicolumn{1}{l|}{-}            & 1.49           & \multicolumn{1}{l|}{-}            & 2.44   \\ \hline
\multicolumn{1}{c|}{$\mathcal{L}$}                & Luminosity            & \multicolumn{1}{l|}{1.4}            & 0.5            & \multicolumn{1}{l|}{1.4}            & 0.5            \\ \hline
\multicolumn{1}{c|}{$\sigma_{\tau\tau}$}                  & Tau pair cross-section & \multicolumn{1}{l|}{0.3}            & 0.3            & \multicolumn{1}{l|}{0.3}            & 0.3            \\ 
\end{tabular}
\label{tab:sys}
\end{table}

%% file: 05-UL.tex
\section{Result}
\label{sec:results}
\begin{figure}
    \centering
    \includegraphics[page=1,width=0.48\columnwidth]{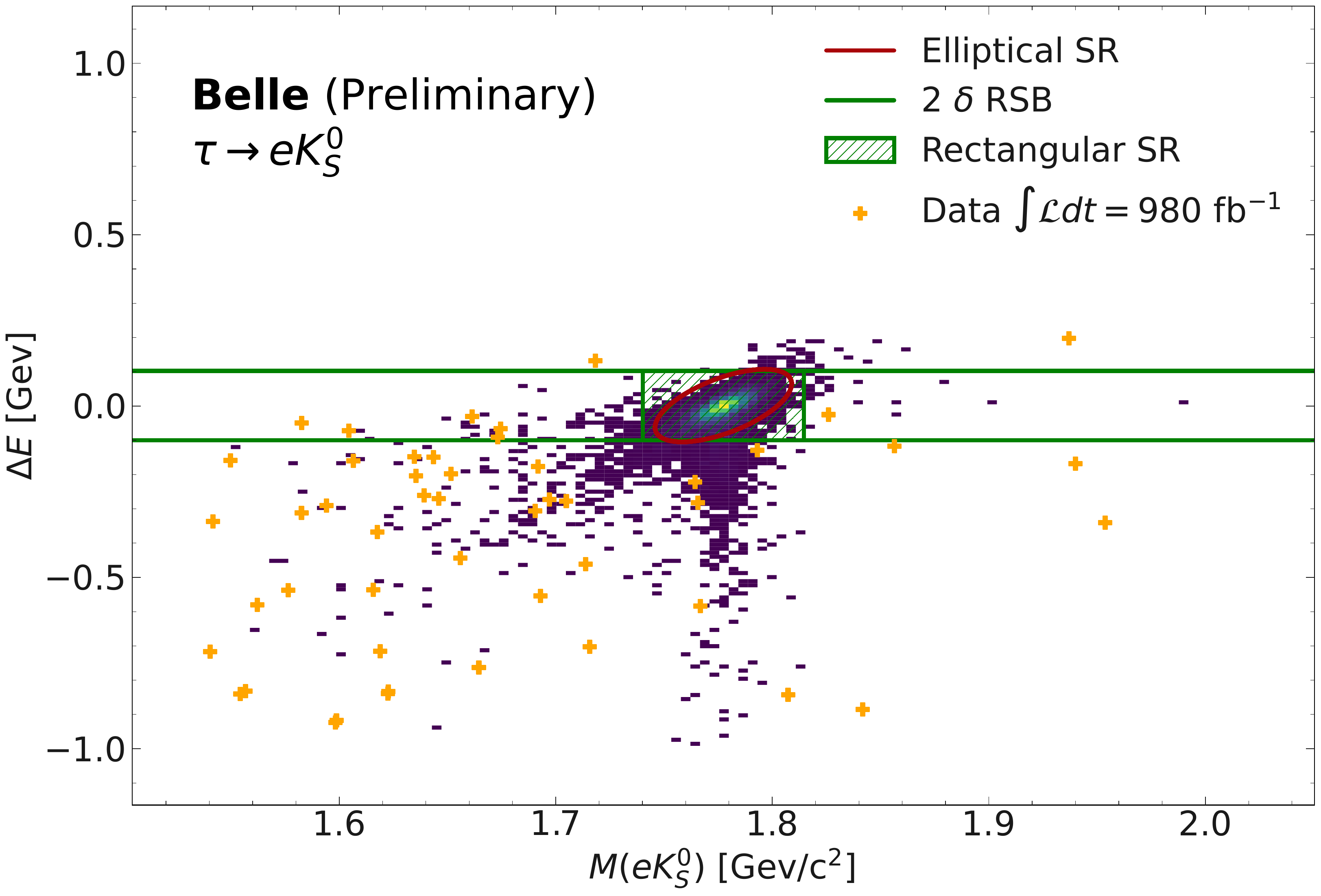}
    \includegraphics[page=1,width=0.48\columnwidth]{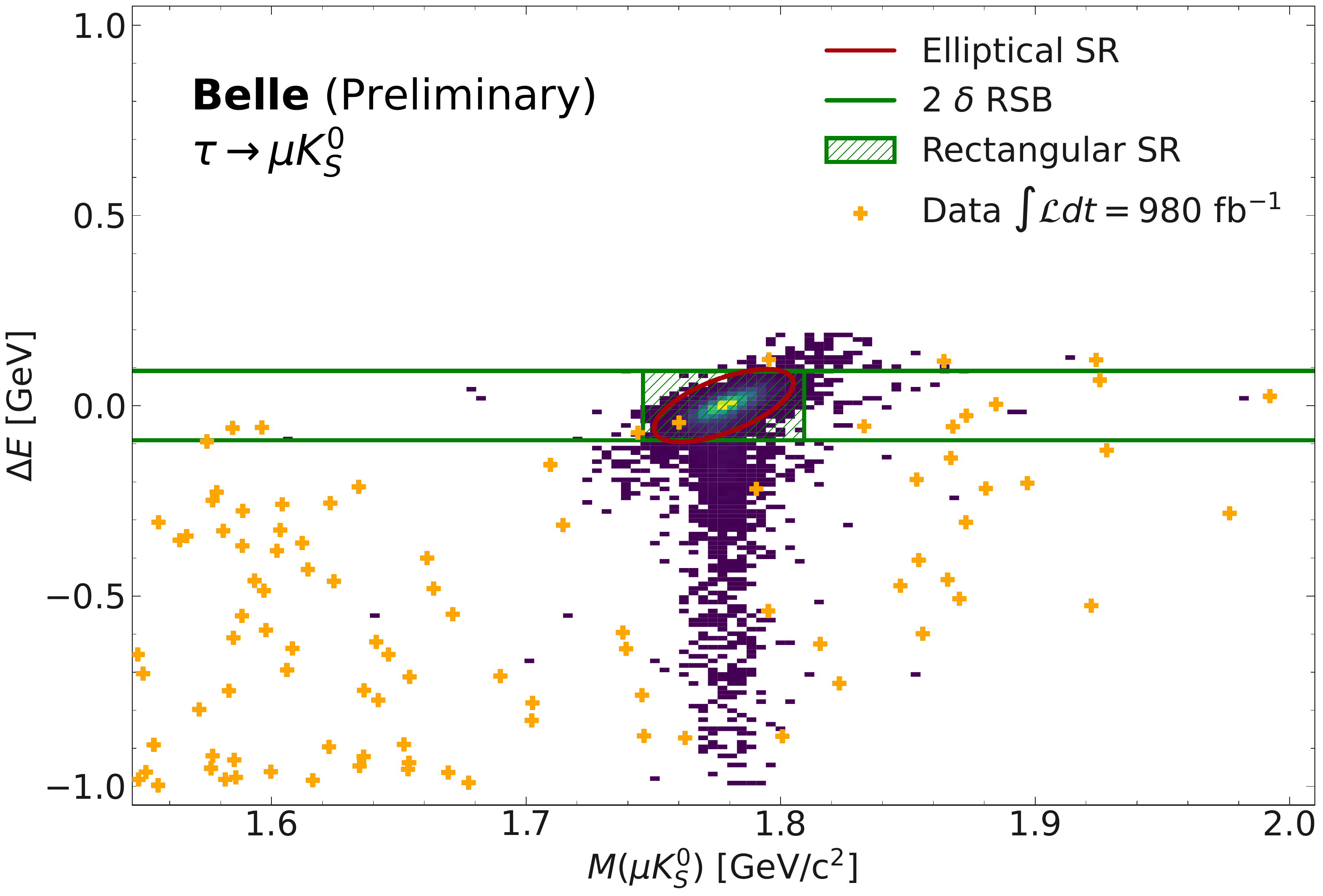}\\
    
    \includegraphics[page=1,width=0.48\columnwidth]{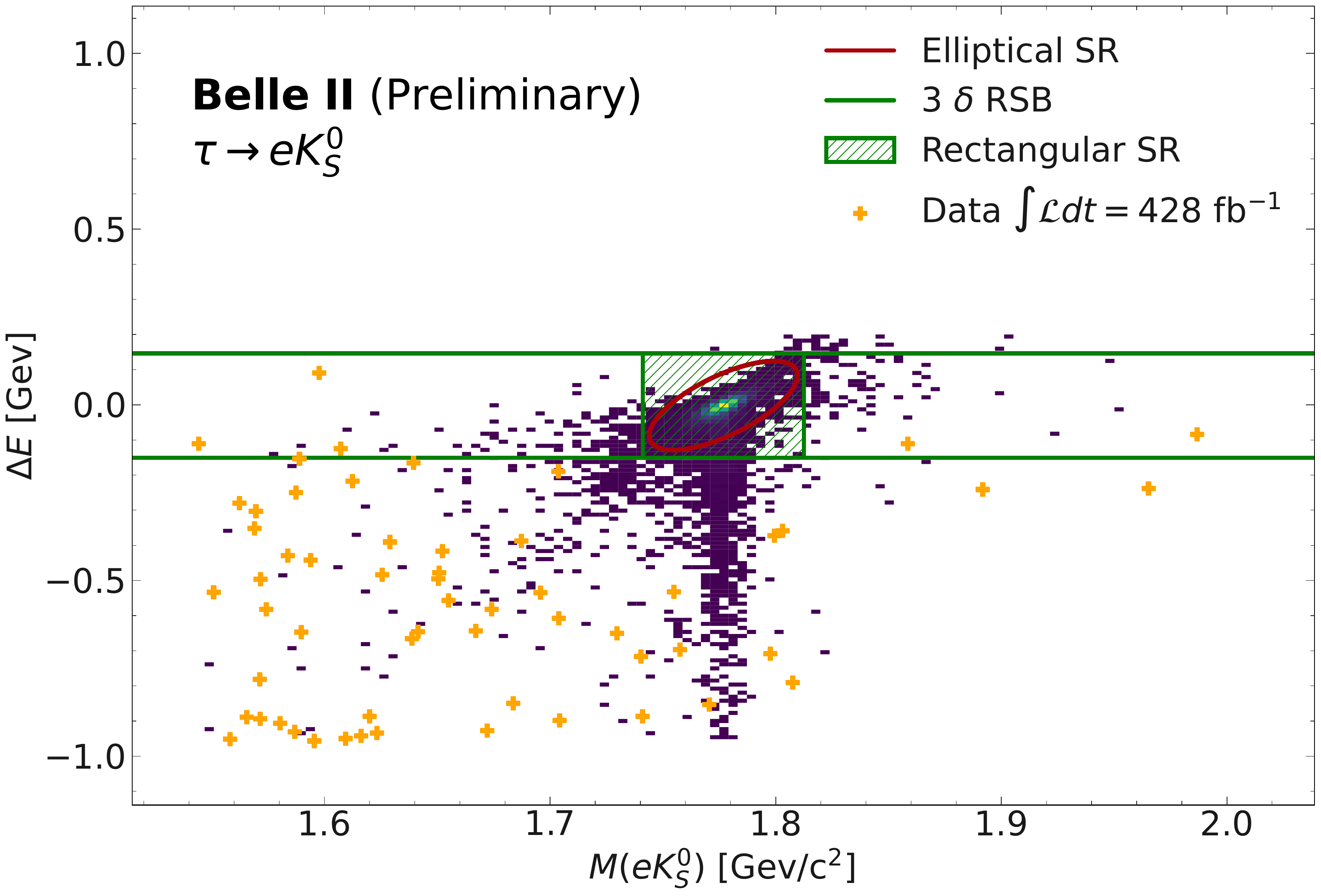}
    \includegraphics[page=1,width=0.48\columnwidth]{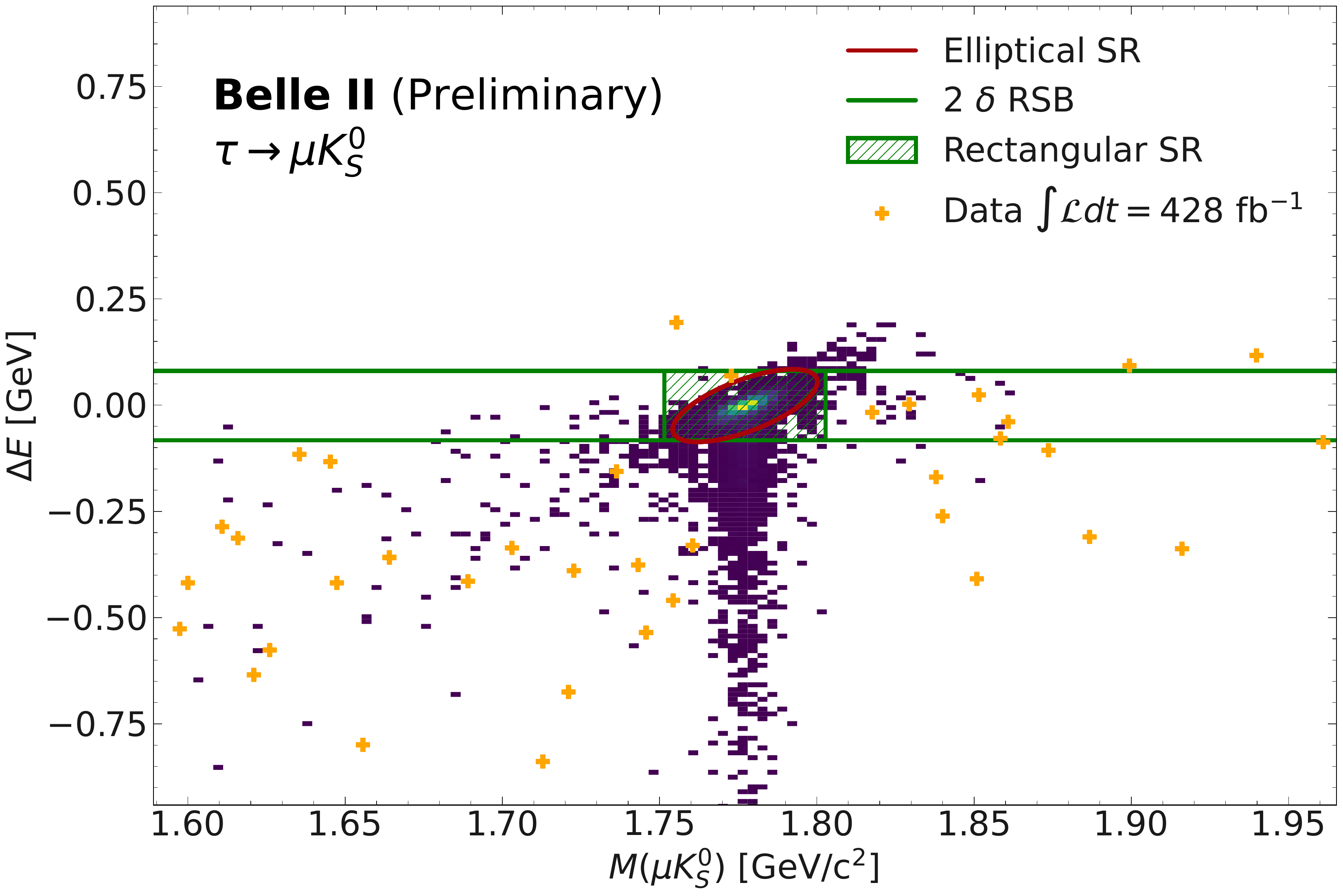}
    
    \caption{Scatter plots of selected events in the (\Mlk, \dE) plane for signal simulation (violet) and data (orange). 
 %     RVK: IT'S NOT A GOOD IDEA TO USE ONLY COLOR TO DISTINGUISH DATA SETS - SOME PEOPLE CAN'T SEE COLOR DIFFERENCES.  CONSIDER USING DIFFERENT SYMBOL SHAPES AS WELL AS COLOR.  THE EVENTS ARE SPARSE SO THIS WILL ALSO IMPROVE VISIBILITY.
    The elliptical SR is shown in red, 
    the rectangular SR as a hatched green area and the RSB is indicated as green horizontal lines. Plots show distributions for the electron (left) and 
    muon (right) channels for Belle (upper-row) and Belle II (lower-row). For the muon mode in Belle
    one data event is observed in the SR.}
    \label{fig:final_2D}
\end{figure}

The distribution of events in the (\Mlk, \dE) plane is shown in Fig.~\ref{fig:final_2D} for the 428\invfb Belle II and 980\invfb Belle data samples. When examining the signal region in data (unboxing), we observe 0(0) events in the Belle II electron (muon) channel and 0(1) events for 
the Belle electron (muon) channel in the signal region.

%\textcolor{red}{
The \taulks branching 
fractions are obtained from the number of signal events $N_{\mathrm{sig}}$, the signal efficiencies $\varepsilon_{\mathrm{\ell K_s^0}}$ and the number of tau leptons produced $N_\tau$:
\begin{equation}\label{eq:result}
 \mathcal{B}(\taulks) = \frac{N_{\mathrm{sig}}}{N_{\tau}\times\varepsilon_{\mathrm{\ell K_s^0}}}=\frac{N_{\mathrm{obs}}-N_{\mathrm{exp}}}{\mathcal{L} \times 2\sigma_{\tau\tau} \times \varepsilon_{\mathrm{\ell K_s^0}}},
\end{equation}
where $\mathcal{L}$ is the integrated luminosity of the data sets,
$N_{\mathrm{exp}}$ is the number of expected background events and $N_{\mathrm{obs}}$ is the number of observed events. The $\tau$-pair production cross-section $\sigma_{\tau\tau}$, determined from the weighted average of the cross-sections at the different center-of-mass energies at which the data were taken, is 
$0.919\pm 0.003\,\nb$ for Belle II data and  $0.916\pm 0.003\,\nb$ for Belle data. 

As we do not observe  any significant excess above the expected background within the signal region, 
we calculate $90\,\%$ C.L. upper limits on the \taulks branching 
fractions using
%:% Eq.~\ref{eq:result}: 
%\begin{equation}\label{eq:result}
% \mathcal{B}(\taulks) = \frac{N_{\mathrm{obs}}-N_{\mathrm{exp}}}{\mathcal{L} \times 2\sigma_{\tau\tau} \times \varepsilon_{\mathrm{\ell K_s^0}}},
%\end{equation}
%where $\mathcal{L}$ in the integrated luminosity of the data sets, $ \varepsilon_{\mathrm{\ell K_s^0}}$ the signal efficiency, $N_{\mathrm{exp}}$ is the number of expected background events and $N_{\mathrm{obs}}$ is the number of observed events. The $\tau$-pair production cross-section $\sigma_{\tau\tau}$, determined from the weighted average of the cross-sections at the different center-of-mass energies at which the data were taken, is 
%$0.919\pm 0.003\,\nb$ for Belle II data and  $0.916\pm 0.003\,\nb$ for Belle data.  
%The UL is estimated using 
the $\mathrm{CL}_s$ method \cite{Junk:1999kv, Read:2002hq} in a frequentist approach implemented in the \texttt{pyhf} library~\cite{pyhf, pyhf_joss}.

To determine the expected limit sensitivity, 
%Since the calculation is performed for two experiments, Belle and Belle II,
we generate 10000 pseudo-experiments at 50  points uniformly distributed 
in the branching ratio range of $(0-4)\times 10^{-8}$ in two bins, one for each experiment, each with their respective signal efficiencies and expected 
background yields. The total statistical and systematic uncertainties affecting each experimental input, as discussed in Section~\ref{sec:systematics}, 
are combined in quadrature.

\begin{figure}[htb]
    \centering
   \includegraphics[width=0.48\columnwidth]{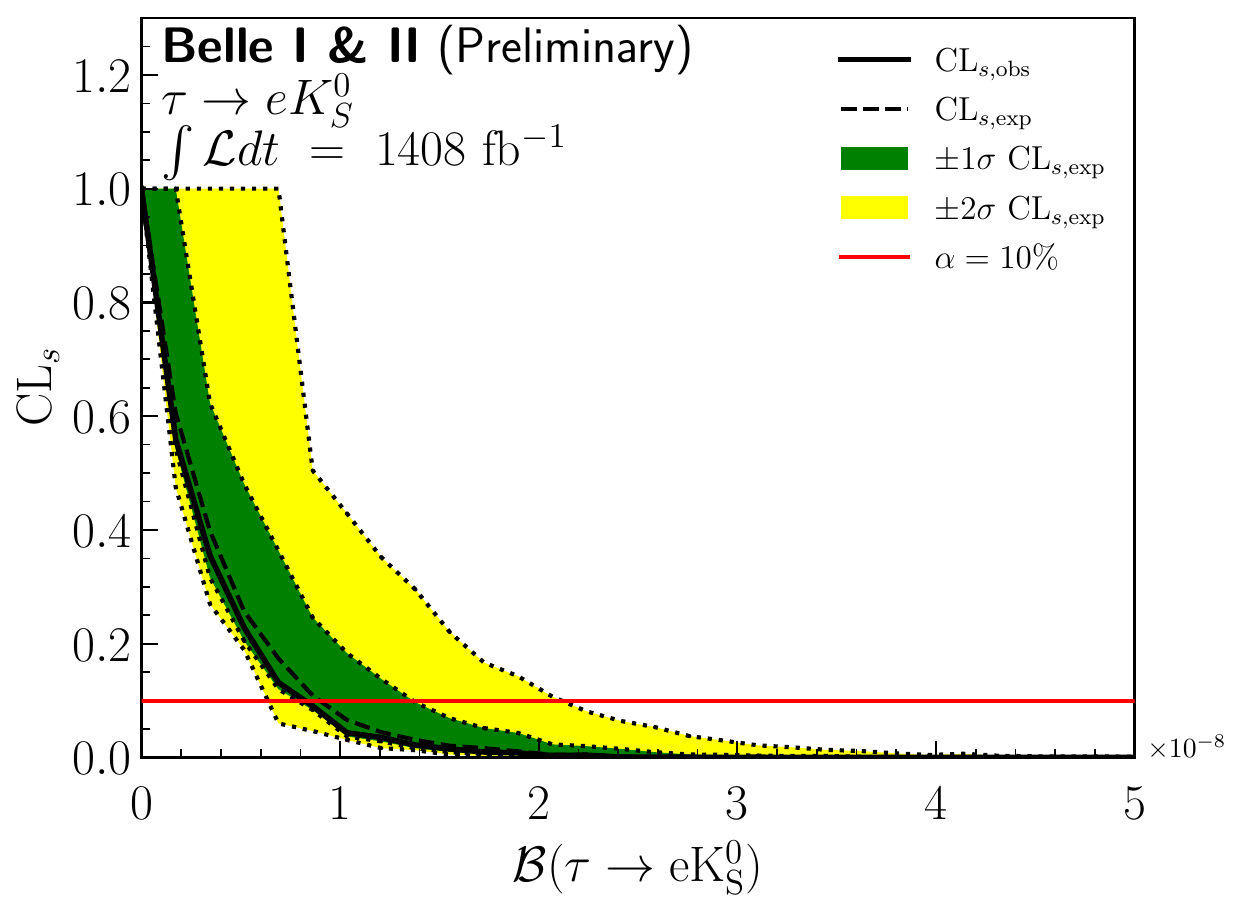}
   \includegraphics[width=0.48\columnwidth]{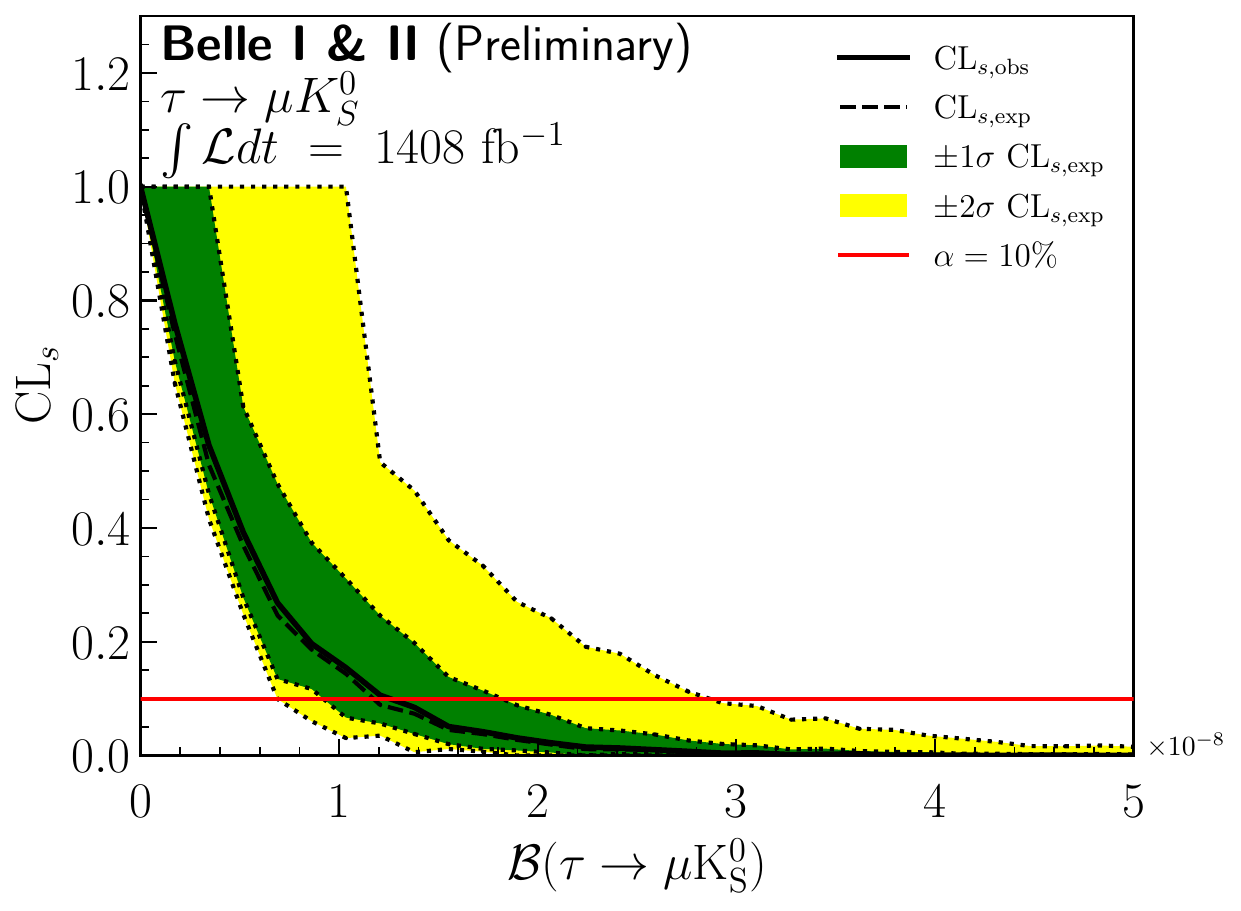}
    \caption{Observed (solid black curve) and expected (dashed black curve) $\mathrm{CL}_s$ as a function of the assumed branching fractions for 
    $\tau^-\to e^- K_S^0$ (left) and $\tau^-\to \mu^- K_S^0$ (right) decays. The red line corresponds to the 90~\% C.L.}
    \label{fig:CLs}
\end{figure}

Figure \ref{fig:CLs} displays the $\mathrm{CL}_s$ curves computed as a function of the branching fractions for the combined Belle and Belle II 
datasets for the $\tau^-\to e^-K_S^0$ and $\tau^-\to \mu^- K_S^0$ decays. The dashed black line represents the expected $\mathrm{CL}_s$, while the green and yellow bands 
show the $\pm1\sigma$ and $\pm2\sigma$ contours, respectively.

The expected limits, assuming an observed number of events consistent with the background estimation in Table \ref{tab:expected_data}, 
are $0.9\times 10^{-8}$ and $1.2\times 10^{-8}$ at 90~\% C.L.\ for the electron and muon channel, respectively. The observed limits in data after unboxing are $0.8\times 10^{-8}$ and $1.2\times 10^{-8}$ at 90~\% C.L for the electron and muon channels, respectively.

%% file: 06-summ.tex
\section{Summary}
\label{sec:summ}
We present a search for the LFV decays $\tau^-\to e^-K_S^0$ and $\tau^-\to \mu^- K_S^0$
using
428\invfb of data collected by the Belle II experiment and 
980\invfb of data collected by the Belle experiment, which is the world's largest tau pair data set.
A set of dedicated Boosted Decision Tree classifiers are used to discriminate signal decays from background processes. The signal yield is determined
in the plane of the reconstructed $\tau$ mass and the difference between the reconstructed and expected $\tau$ energy, two variables in which the signal decays peak.
%, using a %novel selection strategy exploiting a BDT-based background rejection, on
We observe 0(0) events for the electron channel and 0(1) events for the muon channel in the signal region for Belle II(Belle) and thus set 90~\% C.L. upper limits on both channels computed in a frequentist approach.
The observed limits at 90\% C.L. %are 0.804(1.06)$\times10^{-8}$ for the electron and 1.20(1.24)$\times10^{-8}$ for the muon channel, 
are 0.8$\times10^{-8}$ for the electron and 1.2$\times10^{-8}$ for the muon channel, 3.3 and 1.9 times more stringent, respectively, than
the previous best limits of 2.6$\times10^{-8}$ and 2.3$\times10^{-8}$~\cite{Belle:2010rxj}.
%These are the most restrictive bounds on $\tau \to \ell K_S^{0}$ decays to date.

%% file: acknowledgements.tex
% Policy from October 20, 2022
This work, based on data collected using the Belle II detector, which was built and commissioned prior to March 2019,
and data collected using the Belle detector, which was operated until June 2010,
was supported by
%Armenia
Higher Education and Science Committee of the Republic of Armenia Grant No.~23LCG-1C011;
%Australia
Australian Research Council and Research Grants
No.~DP200101792, % Jackson
No.~DP210101900, % Urquijo
No.~DP210102831, % Sevior
No.~DE220100462, % Hsu
No.~LE210100098, % Infrastructure
and
No.~LE230100085; % Infrastructure
%Austria
Austrian Federal Ministry of Education, Science and Research,
Austrian Science Fund (FWF) Grants
DOI:~10.55776/P34529,
DOI:~10.55776/J4731,
DOI:~10.55776/J4625,
DOI:~10.55776/M3153,
and
DOI:~10.55776/PAT1836324,
and
Horizon 2020 ERC Starting Grant No.~947006 ``InterLeptons'';
%Canada
Natural Sciences and Engineering Research Council of Canada, Compute Canada and CANARIE;
%China
National Key R\&D Program of China under Contract No.~2022YFA1601903,
National Natural Science Foundation of China and Research Grants
No.~11575017,
No.~11761141009,
No.~11705209,
No.~11975076,
No.~12135005,
No.~12150004,
No.~12161141008,
No.~12475093,
and
No.~12175041,
and Shandong Provincial Natural Science Foundation Project~ZR2022JQ02;
%Czech Republic
the Czech Science Foundation Grant No.~22-18469S 
and
Charles University Grant Agency project No.~246122;
%EU
European Research Council, Seventh Framework PIEF-GA-2013-622527,
Horizon 2020 ERC-Advanced Grants No.~267104 and No.~884719,
Horizon 2020 ERC-Consolidator Grant No.~819127,
Horizon 2020 Marie Sklodowska-Curie Grant Agreement No.~700525 ``NIOBE''
and
No.~101026516,
and
Horizon 2020 Marie Sklodowska-Curie RISE project JENNIFER2 Grant Agreement No.~822070 (European grants);
%France
L'Institut National de Physique Nucl\'{e}aire et de Physique des Particules (IN2P3) du CNRS
and
L'Agence Nationale de la Recherche (ANR) under Grant No.~ANR-21-CE31-0009 (France);
%Germany
BMBF, DFG, HGF, MPG, and AvH Foundation (Germany);
%India
Department of Atomic Energy under Project Identification No.~RTI 4002,
Department of Science and Technology,
and
UPES SEED funding programs
No.~UPES/R\&D-SEED-INFRA/17052023/01 and
No.~UPES/R\&D-SOE/20062022/06 (India);
%Israel
Israel Science Foundation Grant No.~2476/17,
U.S.-Israel Binational Science Foundation Grant No.~2016113, and
Israel Ministry of Science Grant No.~3-16543;
%Italy
Istituto Nazionale di Fisica Nucleare and the Research Grants BELLE2,
and
the ICSC – Centro Nazionale di Ricerca in High Performance Computing, Big Data and Quantum Computing, funded by European Union – NextGenerationEU;
%Japan
Japan Society for the Promotion of Science, Grant-in-Aid for Scientific Research Grants
No.~16H03968,
No.~16H03993,
No.~16H06492,
No.~16K05323,
No.~17H01133,
No.~17H05405,
No.~18K03621,
No.~18H03710,
No.~18H05226,
No.~19H00682, % Niigata
No.~20H05850,
No.~20H05858,
No.~22H00144,
No.~22K14056,
No.~22K21347,
No.~23H05433,
No.~26220706,
and
No.~26400255,
%the National Institute of Informatics, and Science Information NETwork 5 (SINET5), 
and
the Ministry of Education, Culture, Sports, Science, and Technology (MEXT) of Japan;  
%Korea
National Research Foundation (NRF) of Korea Grants
No.~2016R1-D1A1B-02012900,
No.~2018R1-A6A1A-06024970,
No.~2021R1-A6A1A-03043957,
No.~2021R1-F1A-1060423,
No.~2021R1-F1A-1064008,
No.~2022R1-A2C-1003993,
No.~2022R1-A2C-1092335,
No.~RS-2023-00208693,
No.~RS-2024-00354342
and
No.~RS-2022-00197659,
Radiation Science Research Institute,
Foreign Large-Size Research Facility Application Supporting project,
the Global Science Experimental Data Hub Center, the Korea Institute of
Science and Technology Information (K24L2M1C4)
and
KREONET/GLORIAD;
%Malaysia
Universiti Malaya RU grant, Akademi Sains Malaysia, and Ministry of Education Malaysia;
%Mexico
% CINVESTAV-IPN, UNAM, UAS, BUAP and CONACYT are funded under
Frontiers of Science Program Contracts
No.~FOINS-296,
No.~CB-221329,
No.~CB-236394,
No.~CB-254409,
and
No.~CB-180023, and SEP-CINVESTAV Research Grant No.~237 (Mexico);
%Poland
the Polish Ministry of Science and Higher Education and the National Science Center;
%Russia
the Ministry of Science and Higher Education of the Russian Federation
and
the HSE University Basic Research Program, Moscow;
%Saudi Arabia
University of Tabuk Research Grants
No.~S-0256-1438 and No.~S-0280-1439 (Saudi Arabia), and
Researchers Supporting Project number (RSPD2025R873), King Saud University, Riyadh,
Saudi Arabia;
%Slovenia
Slovenian Research Agency and Research Grants
No.~J1-9124
and
No.~P1-0135;
%Spain
Ikerbasque, Basque Foundation for Science,
the State Agency for Research of the Spanish Ministry of Science and Innovation through Grant No. PID2022-136510NB-C33,
Agencia Estatal de Investigacion, Spain
Grant No.~RYC2020-029875-I
and
Generalitat Valenciana, Spain
Grant No.~CIDEGENT/2018/020;
%Swiss (Belle 1)
the Swiss National Science Foundation;
%Sweden
The Knut and Alice Wallenberg Foundation (Sweden), Contracts No.~2021.0174 and No.~2021.0299;
%Taiwan
National Science and Technology Council,
and
Ministry of Education (Taiwan);
%Thailand
Thailand Center of Excellence in Physics;
%Turkey
TUBITAK ULAKBIM (Turkey);
%Ukraine
National Research Foundation of Ukraine, Project No.~2020.02/0257,
and
Ministry of Education and Science of Ukraine;
%USA
the U.S. National Science Foundation and Research Grants
No.~PHY-1913789 % Indiana CEEM
and
No.~PHY-2111604, % Luther
and the U.S. Department of Energy and Research Awards
No.~DE-AC06-76RLO1830, % PNNL
No.~DE-SC0007983, % Wayne State
No.~DE-SC0009824, % Florida
No.~DE-SC0009973, % VPI
No.~DE-SC0010007, % Duke
No.~DE-SC0010073, % South Carolina
No.~DE-SC0010118, % Carnegie Mellon
No.~DE-SC0010504, % Hawaii
No.~DE-SC0011784, % Cincinnati
No.~DE-SC0012704, % BNL
No.~DE-SC0019230, % Duke
No.~DE-SC0021274, % Mississippi
No.~DE-SC0021616, % Mississippi
No.~DE-SC0022350, % Louisville
No.~DE-SC0023470; % South Alabama
%last group
and
%Vietnam
the Vietnam Academy of Science and Technology (VAST) under Grants
No.~NVCC.05.12/22-23
and
No.~DL0000.02/24-25.

% Policy from October 20, 2022
These acknowledgements are not to be interpreted as an endorsement of any statement made
by any of our institutes, funding agencies, governments, or their representatives.

We thank the SuperKEKB team for delivering high-luminosity collisions;
the KEK cryogenics group for the efficient operation of the detector solenoid magnet and IBBelle on site;
the KEK Computer Research Center for on-site computing support; the NII for SINET6 network support;
and the raw-data centers hosted by BNL, DESY, GridKa, IN2P3, INFN, 
PNNL/EMSL, 
and the University of Victoria.

%% file: 07-appendix.tex
\clearpage
\begin{appendices}
\section{Simulated MC samples in Belle and Belle II}
\label{sec:appA}
\begin{table}[h]
\caption{MC samples simulated in Belle and Belle II (ab$^{-1}$).}
\begin{center}
\begin{tabular}{ccc}
\hline
\multicolumn{1}{|c|}{\textbf{MC sample}}                   & \multicolumn{1}{c|}{\textbf{Belle}} & \multicolumn{1}{c|}{\textbf{Belle II}} \\ \hline
\multicolumn{1}{|c|}{$e^+e^-\to \tau^+\tau^-$}             & \multicolumn{1}{c|}{5.79}              & \multicolumn{1}{c|}{7}                 \\ \hline
\multicolumn{1}{|c|}{$e^+e^-\to \tau^+\tau^-\tau^+\tau^-$} & \multicolumn{1}{c|}{}               & \multicolumn{1}{c|}{10}                 \\ \hline
\multicolumn{1}{|c|}{$e^+e^-\to \mu^+\mu^-(\gamma)$}               & \multicolumn{1}{c|}{4.685}              & \multicolumn{1}{c|}{1}                 \\ \hline
\multicolumn{1}{|c|}{$e^+e^-\to \mu^+\mu^-\mu^+\mu^-$}     & \multicolumn{1}{c|}{}               & \multicolumn{1}{c|}{2}                 \\ \hline
\multicolumn{1}{|c|}{$e^+e^-\to \mu^+\mu^-\tau^+\tau^-$}   & \multicolumn{1}{c|}{}               & \multicolumn{1}{c|}{2}                 \\ \hline
\multicolumn{1}{|c|}{$e^+e^-\to e^+e^-(\gamma)$}           & \multicolumn{1}{c|}{0.52}              & \multicolumn{1}{c|}{0.12}                 \\ \hline
\multicolumn{1}{|c|}{$e^+e^-\to e^+e^-\mu^+\mu^-$}          & \multicolumn{1}{c|}{4.685}              & \multicolumn{1}{c|}{0.2}                 \\ \hline
\multicolumn{1}{|c|}{$e^+e^-\to e^+e^-\tau^+\tau^-$}        & \multicolumn{1}{c|}{}               & \multicolumn{1}{c|}{2}                 \\ \hline
\multicolumn{1}{|c|}{$e^+e^-\to e^+e^-e^+e^-$}              & \multicolumn{1}{c|}{4.685}              & \multicolumn{1}{c|}{0.2}                 \\ \hline
\multicolumn{1}{|c|}{$e^+e^-\to e^+e^-u\bar{u}$}           & \multicolumn{1}{c|}{5.79}              & \multicolumn{1}{c|}{}                  \\ \hline
\multicolumn{1}{|c|}{$e^+e^-\to e^+e^-d\bar{d}$}           & \multicolumn{1}{c|}{5.79}              & \multicolumn{1}{c|}{}                  \\ \hline
\multicolumn{1}{|c|}{$e^+e^-\to e^+e^-s\bar{s}$}           & \multicolumn{1}{c|}{5.79}              & \multicolumn{1}{c|}{}                  \\ \hline
\multicolumn{1}{|c|}{$e^+e^-\to e^+e^-c\bar{c}$}           & \multicolumn{1}{c|}{5.79}              & \multicolumn{1}{c|}{}                  \\ \hline
\multicolumn{1}{|c|}{$e^+e^-\to e^+e^-\pi^+\pi^-$}          & \multicolumn{1}{c|}{}               & \multicolumn{1}{c|}{1}                 \\ \hline
\multicolumn{1}{|c|}{$e^+e^-\to e^+e^-K^+K^-$}              & \multicolumn{1}{c|}{}               & \multicolumn{1}{c|}{2}                 \\ \hline
\multicolumn{1}{|c|}{$e^+e^-\to e^+e^-p\bar{p}$}              & \multicolumn{1}{c|}{}               & \multicolumn{1}{c|}{2}                 \\ \hline
\multicolumn{1}{|c|}{$e^+e^-\to q\bar{q}$}                 & \multicolumn{1}{c|}{4.155}              & \multicolumn{1}{c|}{7}                 \\ \hline
\multicolumn{1}{|c|}{$B\bar{B}$}                           & \multicolumn{1}{c|}{6.952}              & \multicolumn{1}{c|}{1}                 \\ \hline
\multicolumn{1}{l}{}                                       & \multicolumn{1}{l}{}                & \multicolumn{1}{l}{}           
\end{tabular}

\label{tab:simMC}
\end{center}
\end{table}

\section{Preselections for $\tau\to e K_S^0$ decay}
\label{sec:electron_presel}
Here we list the different requirements applied to the Belle and Belle~II data as pre-selections for the electron channel, $\tau\to e K_S^0$.
\clearpage

\begin{longtable}{|p{7cm}|p{7cm}|}
\hline
\textbf{Belle Selections} & \textbf{Belle II Selections} \\
\hline
\endfirsthead
\hline
\textbf{Belle Selections} & \textbf{Belle II Selections} \\
\hline
\endhead
\hline
\endfoot

\multicolumn{2}{|l|}{\textbf{Electron-tag}} \\
\hline
\begin{itemize}
    \item $\cos(p_{\mathrm{miss}}^ *, p_{\mathrm{tag}}^*) > 0$ 
    \item $ N^{\rm{tot}}_{\pi^0} = 0$
    \item $\Delta E_{\mathrm{tag}} < -1$ GeV
    \item $M^{ee}(K_S^0) > 0.2$ GeV/c$^2$ 
    \item $M_{\mathrm{tag}}(e\gamma) < 7$ GeV/c$^2$
\end{itemize}
& 
\begin{itemize}
    \item $0.3 < \theta_{\mathrm{miss}} < 2.7$ 
    \item $0.49 < M_{\pi^{+}\pi^{-}} < 0.505$ GeV/c$^2$
    \item $\cos(\theta_{K_S}-\theta_{\hat{t}}) > 0.8$ 
    \item $E_{K_s} > 1$ GeV
    \item $| p^*_{z\mathrm{tag}} | < 2.5$ GeV/c
\end{itemize} \\
\hline
\multicolumn{2}{|l|}{\textbf{Muon-tag}} \\
\hline
\begin{itemize}
    \item $ N^{\rm{tag}}_{\pi^0} = 0$
    \item $M_{\mathrm{tag}}(e\gamma) < 0.2$ GeV/c$^2$
\end{itemize}
&
\begin{itemize}
    \item $ N^{\rm{tag}}_{\pi^0} = 0$
    \item $p^{*}_{T, \mathrm{third}} > 0.1$ GeV/c 
\end{itemize} \\
\hline
\multicolumn{2}{|l|}{\textbf{Pion-tag}} \\
\hline
\begin{itemize}
    \item $M^{ee}(K_S^0) > 0.2$ GeV/c$^2$
\end{itemize}
&
\begin{itemize}
    \item $-3 < p^*_{z}(\ell) < 3$ GeV/c
\end{itemize} \\
\hline
\multicolumn{2}{|l|}{\textbf{All tags}} \\
\hline
\begin{itemize}
    \item $p_{\mathrm{miss}} > (- 4 \times M_{\mathrm{miss}} - 1)$
    \item $p_{\mathrm{miss}} > (1.3 \times M_{\mathrm{miss}} - 0.8)$
    \item $N^{\rm{tot}}_{\gamma} < 3$
\end{itemize}
&
\begin{itemize}
    \item $M^{ee}(K_S^0) > 0.25$ GeV/c$^2$
    \item $\cos(\theta_{\mathrm{miss}}^ *,\theta^*_{\mathrm{tag}}) > 0$ 
    \item $\Delta E_{\mathrm{tag}} < 1$ GeV
    \item $0.85 < thrust < 0.98$
\end{itemize} \\
\hline
\end{longtable}
\clearpage

\section{Preselections for $\tau\to \mu K_S^0$ decay}
\label{sec:muon_presel}

Here we list the different requirements applied to the Belle and Belle~II data as pre-selections for the muon channel, $\tau\to \mu K_S^0$.

\begin{longtable}{|p{7cm}|p{7cm}|}
\hline
\textbf{Belle Selections} & \textbf{Belle II Selections} \\
\hline
\endfirsthead

\hline
\textbf{Belle Selections} & \textbf{Belle II Selections} \\
\hline
\endhead

\hline
\endfoot

\multicolumn{2}{|l|}{\textbf{Electron-tag}} \\
\hline
\begin{itemize}
    \item $N^{\rm{tag}}_{\gamma} < 2$
\end{itemize}
& 
\begin{itemize}
    \item $E_\ell > 0.5$ GeV
    \item $ N^{\rm{tot}}_{\pi^0} = 0$
    \item $\chi_{POCA}(\tau) <$ 10 
\end{itemize} \\
\hline
\multicolumn{2}{|l|}{\textbf{Muon-tag}} \\
\hline
\begin{itemize}
    \item $M^{ee}(K_S^0) > 0.2$ GeV/c$^2$
    \item $\cos(p_{\mathrm{miss}}^ *,p_{\mathrm{tag}}^*) > 0$ 
    \item $M^{\mathrm{tag}}_{\mathrm{bc}} > 3$ GeV/c$^2$ 
     \item $ N^{\rm{tot}}_{\pi^0} = 0$
\end{itemize}
&
\\
\hline
\multicolumn{2}{|l|}{\textbf{Pion-tag}} \\
\hline
\begin{itemize}
     \item $ N^{\rm{sig}}_{\pi^0} = 0$
\end{itemize}
&
\begin{itemize}
    \item $E_\ell > 0.8$ GeV
    \item $\theta(\ell, K_s) < 1$ (angle between lepton and $K_S^0$)
    \item $p^{*}_{T,\mathrm{sub}} > 0.3$ GeV/c 
    \item $\cos(\theta_{\mathrm{miss}}^ * - \theta^*_{\mathrm{tag}}) > 0$ 
\end{itemize} \\
\hline
\multicolumn{2}{|l|}{\textbf{All tags}} \\
\hline
\begin{itemize}
   \item $ N^{\rm{tot}}_{\pi^0} < 2$
    \item $E^*_{\mathrm{tag}} < 5$ GeV (energy of the tag track in the c.m.)
    \item $0.9 < thrust < 0.975$
\end{itemize}
& \\
\hline
\end{longtable}

\section{BDT input variables}
\label{sec:bdtvar}
\begin{table}[h!]
\caption{Variables used for BDT training. Ordering follows the feature importance in the training.}
\begin{center}
\begin{tabular}{|l|}
\hline
Signal tau variables\\
\hline
Transverse momentum of largest $p_{T, lead}$, middle $p_{T, sub}$ and lowest $p_{T, third}$ signal-side\\ track in c.m.  \\
Signal tau $p_T$ \\ 
Signal tau flight time and its uncertainty \\ 
Signal tau flight distance and its uncertainty\\
Angle between $K_S^0$ and lepton \\
$K_S^0$ mass calculated with proton mass hypothesis for first pion \\ 
$K_S^0$ mass calculated with proton mass hypothesis for second pion \\  
Energy of $K_S^0$ \\ 
Momentum of $K_S^0$ in c.m. \\ 
$K_S^0$ Error of the flight distance \\ 
$K_S^0$ flight distance \\ 
$K_S^0$ Error of flight Time \\ 
$K_S^0$ flight time \\ 
Momentum of lepton in c.m. \\ 
\hline
Tag tau variables\\
\hline
Mass of tag-side tau \\ 
Beam constrained mass of tag-side tau \\ 
Invariant mass of tag side track and photons  \\ 
\hline
Event based variables\\
\hline
Transverse missing momentum of event in c.m. \\ 
Visible energy of event in c.m. \\ 
Missing mass squared of event \\ 
Thrust value\\ 
Total energy of photons in event \\ 
Number of photons on the tag side \\ 
Total energy of photons on tag side  \\ 
Number of  neutral pions \\ 
Missing momentum of event in c.m. \\
Missing energy of event in c.m. \\ 
Cosine of the angle between missing momentum  and tag-side track in c.m. \\ 
Cosine of the angle between missing momentum and signal-side lepton \\ 
Number of photons  in the event                \\ 
\hline
\end{tabular}

\label{tab:BDTvars}
\end{center}
\end{table}

\end{appendices}

%% file: jhep.bbl
\providecommand{\href}[2]{#2}\begingroup\raggedright\begin{thebibliography}{10}

\bibitem{ref:SM_cLFV}
T.~Li, M.A.~Schmidt, C.-Y.~Yao and M.~Yuan, \emph{{Charged lepton flavor violation in light of the muon magnetic moment anomaly and colliders}}, \href{https://doi.org/10.1140/epjc/s10052-021-09569-9}{\emph{Eur. Phys. J. C} {\bfseries 81} (2021) 811} [\href{https://arxiv.org/abs/2104.04494}{{\ttfamily 2104.04494}}].

\bibitem{ref:tau_lfv}
G.~Hern\'andez-Tom\'e, G.~L\'opez~Castro and P.~Roig, \emph{{Flavor violating leptonic decays of $\tau$ and $\mu$ leptons in the Standard Model with massive neutrinos}}, \href{https://doi.org/10.1140/epjc/s10052-019-6563-4}{\emph{Eur. Phys. J. C} {\bfseries 79} (2019) 84} [\href{https://arxiv.org/abs/1807.06050}{{\ttfamily 1807.06050}}].

\bibitem{Blackstone:2019njl}
P.~Blackstone, M.~Fael and E.~Passemar, \emph{{$\tau \rightarrow \mu \mu \mu $ at a rate of one out of $10^{14}$ tau decays?}}, \href{https://doi.org/10.1140/epjc/s10052-020-8059-7}{\emph{Eur. Phys. J. C} {\bfseries 80} (2020) 506} [\href{https://arxiv.org/abs/1912.09862}{{\ttfamily 1912.09862}}].

\bibitem{Petrov:2013vka}
A.A.~Petrov and D.V.~Zhuridov, \emph{{Lepton flavor-violating transitions in effective field theory and gluonic operators}}, \href{https://doi.org/10.1103/PhysRevD.89.033005}{\emph{Phys. Rev. D} {\bfseries 89} (2014) 033005} [\href{https://arxiv.org/abs/1308.6561}{{\ttfamily 1308.6561}}].

\bibitem{Husek:2020fru}
T.~Husek, K.~Monsalvez-Pozo and J.~Portoles, \emph{{Lepton-flavour violation in hadronic tau decays and $\mu-\tau$ conversion in nuclei}}, \href{https://doi.org/10.1007/JHEP01(2021)059}{\emph{JHEP} {\bfseries 01} (2021) 059} [\href{https://arxiv.org/abs/2009.10428}{{\ttfamily 2009.10428}}].

\bibitem{Husek:2021isa}
T.~Husek, K.~Monsalvez-Pozo and J.~Portoles, \emph{{Constraints on leptoquarks from lepton-flavour-violating tau-lepton processes}}, \href{https://doi.org/10.1007/JHEP04(2022)165}{\emph{JHEP} {\bfseries 04} (2022) 165} [\href{https://arxiv.org/abs/2111.06872}{{\ttfamily 2111.06872}}].

\bibitem{Carpentier:2010ue}
M.~Carpentier and S.~Davidson, \emph{{Constraints on two-lepton, two quark operators}}, \href{https://doi.org/10.1140/epjc/s10052-010-1482-4}{\emph{Eur. Phys. J. C} {\bfseries 70} (2010) 1071} [\href{https://arxiv.org/abs/1008.0280}{{\ttfamily 1008.0280}}].

\bibitem{ref:hflav}
{\scshape Heavy Flavor Averaging Group (HFLAV)} collaboration, \emph{{Averages of $b$-hadron, $c$-hadron, and $\tau$-lepton properties as of 2023}},  \href{https://arxiv.org/abs/2411.18639}{{\ttfamily 2411.18639}}.

\bibitem{Belle:2010rxj}
{\scshape Belle Collaboration} collaboration, \emph{{Search for Lepton Flavor Violating tau- Decays into $\ell K^0_s$ and $\ell K^0_s K^0_s$}}, \href{https://doi.org/10.1016/j.physletb.2010.07.012}{\emph{Phys. Lett. B} {\bfseries 692} (2010) 4} [\href{https://arxiv.org/abs/1003.1183}{{\ttfamily 1003.1183}}].

\bibitem{ABASHIAN2002117}
A.~Abashian et~al., \emph{The belle detector}, \href{https://doi.org/https://doi.org/10.1016/S0168-9002(01)02013-7}{\emph{Nuclear Instruments and Methods in Physics Research Section A: Accelerators, Spectrometers, Detectors and Associated Equipment} {\bfseries 479} (2002) 117}.

\bibitem{Akai:2018mbz}
K.~Akai, K.~Furukawa and H.~Koiso, \emph{{SuperKEKB collider}}, \href{https://doi.org/10.1016/j.nima.2018.08.017}{\emph{Nucl. Instrum. Meth.} {\bfseries A907} (2018) 188} [\href{https://arxiv.org/abs/1809.01958}{{\ttfamily 1809.01958}}].

\bibitem{Abe:2010gxa}
{\scshape Belle II} collaboration, \emph{{Belle II technical design report}},  \href{https://arxiv.org/abs/1011.0352}{{\ttfamily 1011.0352}}.

\bibitem{Belle-IISVD:2022upf}
{\scshape Belle II SVD} collaboration, \emph{{The design, construction, operation and performance of the Belle~II silicon vertex detector}}, \href{https://doi.org/10.1088/1748-0221/17/11/P11042}{\emph{JINST} {\bfseries 17} (2022) P11042} [\href{https://arxiv.org/abs/2201.09824}{{\ttfamily 2201.09824}}].

\bibitem{Kotchetkov:2018qzw}
D.~Kotchetkov et~al., \emph{{Front-end electronic readout system for the Belle II imaging Time-Of-Propagation detector}}, \href{https://doi.org/10.1016/j.nima.2019.162342}{\emph{Nucl. Instrum. Meth. A} {\bfseries 941} (2019) 162342} [\href{https://arxiv.org/abs/1804.10782}{{\ttfamily 1804.10782}}].

\bibitem{Kurokawa:2001nw}
S.~Kurokawa and E.~Kikutani, \emph{{Overview of the KEKB accelerators}}, \href{https://doi.org/10.1016/S0168-9002(02)01771-0}{\emph{Nucl. Instrum. Meth. A} {\bfseries 499} (2003) 1}.

\bibitem{Banerjee:2007is}
S.~Banerjee, B.~Pietrzyk, J.M.~Roney and Z.~W\c{a}s, \emph{{Tau and muon pair production cross-sections in electron-positron annihilations at $\sqrt{s} = 10.58$ GeV}}, \href{https://doi.org/10.1103/PhysRevD.77.054012}{\emph{Phys. Rev. D} {\bfseries 77} (2008) 054012} [\href{https://arxiv.org/abs/0706.3235}{{\ttfamily 0706.3235}}].

\bibitem{Jadach:1999vf}
S.~Jadach, B.F.L.~Ward and Z.~W\c{a}s, \emph{{The precision Monte Carlo event generator KK for two-fermion final states in $e^+e^-$ collisions}}, \href{https://doi.org/10.1016/S0010-4655(00)00048-5}{\emph{Comput. Phys. Commun.} {\bfseries 130} (2000) 260} [\href{https://arxiv.org/abs/hep-ph/9912214}{{\ttfamily hep-ph/9912214}}].

\bibitem{Jadach_2000}
S.~Jadach, B.~Ward and Z.~W\c{a}s, \emph{The precision monte carlo event generator $\mathcal{KK}$ for two-fermion final states in $e^+e^-$ collisions}, \href{https://doi.org/10.1016/s0010-4655(00)00048-5}{\emph{Computer Physics Communications} {\bfseries 130} (2000) 260} [\href{https://arxiv.org/abs/hep-ph/9912214}{{\ttfamily hep-ph/9912214}}].

\bibitem{Jadach:1990mz}
S.~Jadach, J.H.~Kuhn and Z.~W\c{a}s, \emph{{TAUOLA: A library of Monte Carlo programs to simulate decays of polarized tau leptons}}, \href{https://doi.org/10.1016/0010-4655(91)90038-M}{\emph{Comput. Phys. Commun.} {\bfseries 64} (1990) 275}.

\bibitem{Barberio:1990ms}
E.~Barberio, B.~van Eijk and Z.~W\c{a}s, \emph{{PHOTOS: A universal Monte Carlo for QED radiative corrections in decays}}, \href{https://doi.org/10.1016/0010-4655(91)90012-A}{\emph{Comput. Phys. Commun.} {\bfseries 66} (1991) 115}.

\bibitem{Sjostrand:2014zea}
T.~Sj\"{o}strand, S.~Ask, J.R.~Christiansen, R.~Corke, N.~Desai, P.~Ilten et~al., \emph{{An Introduction to PYTHIA 8.2}}, \href{https://doi.org/10.1016/j.cpc.2015.01.024}{\emph{Comput. Phys. Commun.} {\bfseries 191} (2015) 159} [\href{https://arxiv.org/abs/1410.3012}{{\ttfamily 1410.3012}}].

\bibitem{Lange:2001uf}
D.J.~Lange, \emph{{The EvtGen particle decay simulation package}}, \href{https://doi.org/10.1016/S0168-9002(01)00089-4}{\emph{Nucl. Instrum. Meth.} {\bfseries A462} (2001) 152}.

\bibitem{Jadach:1991by}
S.~Jadach, E.~Richter-Was, B.F.L.~Ward and Z.~W\c{a}s, \emph{{Monte Carlo program BHLUMI-2.01 for Bhabha scattering at low angles with Yennie-Frautschi-Suura exponentiation}}, \href{https://doi.org/10.1016/0010-4655(92)90196-6}{\emph{Comput. Phys. Commun.} {\bfseries 70} (1992) 305}.

\bibitem{Balossini:2006wc}
G.~Balossini, C.M.~Carloni~Calame, G.~Montagna, O.~Nicrosini and F.~Piccinini, \emph{{Matching perturbative and parton shower corrections to Bhabha process at flavour factories}}, \href{https://doi.org/10.1016/j.nuclphysb.2006.09.022}{\emph{Nucl. Phys.} {\bfseries B 758} (2006) 227}.

\bibitem{Balossini:2008xr}
G.~Balossini, C.~Bignamini, C.M.C.~Calame, G.~Montagna, O.~Nicrosini and F.~Piccinini, \emph{{Photon pair production at flavour factories with per mille accuracy}}, \href{https://doi.org/10.1016/j.physletb.2008.04.007}{\emph{Phys. Lett.} {\bfseries B 663} (2008) 209}.

\bibitem{CarloniCalame:2003yt}
C.M.~Carloni~Calame, G.~Montagna, O.~Nicrosini and F.~Piccinini, \emph{{The BABAYAGA event generator}}, \href{https://doi.org/10.1016/j.nuclphysbps.2004.02.008}{\emph{Nucl. Phys. {\bf B} Proc. Suppl.} {\bfseries 131} (2004) 48}.

\bibitem{CarloniCalame:2001ny}
C.M.~Carloni~Calame, \emph{{An improved parton shower algorithm in QED}}, \href{https://doi.org/10.1016/S0370-2693(01)01108-X}{\emph{Phys. Lett.} {\bfseries B 520} (2001) 16}.

\bibitem{CarloniCalame:2000pz}
C.M.~Carloni~Calame, C.~Lunardini, G.~Montagna, O.~Nicrosini and F.~Piccinini, \emph{{Large angle Bhabha scattering and luminosity at flavor factories}}, \href{https://doi.org/10.1016/S0550-3213(00)00356-4}{\emph{Nucl. Phys.} {\bfseries B 584} (2000) 459}.

\bibitem{BERENDS1985421}
F.~Berends, P.~Daverveldt and R.~Kleiss, \emph{Radiative corrections to the process $e^+e^- \to e^+ e^-\mu^+ \mu^-$}, \href{https://doi.org/https://doi.org/10.1016/0550-3213(85)90540-1}{\emph{Nucl. Phys.} {\bfseries B 253} (1985) 421}.

\bibitem{BERENDS1985441}
F.~Berends, P.~Daverveldt and R.~Kleiss, \emph{Complete lowest-order calculations for four-lepton final states in electron-positron collisions}, \href{https://doi.org/https://doi.org/10.1016/0550-3213(85)90541-3}{\emph{Nucl. Phys.} {\bfseries B 253} (1985) 441}.

\bibitem{BERENDS1986285}
F.~Berends, P.~Daverveldt and R.~Kleiss, \emph{{Monte Carlo simulation of two-photon processes: II: Complete lowest order calculations for four-lepton production processes in electron-positron collisions}}, \href{https://doi.org/https://doi.org/10.1016/0010-4655(86)90115-3}{\emph{Comp. Phys. Commun.} {\bfseries 40} (1986) 285}.

\bibitem{Uehara:1996bgt}
S.~Uehara, \emph{{TREPS: A Monte-Carlo Event Generator for Two-photon Processes at $e^+e^-$ Colliders using an Equivalent Photon Approximation}},  \href{https://arxiv.org/abs/1310.0157}{{\ttfamily 1310.0157}}.

\bibitem{Kuhr:2018lps}
{\scshape Belle II Framework Software Group} collaboration, \emph{{The Belle II Core Software}}, \href{https://doi.org/10.1007/s41781-018-0017-9}{\emph{Comput. Softw. Big Sci.} {\bfseries 3} (2019) 1} [\href{https://arxiv.org/abs/1809.04299}{{\ttfamily 1809.04299}}].

\bibitem{basf2-zenodo}
{Belle II collaboration}, ``{Belle II Analysis Software Framework (basf2)}.'' \url{https://doi.org/10.5281/zenodo.5574115}.

\bibitem{Agostinelli:2002hh}
{\scshape GEANT4} collaboration, \emph{{GEANT4: A simulation toolkit}}, \href{https://doi.org/10.1016/S0168-9002(03)01368-8}{\emph{Nucl.Instrum.Meth.} {\bfseries A506} (2003) 250}.

\bibitem{Brun:1994aa}
R.~Brun, F.~Bruyant, F.~Carminati, S.~Giani, M.~Maire, A.~McPherson et~al., \emph{{GEANT Detector Description and Simulation Tool, CERN Report No. DD/EE/84-1 (1984)}}, .

\bibitem{Gelb:2018agf}
M.~Gelb et~al., \emph{{B2BII: Data Conversion from Belle to Belle II}}, \href{https://doi.org/10.1007/s41781-018-0016-x}{\emph{Comput. Softw. Big Sci.} {\bfseries 2} (2018) 9} [\href{https://arxiv.org/abs/1810.00019}{{\ttfamily 1810.00019}}].

\bibitem{Brandt:1964sa}
S.~Brandt, C.~Peyrou, R.~Sosnowski and A.~Wroblewski, \emph{{The Principal axis of jets. An Attempt to analyze high-energy collisions as two-body processes}}, \href{https://doi.org/10.1016/0031-9163(64)91176-X}{\emph{Phys. Lett.} {\bfseries 12} (1964) 57}.

\bibitem{Farhi:1977sg}
E.~Farhi, \emph{{A QCD Test for Jets}}, \href{https://doi.org/10.1103/PhysRevLett.39.1587}{\emph{Phys. Rev. Lett.} {\bfseries 39} (1977) 1587}.

\bibitem{Milesi:2020esq}
M.~Milesi, J.~Tan and P.~Urquijo, \emph{{Lepton identification in Belle II using observables from the electromagnetic calorimeter and precision trackers}}, \href{https://doi.org/10.1051/epjconf/202024506023}{\emph{EPJ Web Conf.} {\bfseries 245} (2020) 06023}.

\bibitem{ABASHIAN200269}
A.~Abashian, K.~Abe, K.~Abe, P.~Behera, F.~Handa, T.~Iijima et~al., \emph{Muon identification in the belle experiment at kekb}, \href{https://doi.org/https://doi.org/10.1016/S0168-9002(02)01164-6}{\emph{Nuclear Instruments and Methods in Physics Research Section A: Accelerators, Spectrometers, Detectors and Associated Equipment} {\bfseries 491} (2002) 69}.

\bibitem{Hanagaki_2002}
K.~Hanagaki, H.~Kakuno, H.~Ikeda, T.~Iijima and T.~Tsukamoto, \emph{Electron identification in belle}, \href{https://doi.org/10.1016/s0168-9002(01)02113-1}{\emph{Nuclear Instruments and Methods in Physics Research Section A: Accelerators, Spectrometers, Detectors and Associated Equipment} {\bfseries 485} (2002) 490–503}.

\bibitem{Krohn:2019dlq}
{\scshape Belle II Analysis Software Group} collaboration, \emph{{Global decay chain vertex fitting at Belle II}}, \href{https://doi.org/10.1016/j.nima.2020.164269}{\emph{Nucl. Instrum. Meth.} {\bfseries A976} (2020) 164269} [\href{https://arxiv.org/abs/1901.11198}{{\ttfamily 1901.11198}}].

\bibitem{ParticleDataGroup:2022pth}
{\scshape Particle Data Group} collaboration, \emph{{Review of Particle Physics}}, \href{https://doi.org/10.1093/ptep/ptac097}{\emph{PTEP} {\bfseries 2022} (2022) 083C01}.

\bibitem{Skwarnicki:1986xj}
T.~Skwarnicki, \emph{{A study of the radiative CASCADE transitions between the Upsilon-Prime and Upsilon resonances}}, Ph.D. thesis, Cracow, INP, 1986.

\bibitem{XGBoostPaper}
T.~Chen and C.~Guestrin, \emph{Xgboost: A scalable tree boosting system},  in \emph{Proceedings of the 22nd ACM SIGKDD International Conference on Knowledge Discovery and Data Mining}, KDD '16, (New York, NY, USA), pp.~785--794, ACM, 2016, \href{https://doi.org/10.1145/2939672.2939785}{DOI} [\href{https://arxiv.org/abs/1603.02754}{{\ttfamily 1603.02754}}].

\bibitem{Akiba:2019lwq}
T.~Akiba, S.~Sano, T.~Yanase, T.~Ohta and M.~Koyama, \emph{{Optuna: A Next-generation Hyperparameter Optimization Framework}},  \href{https://arxiv.org/abs/1907.10902}{{\ttfamily 1907.10902}}.

\bibitem{Punzi:2003bu}
G.~Punzi, \emph{{Sensitivity of searches for new signals and its optimization}}, {\emph{eConf} {\bfseries C030908} (2003) MODT002} [\href{https://arxiv.org/abs/physics/0308063}{{\ttfamily physics/0308063}}].

\bibitem{Belle:2023ziz}
{\scshape Belle} collaboration, \emph{{Search for lepton-flavor-violating \ensuremath{\tau} decays into a lepton and a vector meson using the full Belle data sample}}, \href{https://doi.org/10.1007/JHEP06(2023)118}{\emph{JHEP} {\bfseries 06} (2023) 118} [\href{https://arxiv.org/abs/2301.03768}{{\ttfamily 2301.03768}}].

\bibitem{ref:lumi_new}
{\scshape Belle II} collaboration, \emph{{Measurement of the integrated luminosity of data samples collected during 2019-2022 by the Belle II experiment}},  \href{https://arxiv.org/abs/2407.00965}{{\ttfamily 2407.00965}}.

\bibitem{ref:belle_lumi}
{\scshape Belle} collaboration, \emph{{Physics Achievements from the Belle Experiment}}, \href{https://doi.org/10.1093/ptep/pts072}{\emph{PTEP} {\bfseries 2012} (2012) 04D001} [\href{https://arxiv.org/abs/1212.5342}{{\ttfamily 1212.5342}}].

\bibitem{Banerjee_2008}
S.~Banerjee, B.~Pietrzyk, J.M.~Roney and Z.~W\c{a}s, \emph{Tau and muon pair production cross sections in electron-positron annihilations at $\sqrt{s}=10.58$ {GeV}}, \href{https://doi.org/10.1103/physrevd.77.054012}{\emph{Physical Review D} {\bfseries 77} (2008) } [\href{https://arxiv.org/abs/0706.3235}{{\ttfamily 0706.3235}}].

\bibitem{Junk:1999kv}
T.~Junk, \emph{{Confidence level computation for combining searches with small statistics}}, \href{https://doi.org/10.1016/S0168-9002(99)00498-2}{\emph{Nucl. Instrum. Meth. A} {\bfseries 434} (1999) 435} [\href{https://arxiv.org/abs/hep-ex/9902006}{{\ttfamily hep-ex/9902006}}].

\bibitem{Read:2002hq}
A.L.~Read, \emph{{Presentation of search results: The CL(s) technique}}, \href{https://doi.org/10.1088/0954-3899/28/10/313}{\emph{J. Phys. G} {\bfseries 28} (2002) 2693}.

\bibitem{pyhf}
L.~Heinrich, M.~Feickert and G.~Stark, ``{pyhf: v0.7.6}.''
\newblock 10.5281/zenodo.1169739.

\bibitem{pyhf_joss}
L.~Heinrich, M.~Feickert, G.~Stark and K.~Cranmer, \emph{pyhf: pure-python implementation of histfactory statistical models}, \href{https://doi.org/10.21105/joss.02823}{\emph{Journal of Open Source Software} {\bfseries 6} (2021) 2823}.

\end{thebibliography}\endgroup
